\documentclass[aps,prx,amsmath,nofootinbib,longbibliography,amssymb,superscriptaddress,twocolumn,10pt]{revtex4-2}

\usepackage[english]{babel}
\usepackage{graphicx}
\usepackage{ulem}
\usepackage{soul}
\usepackage{float} 
\usepackage[svgnames]{xcolor} 
\usepackage{comment}
\usepackage[colorlinks, linkcolor=NavyBlue , citecolor= NavyBlue,urlcolor = NavyBlue]{hyperref}
\usepackage{verbatim}
\usepackage{esint}
\usepackage{marginnote}

\renewcommand{\vec}[1]{\boldsymbol{#1}}
\newcommand{\RNum}[1]{\uppercase\expandafter{\romannumeral #1\relax}}

\newcommand{\evyatar}[1]{{\color{green}{\bf ET: #1}}}

\def \bea {\begin{eqnarray}}
\def \eea {\end{eqnarray}}

\def \tanh {\textnormal{tanh}}
\def \coth {\textnormal{coth}}

\begin{document}

\title{Solvable models of two-level systems coupled to itinerant electrons: {\\}Robust non-Fermi liquid and quantum critical pairing}

\author{Evyatar Tulipman}\thanks{These authors contributed equally to this work.}
\affiliation{Department of Condensed Matter Physics, Weizmann Institute of Science, Rehovot 76100, Israel}

\author{Noga Bashan}\thanks{These authors contributed equally to this work.}
\affiliation{Department of Condensed Matter Physics, Weizmann Institute of Science, Rehovot 76100, Israel}

\author{J\"{o}rg Schmalian}
\affiliation{Karlsruher Institut für Technologie, Institut f\"{u}r Theorie der Kondensierten Materie,  76049, Karlsruhe, Germany}

\affiliation{Karlsruher Institut für Technologie, Institut f\"{u}r Quantenmaterialien und Technologien,  76021, Karlsruhe, Germany}

\author{Erez Berg}
\affiliation{Department of Condensed Matter Physics, Weizmann Institute of Science, Rehovot 76100, Israel}

\date{\today}

\begin{abstract}

    Strange metal behavior is traditionally associated with an underlying putative quantum critical point at zero temperature. However, in many correlated metals, e.g., high-$T_c$ cuprate superconductors, strange metallicity persists at low temperatures over an extended range of microscopic parameters, suggesting the existence of an underlying \textit{quantum critical phase}, whose possible physical origins remain poorly understood. Systematic investigations of physical scenarios giving rise to such a critical, non-Fermi liquid (NFL) phase are therefore crucial to better understand this puzzling behavior. In a previous work \cite{bashan2023tunable}, we considered a solvable large-$N$ model consisting of itinerant electrons coupled to local two-level systems (TLSs) via spatially random interactions, inspired by the possibility of emergent metallic glassiness due to frustrated competing orders, and found that the system hosts an NFL phase with tunable exponents at intermediate couplings.
    In this work, we expand our investigation to the following: (i) We study the extent to which this NFL phase is generic by considering various deformations of our theory, including coupling of electrons to multiple operators of the TLSs and arbitrarily directed TLS-fields. We find that the physical picture obtained in \cite{bashan2023tunable} qualitatively persist in a wide region of parameter space, showcasing the robustness of the NFL phase; (ii) We analyze the superconducting instability due to coupling of TLSs to electrons, and find a rich structure, including quantum critical pairing associated with the NFL phase and conventional BCS pairing in the weak and strong coupling limits; (iii) We elaborate on the analysis of Ref.~\cite{bashan2023tunable}, including single-particle, transport and thermodynamic properties. 
\end{abstract}
\maketitle
\section{Introduction}
One of the central problems in condensed matter physics concerns the low-temperature anomalous normal-state transport properties of correlated metals such as high-$T_c$ cuprate superconductors and others \cite{varma_colloquium_2020,proust_remarkable_2019,bruin_similarity_2013,legros_universal_2019,cao_strange_2020}. A hallmark of the anomalous behavior is the linear-in-temperature scaling of the dc resistivity, known as `strange metal' behavior. Such behavior stands at odds with conventional Fermi-liquid theory, where a $T^2$-scaling is predicted, and is believed to indicate that quantum fluctuations are  so pronounced as to completely invalidate the Landau Fermi-liquid quasiparticle picture \cite{chowdhury_sachdev-ye-kitaev_2022,hartnoll_planckian_2021,phillips_stranger_2022,sachdev_quantum_2023}.


The traditional theoretical approach to describe strange metals and other non-Fermi liquids (NFLs)
involves coupling a Fermi surface to bosonic collective fluctuations of an order parameter, sometimes leading to NFL behavior when tuning the system to a quantum critical point (QCP) \cite{lohneysen_fermi-liquid_2007,varma_singular_2002,lee_recent_2018}. In this case, the NFL behavior manifests in a critical fan, emanating from a single (critical) point at $T\to 0$. Interestingly, while consistent with some materials, e.g. heavy fermion systems \cite{gegenwart_quantum_2008}, there are numerous examples, e.g. cuprates \cite{cooper_anomalous_2009,hussey_tale_2018,hussey_generic_2013,greene_strange_2020}  as well as twisted bilayer graphene, organic superconductors, and other systems~\cite{taillefer2010scattering,cao_strange_2020,jaoui_quantum_2022,ghiotto_quantum_2021,pfleiderer_non-fermi_2003,pfleiderer_partial_2004,doiron-leyraud_fermi-liquid_2003,paschen_quantum_2021,zhao_quantum-critical_2019}, where NFL behavior persists over an extended region of non-thermal parameters at $T\to0$ and thus cannot be ascribed to a single QCP. Rather, the extended NFL behavior raises the possibility of a quantum critical {\it phase}. Since no general prescription exists for this scenario, capturing such behavior within a controlled, physically motivated theory is considerably challenging.


 One potential route to realize an extended critical NFL phase requires an efficient source of scattering for itinerant electrons over an extended region of non-thermal parameters, e.g., by identifying a physical setting where ``critical'' low-energy excitations (i.e., with support at the lowest energies) exist. To this end, in a recent work \cite{bashan2023tunable}, we demonstrated that an NFL phase can arise when itinerant electrons are interacting with fluctuations of a metallic glass (e.g., charge or stripe glass), described as a collection of two-level systems (TLSs) that correspond to quasi-local collective excitations, analogously to the excitations in structural glasses\footnote{For a recent detailed analysis of the density of tunneling defects in metallic glasses, see Ref.~\cite{mocanu_microscopic_2023}.}. Physically, our theory is motivated by the complex phase diagrams of such strongly correlated materials that often host multiple frustrated, competing orders which can give rise to glassiness, even in the absence of impurity disorder \cite{schmalian_stripe_2000}. The presence of disorder could further stabilize such extended NFL behavior, as was observed in Ref.~\cite{patel2023localization}.



In fact, it has been long recognized that inelastic scattering of electrons off of local TLSs can result in a $T$-linear resistivity in the weak coupling limit \cite{black_resistivity_1979}. Nonetheless, going beyond weak coupling, the interaction with electrons may dramatically alter the properties of the TLSs, e.g., by renormalization of the bare TLS parameters or inducing inter-TLS correlations analogously to the Ruderman–Kittel–Kasuya–Yosida (RKKY) mechanism \cite{ruderman_indirect_1954,kasuya_theory_1956,yosida_magnetic_1957}. Clearly these effects can further affect the electrons themselves. 

To study the evolution of the physical picture beyond weak coupling, in Ref.~\cite{bashan2023tunable}, we considered a large-$N$ theory consisting of itinerant electrons interacting with local TLSs via spatially random couplings. We found that the non-Gaussian saddle point of the theory hosts a robust NFL phase with tunable exponents, which is not associated with quantum criticality; see Sec.~\ref{x-model section} for a summary of our findings. Roughly speaking, the electrons constitute an ohmic bath for the TLSs which results in a renormalization of the TLS energy splitting towards lower energies. At sufficiently strong couplings, a significant portion of the TLSs are renormalized to low energies, which in turn provides an efficient source of scattering for the electrons, resulting in an NFL behavior over a finite range of coupling strengths. 

In Ref.~\cite{bashan2023tunable}, we have considered the special case where electrons are coupled to a single operator of TLSs (i.e., $\sigma_x$ of spin-$1/2$ Pauli operators), which allowed us to focus on the effects of inelastic electron-TLS scattering in a simple setting. Specifically, we considered the case where electrons are coupled to $\sigma_x$ of the TLSs, which we shall refer to as the `$x$-model' henceforth. Aiming for a broader understanding, it is natural to ask whether the behavior of the $x$-model persist to the generic case, where electrons may couple to all three operators of the TLSs. Another important aspect of the physical picture concerns TLS-induced pairing, which is expected to take over at sufficiently low temperatures.  

In this work, we investigate a class of large-$N$ models, generalizing our study of the $x$-model of Ref.~\cite{bashan2023tunable} to more generic settings. We begin by considering the effect of coupling of spinless electrons to all operators of the TLSs. Importantly, the low-energy behavior is qualitatively identical to that of the $x$-model within a wide range of parameters, showcasing the robustness of the NFL phase found in Ref.~\cite{bashan2023tunable}. Towards a more realistic scenario, we generalize our analysis to spinful electrons and find a transition to a superconducting ground state due to TLS-induced pairing. Remarkably, the rich phenomenology of the normal state in our model is also manifested in the form of the critical temperature $T_{c}$, exhibiting various crossovers, where in particular the transition from the NFL phase assumes a quantum critical form. In addition to these key finding, we further study various aspects of the model, including transport and thermodynamic properties and $1/N$ corrections.

The structure of the paper is as follows. In Sec.~\ref{model} we present the model and discuss physically motivated choices of parameters. In Sec.~\ref{sum of res} we provide a brief summary of our main results. In Sec.~\ref{mappingtoTLSboson} we provide a mapping between our model to a set of decoupled spin-boson (SB) models. In Sec.~\ref{x-model section} we provide an extensive review of the $x$-model, previously studied in Ref.~\cite{bashan2023tunable}, which corresponds to the special case where the electrons interact only with the $\hat{x}$ component of the TLSs. In Sec.~\ref{multibath-model section} we analyze the general case where the electrons interact with all operators of the TLSs (dubbed the $xyz-$model), and in Sec.\ref{biased section}, to the most general case where the field acting on the TLS is allowed to point in arbitrary directions. In Sec.~\ref{transport section}, Sec.~\ref{thermodynamics section} and Sec.~\ref{superconductivity section} we use our results to study transport, thermodynamic and superconducting properties of the model, respectively. In Sec.~\ref{finite N section} we consider several finite-$N$ corrections.  Sec.~\ref{discussion section} contains a summary of our work and discuss possible implications and further directions.

\section{Model}
\label{model}

We consider the following Hamiltonian, defined on a $d$-dimensional hypercubic lattice: 
\begin{equation}
    H = \sum_{\boldsymbol{k};i=1}^{N} \left(\varepsilon_{\boldsymbol{k}} - \mu \right) c_{i\boldsymbol{k}}^{\dagger}c_{i\boldsymbol{k}}-\sum_{\boldsymbol{r};l=1}^{M}\boldsymbol{h}_{l,\boldsymbol{r}}\cdot\boldsymbol{\sigma}_{l,\boldsymbol{r}} + H_{\rm int},
    \label{Hamiltonian}
\end{equation}
where 
\begin{equation}
\label{H int}
    H_{\rm int} =  \sum_{\boldsymbol{r};i,j=1}^{N}\left[ \frac{1}{N^{1/2}}V_{ij,\boldsymbol{r}} + \frac{1}{N}\sum_{l=1}^{M}\vec{g}_{ijl,\boldsymbol{r}}\cdot \vec{\sigma}_{l,\boldsymbol{r}} \right] c_{i\boldsymbol{r}}^{\dagger}c_{j\boldsymbol{r}}.
\end{equation}
Each site contains $N$ 
electronic ``orbitals'' $i=1,...,N$, and $M$ species of TLSs $l=1,...,M$ where $\vec{\sigma}_{l,\boldsymbol{r}}$ is a vector of spin-$1/2$ Pauli operators at position $\boldsymbol{r}$. $\varepsilon_{\boldsymbol{k}}$ and $\mu$ denote the electronic dispersion and chemical potential, respectively, and are assumed to be diagonal  and $i-$independent in orbital space. For simplicity, we consider spinless fermions. We will reintroduce the spin index later when we discuss superconductivity. 
Thus, electrons create a dynamic local field that acts on the TLSs while, at the same time, they scatter off those local degrees of freedom. 

Each TLS is subjected to a field 
$\boldsymbol{h}_{l,\boldsymbol{r}}=(h_l^{x},0,h_l^{z})$, with $h^z$ being the asymmetry and $h^x$ the tunneling rate between the two states associated with each TLS, both taken 
to be independent random variables drawn from a probability distribution $\mathcal{P}_{\beta_{a}}(h^a)$ with $a=x,z$. Below, we refer to this as the physical basis for $\boldsymbol{h}$. Note that $h^y \equiv 0$ in order to respect time-reversal symmetry. We focus on power-law distributions, $\mathcal{P}_{\beta_{a}}(h^a)\propto (h^a)^{\beta_a}$,
supported on the interval $0<h^a<h^a_c$. It suffices to consider positive fields as the sign of $h^a$ can be absorbed into  the definition of $\sigma_z$. $h^a_c$ denote the TLS bare bandwidth and $\beta_a>-1$
is a tunable parameter. The generalization to other distributions is straightforward. In some cases, it will be convenient to rotate to the eigenbasis where $\vec{h} = h \hat{z}$, with
$h=\sqrt{(h^x)^2+(h^z)^2}$ being the energy splitting of the TLS. We call this the diagonal basis of $\boldsymbol{h}$.

While we present throughout this work general results for arbitrary $\beta_x$ and $\beta_z$, there are two physically motivated choices for the splitting distributions.
Working in the ``physical'' basis (i.e. the eigenbasis of $\sigma_z$), we assume that the distribution of the level asymmetry, $h^z$, has finite weight around $h^z=0$ \cite{phillips_tunneling_1972}, which corresponds to a uniform distribution with $\beta_z = 0$. It is then natural to consider the cases where the width of the distribution of the tunneling rate, $h^x$, is either comparable or negligible compared to the asymmetry, corresponding to $\beta_x = 0, h_c^x=h_c^z\equiv h_c$, or setting $h_c^x \equiv 0$, respectively. In the latter the diagonal basis coincides with the physical basis, while in the former, 
changing to the diagonal basis results in a linear distribution of the eigenvalues $h$, i.e. $\beta_z=1$.

The couplings $\vec{g}_{ijl,\boldsymbol{r}}=\left({g}^{x}_{ijl,\boldsymbol{r}},{g}^{y}_{ijl,\boldsymbol{r}},{g}^{z}_{ijl,\boldsymbol{r}}\right)$ are taken to be uncorrelated Gaussian random variables with zero mean and variance $g_a^2$.  For $a=x,z$, we  consider real-valued couplings with 
\begin{eqnarray}
    \overline{g^a_{ijl,\boldsymbol{r}} g^{a}_{i'j'l',\boldsymbol{r}'}} = g_a^2 \delta_{\boldsymbol{r},\boldsymbol{r}'} \delta_{ll'}(\delta_{ii'}\delta_{jj'} + \delta_{ij'}\delta_{ji'}), 
\end{eqnarray}
and to ensure that $H$ is time reversal symmetric, $g^y$ must be purely imaginary with
\begin{eqnarray}     \overline{g^y_{ijl,\boldsymbol{r}} g^{y}_{i'j'l',\boldsymbol{r}'}} = -g_y^2 \delta_{\boldsymbol{r},\boldsymbol{r}'} \delta_{ll'}(\delta_{ii'}\delta_{jj'} - \delta_{ij'}\delta_{ji'}).
\end{eqnarray} 
The $g_a$ are all real valued. The different components are uncorrelated, i.e. $\overline{g^a g^{a'}}=0$ if $a\neq a'$.
$\overline{(\cdot)}$ denotes averaging over realizations of the coupling constants.

\begin{figure*}[t]
\centering

 \includegraphics[width=2\columnwidth]{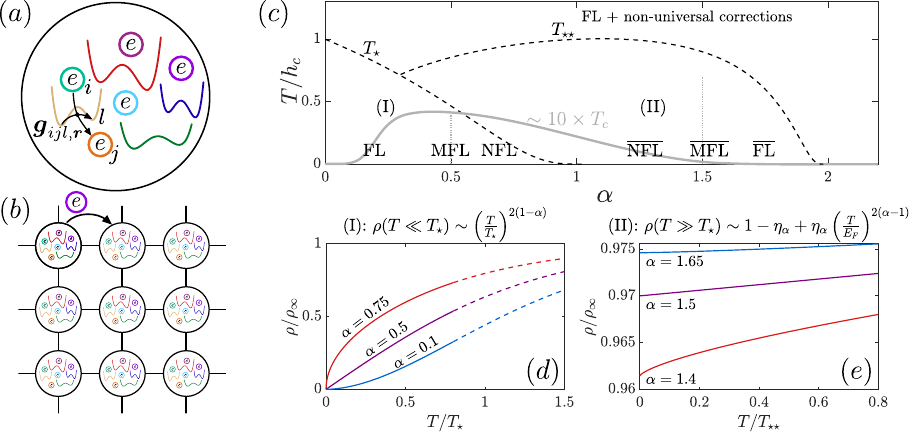}

\caption{ \textbf{(a,b)} Illustration of the lattice model Eqs.~(\ref{Hamiltonian},\ref{H int}). (a) a unit cell containing a large number of electronic states and local two-level systems interacting via random couplings; (b) the electrons hop between unit cells.  \textbf{(c)} Phase diagram of the $x-$model in $\alpha - T$ plane ($\alpha = \alpha_x$ is the dimensionless coupling strength) for a linear splitting distribution ($\beta=1$). Region (I), defined by $T\ll T_{\star} \sim h_{c,R}$, $h_{c,R}$ being the renormalized cutoff of the TLS-splitting, is characterized by the leading inelastic scattering exponent $2(1-\alpha)$ as manifested in the dc resistivity, see {(d)}; the system crosses over from a Fermi-liquid (FL) for $\alpha<1/2$, to a marginal Fermi-liquid (MFL) for $\alpha = 1/2$, and a non Fermi-liquid (NFL) for $1/2<\alpha <1$. At $\alpha \approx 1$, the TLSs undergo a freezing transition at $T=0$. In Region (II), for $\alpha\gtrsim 1$ at finite $T$, scattering off of the TLSs is mainly elastic with small inelastic corrections that scale $\sim T^{2(\alpha -1)}$, namely, another set of crossovers from NFL ($1<\alpha<3/2$) to MFL ($\alpha=3/2$) to FL ($\alpha>2$); see {(e)}.  Region (II) is defined up to $T\sim T_{\star\star}$, where $T_{\star\star}$ is the scale at which standard FL behavior becomes dominant. The gray line denotes the transition temperature to the superconducting state, $T_{c}$, in the spinful version of the $x$-model. \textbf{(d,e)} Resistivity as a function of $T$ for regions (I) and (II), respsectively, where $\rho_\infty$ is the resistivity due to saturated classical TLSs. Note that, in region (II), the $\alpha$-dependent coefficient $\eta_{\alpha} \propto (h_c/E_F)^2 \ll 1$, corresponding to the weak $T$-variation of the dc resistivity.
}
\label{fig:main figure}
\end{figure*}
%
Similarly, the onsite potential disorder, $V_{ij,\boldsymbol{r}}$, is normally distributed with zero mean and variance $V^2$. 
 Note that setting the couplings $g_x$ and $g_z$ to be uncorrelated in the physical basis or in the eigenbasis is not equivalent in the cases where both $h^x,h^z>0$. Here we first consider simple variants of the model where $g_x$ and $g_z$ are uncorrelated in the diagonal basis (i.e. setting $h^x \equiv0$) and later show that the qualitative physical picture persists when they are uncorrelated in the physical basis (with $h^x \ne 0$).

The Fermi energy, $E_F$, sets the largest energy scale in theory, and also corresponds to the cutoff energy of the electronic bath, traditionally denoted by $\omega_c (= E_{F})$ in the spin-boson literature \cite{leggett_dynamics_1987,costi_thermodynamics_1999,weiss_quantum_2012}. 

The TLS bandwidth satisfies $h_c \ll E_F$,
and we restrict the on-site disorder strength, $V^2$, such that $\Gamma = 2\pi \rho_F V^2 \ll E_F$ ($\Gamma$ being the elastic scattering rate), therefore considering `good metals.' We do not restrict the interaction strengths $g_a$, namely, our study covers the range from weak to strong coupling. We focus on the low-energy limit of the model, defined by $\omega,T\ll h_{c,R}$, where $
h_{c,R}$ is the renormalized cutoff of the TLSs, to be defined below.

The dimensionless coupling parameters that will be used in the following are related to the interaction strengths by ($a=x,y,z$):
\begin{eqnarray}
    \alpha_a &=& \frac{\rho_F^2g_a^2}{\pi^2}, \\
    \lambda_a &=& \frac{M}{N}\frac{\rho_F g_a^2}{h_{c,R}}.
\end{eqnarray}
The parameters $\alpha_a$ (defined in accordance to the spin-boson literature  conventions) represent the strength of the dissipation acting on the TLSs, while $\lambda_a$ quantify the strength of the scattering of electrons by TLSs at low energies.

Throughout this work we consider the limit  $N,M\to\infty$ with a fixed ratio $M/N$.
We will see below that the limit $M\to\infty$ enables us to (i) reduce the electron's self-energy to a summation over rainbow diagrams containing only two-point correlation functions of the TLSs, which is not clear a priori as Wick's theorem does not hold for the TLSs; and (ii) invoke self-averaging of the TLSs, such that sums over the TLS flavors can be replaced by averages over the splitting distribution: $\sum_{l=1}^M f(h_l)\to M\int f(h)\mathcal{P}_\beta(h)dh$. Importantly, since the splitting distribution is independent of position, the self-averaging assumption translates to statistical translation invariance of the model. Note further that the $N\to \infty$ limit is essential for the mapping of our model to the SB model, where the bosonic bath coupled to the TLSs is composed of particle-hole pairs, see Sec.~\ref{mappingtoTLSboson}. 

\section{Brief summary of results}
\label{sum of res}

In the following sections we expound on the properties of different variants of the model. However, for the benefit of the reader, we first briefly outline the key conclusions of our work.  We first describe the physical picture of the $x$-model and then show that this picture qualitatively persists to generic variants of the model. 

\textbf{Normal state.} Consider the normal state properties at low $T$, corresponding to regions (I) and (II) in Fig.~\ref{fig:main figure}. In region (I), as the dimensionless coupling $\alpha=\alpha_x$ is increased, the system crosses over from a FL, MFL, and NFL, up to a critical value $\alpha \approx 1$ where the TLS freeze at $T=0$. These regimes are defined by the exponent of the single-particle scattering rate: $\Sigma''(\omega) - \Sigma''(0)\propto |\omega|^{\gamma}$, $\gamma (\alpha,\beta)=(1+\beta)(1-\alpha)$ (shown for $\beta=1$ in Fig.~\ref{fig:main figure}). This is also manifested in the dc resistivity: $\rho - \rho_0 \propto T^\gamma$. In region (II), the TLSs are frozen at $T=0$, such that scattering off of TLS is mainly elastic. At finite $T$ for $\alpha\gtrsim 1$, however, residual quantum fluctuations of the TLSs provides a source for inelastic scattering, leading to an additional sequence of NFL-MFL-FL crossovers with an inelastic scattering exponent $2(\alpha - 1)$. This residual contribution corresponds to a weak $T$-variation of the dc resistivity as shown in Fig.~\ref{fig:main figure}. The behavior in the critical region $\alpha \approx 1$  and $T\to 0$, separating regions (I) and (II), is more involved and show logarithmic $T$-dependence of the single-particle scattering rate (and the dc resistivity) that smoothly interpolates between the two regions.  

\textbf{Superconductivity.} Considering the model with spinful electrons, a superconducting transition occurs below a critical temperature $T_{c}$ due to TLS-induced pairing; see Fig.~\ref{fig:main figure}. Interestingly, $T_{c}$ is a non-monotonic function of the coupling $\alpha_x$, with remarkably rich pairing phenomenology. Specifically, at intermediate couplings (corresponding, e.g., to the NFL phase of region (I) in Fig.~\ref{fig:main figure}) $T_{c}$ assumes an algebraic, quantum critical scaling form, and, as the coupling is further increased beyond a certain threshold (but still at intermediate values), crosses over to an Allen-Dynes-like, strong coupling form \cite{AllenDynes1975}. In addition, $T_{c}$ assumes the standard BCS form at weak coupling, but also at very strong coupling (e.g., for $\alpha > 3/2$ in Fig.~\ref{fig:main figure}), which corresponds to pairing due to the residual quantum fluctuations of the nearly frozen spins.  

\begin{figure}[t]
\includegraphics[width=1\columnwidth]{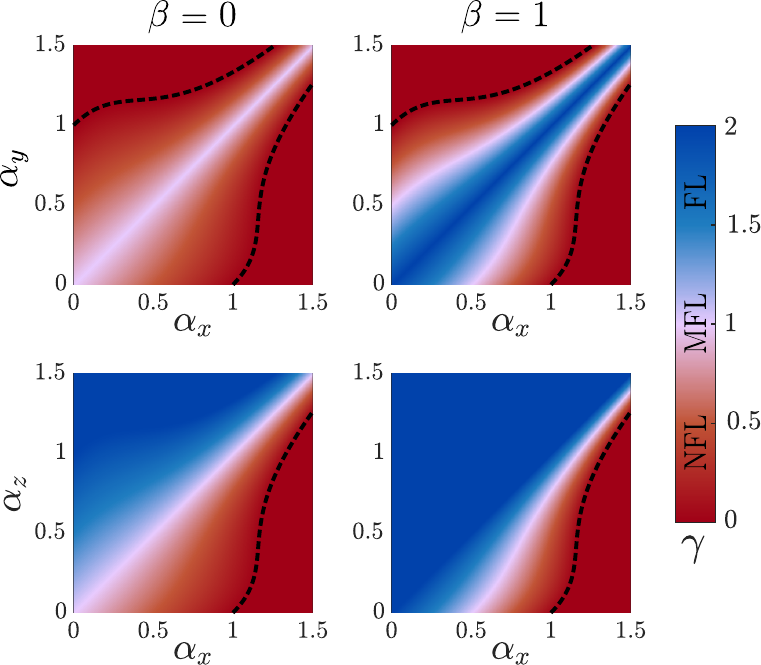}
\caption{$T=0$ normal-state phase diagram of the $xy$-model (top) and $xz$-model (bottom), for constant ($\beta=0$) and linear ($\beta=1$) TLS splitting ditributions. The color represents the exponent in the low-$T$ behavior of the resistivity: $\rho-\rho_0\propto T^\gamma$ (analogous to region (I) in Fig.~\ref{fig:main figure}). The dashed line denotes the BKT transition over which the TLSs freeze, and beyond which there is a non-universal version of the $\overline{\rm NFL}$ phase (similarly to region (II) of Fig.~\ref{fig:main figure} for $\alpha_x > 1$).
}
\label{fig:main figure multibath}
\end{figure}

\textbf{Robustness.} To test the extent to which the physical picture of the $x$-model is generic, we allow for interactions with other operators of the TLSs. In Fig.~\ref{fig:main figure multibath}, we show how the normal state $T = 0$ phase diagram of the $x$-model changes upon introducing coupling to the $y$ (top row) and $z$ (bottom row) operators of the TLSs for constant and linear TLS-splitting distributions (left and right columns, respectively). We shall refer to these variants as the $xy$- and $xz$-models.

 In all cases, the qualitative behavior of the $x$-model persists provided that the largest coupling is orthogonal to the direction of the TLS-field (i.e. to $h_z\sigma_z$), namely, the characterizing exponent $\gamma(\vec{\alpha})$ varies from $1+\beta$ to $0$ as the dominant coupling is increased, up to a critical value at which the TLSs freeze. This sequence of crossovers corresponds to region (I) in Fig.~\ref{fig:main figure}, while the residual crossovers of region (II) are expected to qualitatively change for sufficiently strong perturbations due to non-universal corrections. Note that when the dominant coupling is parallel to the TLS-field (i.e., when $\alpha_z>\alpha_x$), the TLSs are essentially static and the system shows Fermi-liquid behavior with TLS-induced elastic scattering along with weak FL-like corrections. 



\allowdisplaybreaks{
\section{Mapping to Spin-Boson model}
\label{mappingtoTLSboson}

In this section, we use an effective action approach to map our theory to the SB model. We set $V^2=0$ for simplicity. An alternative diagrammatic derivation of the mapping is given in App.~\ref{diagrammatic mapping}. 

We begin by considering the spin coherent-state path integral representation for the TLSs. The partition function is given by 
\begin{eqnarray}
    Z[\boldsymbol{h},\boldsymbol{g}]= \int \mathcal{D}[\boldsymbol{\sigma},c,\overline{c}] e^{-S},
\end{eqnarray}
with the action, $S=S_0 + S_{\rm int}$:
\begin{eqnarray}
    S_0 &=& \sum_{\boldsymbol{r}}\sum_{l=1}^{M}S_{\rm Berry}[\boldsymbol{\sigma}_{l,\boldsymbol{r}}] - \sum_{\boldsymbol{r}}\sum_{l=1}^{M} \int_{\tau} \boldsymbol{h}_{l,\boldsymbol{r}} \cdot \boldsymbol{\sigma}_{l,\boldsymbol{r}} \nonumber\\&+& \sum_{i=1}^{N}\sum_{\boldsymbol{k}}\int_{\tau} \overline{c}_{i\boldsymbol{k}}(\partial_\tau +\varepsilon_{\boldsymbol{k}}- \mu) c_{i\boldsymbol{k}},  \\
    S_{\rm int} &=& \frac{1}{N} \sum_{\boldsymbol{r}}\sum_{i,j=1}^{N}\sum_{l=1}^{M}\int_{\tau} \vec{g}_{ijl,\boldsymbol{r}}\cdot \vec{\sigma}_{l,\boldsymbol{r}}  \overline{c}_{i\boldsymbol{r}}c_{j\boldsymbol{r}}.
\end{eqnarray}
Here we kept the same symbols $\boldsymbol{\sigma}_{l,\boldsymbol{r}}$ for the unit vectors that result from the coherent state representation of the Pauli operators. $S_{\rm Berry}$ denotes the Berry's phase of the TLSs, see e.g. \cite{auerbach_interacting_1998}. 

To proceed, we average over the random couplings using the replica method and introduce the bilocal fields 
\begin{eqnarray}
    G_{\vec{r},\vec{r}'}(\tau,\tau') = \frac{1}{N}\sum_i \overline{c}_{i\vec{r}}(\tau)c_{i\vec{r}'}(\tau') , \label{Gfermion} \\
    \chi_{a,\vec{r}}(\tau,\tau') = \frac{1}{M}\sum_l \sigma_{l,\vec{r}}^{a}(\tau)\sigma_{l,\vec{r}}^{a}(\tau'). \label{Dtls}
\end{eqnarray}
The constraints, \eqref{Gfermion} and \eqref{Dtls}, are enforced via conjugated fields, $\Sigma$ and $\Pi$, respectively. Notice that, for now, we are considering spinless fermions. In this case, there is no pairing instability to leading order in $1/N$. Later on, we shall consider a model of spinful fermions, where the anomalous part of the Green's function must be considered, and an instability towards a superconducting state with an intra-flavor, on-site order parameter occurs~\cite{esterlis_cooper_2019}.

To proceed, we integrate over the fermions and substitute a replica-diagonal \textit{Ansatz}, which allows us to express the partition function as 
$Z[\vec{h}]=\int\mathcal{D}\left[G,\vec{\chi},\Sigma,\vec{\Pi},\vec{\sigma} \right]e^{-S_{\rm eff}}$, where the effective action is given by
}
\begin{widetext}
\begin{eqnarray}
   S_{\rm eff}&=&-N\text{Tr}\ln\left(G_{0}^{-1}-\Sigma\right)-N\int_{\tau,\tau'}\sum_{\boldsymbol{r},\boldsymbol{r}'}\sum_{\sigma}G_{\boldsymbol{r},\boldsymbol{r}'}\left(\tau,\tau'\right)\Sigma_{\boldsymbol{r},\boldsymbol{r}'}\left(\tau,\tau'\right)+\frac{M}{2}\int_{\tau,\tau'}\sum_{\boldsymbol{r},a}\chi_{a,\boldsymbol{r}}\left(\tau,\tau'\right)\Pi_{\boldsymbol{r}}^{a}\left(\tau',\tau\right)\nonumber\\&+&\frac{M}{2}\int_{\tau,\tau'}\sum_{\boldsymbol{r}}\sum_{a}g_{a}^{2}G_{\boldsymbol{r}}\left(\tau,\tau'\right)G_{\boldsymbol{r}}\left(\tau',\tau\right)\chi_{a,\boldsymbol{r}}\left(\tau,\tau'\right)+\sum_{\boldsymbol{r}}\sum_{l=1}^{M}S_{\text{Berry}}\left[\boldsymbol{\sigma}_{l,\boldsymbol{r}}\right]-\int_{\tau}\sum_{\boldsymbol{r}}\sum_{l=1}^{M}\boldsymbol{h}_{l,\boldsymbol{r}} \cdot \boldsymbol{\sigma}_{l,\boldsymbol{r}}\nonumber\\&-&\frac{1}{2}\int_{\tau,\tau'}\sum_{\boldsymbol{r},a}\Pi_{\boldsymbol{r}}^{a}\left(\tau',\tau\right)\sum_{l=1}^{M}\sigma_{l,\boldsymbol{r}}^{a}\left(\tau\right)\sigma_{l,\boldsymbol{r}}^{a}\left(\tau'\right).
\end{eqnarray}
\end{widetext}
In the limit of large $M$ and $N$, with fixed ratio $M/N$, we can
analyze the problem in the saddle point limit. Performing the variation with respect
to $G$ and $\Sigma$ gives
\begin{equation}
\Sigma_{\boldsymbol{r},\boldsymbol{r}'}\left(\tau\right)=\delta_{\boldsymbol{r},\boldsymbol{r}'}\frac{M}{N}\sum_{a}g_{a}^{2}G_{\boldsymbol{r},\boldsymbol{r}}\left(\tau\right)\chi_{a,\boldsymbol{r}}\left(\tau\right)   
\label{eq:self_saddle_1}
\end{equation}
as well as
\begin{equation}
G_{\boldsymbol{r},\boldsymbol{r}'}\left(i\omega\right)=\left.\left(G_{0}^{-1}\left(i\omega\right)-\Sigma\left(i\omega\right)\right)^{-1}\right|_{\boldsymbol{r},\boldsymbol{r}'}.
\label{eq:saddleDyson}
\end{equation}
Here, we have used thermal equilibrium to write the saddle-point equations  with time-translation-invariant  correlation functions and their Fourier transforms.
In addition, the stationary point that follows from the variation
with respect to $\chi$ is
\begin{eqnarray}
\Pi_{a,\boldsymbol{r}}\left(\tau\right) & = & -g_{a}^{2}G_{\boldsymbol{r},\boldsymbol{r}}\left(\tau\right)G_{\boldsymbol{r},\boldsymbol{r}}\left(-\tau\right).
\label{eq:saddlePi}
\end{eqnarray}
The Berry phase term $S_{\rm Berry}$, that reflects the fact that no Wick theorem exists for Pauli operators, implies that  the TLSs cannot be simply integrated over as a Gaussian integral. However,  it allows us to recast the TLS problem to that of $M$ decoupled TLSs per site $\boldsymbol{r}$, 
$\sum_{\boldsymbol{r},l=1}^{M}S_{\boldsymbol{r},l}\left[\boldsymbol{\sigma}_{\boldsymbol{r},l}\right]$, coupled to a bosonic bath of particle-hole excitations.
Each TLS is governed by the spatially local effective action
\begin{widetext}
\begin{eqnarray}  S_{\boldsymbol{r},l}\left[\boldsymbol{\sigma}\right]=S_{{\rm Berry}}\left[\boldsymbol{\sigma}\right]-\int_{\tau}\boldsymbol{h}_{l,\boldsymbol{r}}\cdot\boldsymbol{\sigma}\left(\tau\right)-\int_{\tau,\tau'}\Pi_{\boldsymbol{r}}\left(\tau'-\tau\right)\sigma^{a}\left(\tau\right)\sigma^{a}\left(\tau'\right).
\label{effectiveTLSaction}
\end{eqnarray}
\end{widetext}
 This is indeed the action of the spin-boson model after the bosonic bath degrees of freedom have been integrated out~\cite{weiss_quantum_2012}. The latter give rise to the non-local in time coupling $\Pi_{\boldsymbol{r}}^{a}\left(\tau',\tau\right)$ that is, in general, different for each site.  Of course, in our problem the  origin of the bath function are not bosons but the conduction electrons. For the solution of this local problem this makes, however, no difference. 
 $S_{\boldsymbol{r},l}$ still depends on the random configuration $\boldsymbol{h}_{l,\boldsymbol{r}}$ of the fields.

For a given realization of the fields $\boldsymbol{h}_{l,\boldsymbol{r}}$
the problem is not translation invariant and correlation functions like $\left\langle \sigma_{l,\boldsymbol{r}}^{a}\left(\tau\right)\sigma_{l,\boldsymbol{r}}^{a}\left(0\right)\right\rangle $ fluctuate in space. However, 
to determine the self energy in Eq.~\eqref{eq:self_saddle_1} we only  need to know the average $\chi_{a,\vec{r}}(\tau)$ of this correlation function over the $M$ flavors. 
To proceed we assume that the model is self-averaging in the $M\to \infty$ limit, such that sums over the TLS flavors can be replaced with averaging over the TLS splitting distribution ($\sum_{l=1}^{M} \to M\int \mathcal{P}\left(\vec{h}_{\boldsymbol{r}}\right) d\vec{h}_{\boldsymbol{r}}$). Since the splitting distribution is independent of position, the self-averaging assumption translates to a statistical translation invariance of the model, at least for the average  of interest. Hence, $\chi_{a,\boldsymbol{r}}(\tau)=\chi_{a}(\tau)$ is independent on $\boldsymbol{r}$. The same must then hold for the bath function $\Pi_{a,\boldsymbol{r}}(\tau)=\Pi_{a}(\tau)$. From the  saddle point equations \eqref{eq:saddlePi} it follows that  the local fermionic Green's function and through Eq.~\eqref{eq:saddleDyson} the self energy are both space independent. 
Hence we can go to momentum space and find that
the theory is goverened by a momentum-independent self-energy and the Dyson equation for the electrons read
\begin{eqnarray}    
\Sigma\left(\tau\right)&=&\frac{M}{N}\sum_{a} g^{2}_{a}\chi_{a}\left(\tau\right)G\left(\tau\right),\\
G_{\boldsymbol{k}}\left(i\omega\right)&=&\frac{1}{i\omega-\varepsilon_{\boldsymbol{k}}-\Sigma\left(i\omega\right)},
\label{Dyson}
\end{eqnarray}
where $G\left(\tau\right)=\int_{\boldsymbol{k}}G_{\boldsymbol{k}}\left(\tau\right)$ is the local Green's function. 
For a momentum independent fermionic self energy we obtain in the limit of large electron bandwidth
\begin{eqnarray}
G\left(i\omega\right)=\int_{\boldsymbol{k}}G_{\boldsymbol{k}}\left(i\omega\right)\approx-i\pi\rho_F{\rm sgn}\left(\omega\right).
\label{GF_av}
\end{eqnarray}
The particle-hole correlation function can now be evaluated. We find
\begin{equation}
    \Pi_a(\omega) = \frac{\rho_F^2 g_a^2 }{2\pi}|\omega|,
\end{equation} 
irrespective of the electronic self-energy. We thus conclude that each TLS is coupled to an Ohmic bath of particle-hole excitations that is independent of the back reaction of the TLS on the electronic degrees of freedom. This is a consequence of the fact that  the $\Pi_a$ are independent of $\Sigma$. Thus, we have shown that the (spatially local) TLS-correlator
\begin{eqnarray}
    \chi_a\left(\tau-\tau'\right)&=&\frac{1}{M}\sum_{l}\left\langle \sigma_{l}^{a}\left(\tau\right)\sigma_{l}^{a}\left(\tau'\right)\right\rangle, 
    \label{DysonTLS}
\end{eqnarray}
is determined by the behavior of $M$ decoupled SB models. 

The strategy of the solution of our model in the large-$N$ limit is therefore: 
(i) solve the spin-boson problem  with ohmic bath for a given realization of the field $\boldsymbol{h}$, (ii) average over the TLS distribution function of the fields, and (iii) use the resulting propagator $\chi_a(\omega)$ of the TLSs to determine the fermionic self energy from Eq.\eqref{Dyson}. The non-linear character of the problem is rooted in the rich physics of the spin-boson problem, along with the averaging over the distribution functions of the fields $\boldsymbol{h}$.
Given the importance of the spin-boson model for our analysis we will give a summary of this model in the next section.

\section{The $x-$model}
\label{x-model section}

We begin with a review of the solution of the model for the simplest case where the electrons interact only with $\sigma_x$ of the TLSs' pseudospins, i.e., we are working in the diagonal basis and setting $g^z,g^y=0$, as in Ref.~\cite{bashan2023tunable}. 
This special case allows for a more transparent discussion of the key steps of our analysis. In addition, we will see that the more general problem reduces in many cases to this model in the limit of sufficiently low energies. 
In terms of the mapping provided in Sec.~\ref{mappingtoTLSboson}, the model is mapped into the SB model with one bath. Throughout this section, since $\alpha_y,\alpha_z=0$ we will use the notation $\alpha=\alpha_x$ for simplicity.

\subsection{A simple view of the physical picture for $\alpha < 1$}


\label{simple view section}
Consider first the weak coupling limit, $\alpha \to 0$, where the effect of interactions can be studied perturbatively. To leading order in $g^2$ the decay rate of an electron with energy $\omega$ is proportional 
to the amount of TLSs at accessible energies, namely, 
\begin{eqnarray}
\Sigma''\left(\omega\right)\propto \int_0^\omega \mathcal{P}_\beta \left(h \right) dh \propto |\omega|^{1+\beta},  
\end{eqnarray}
 where $\Sigma''$ denotes the imaginary part of the electronic retarded self energy ($F''$ will denote the imaginary part of the retarded function $F$ throughout the paper). For $\beta=0$ this weak coupling analysis yields marginal Fermi liquid behavior. On the other hand, for any $\beta>0$, i.e. for distribution functions that vanish  for $h\rightarrow 0$ one only finds Fermi liquid behavior.
 Increasing the strength of the interaction modifies the behavior of the TLSs in two main aspects: it renormalizes the energy splitting of each TLS, such that $h_\ell\to h_R(h_\ell)$; and broadens the TLS spectral function. The broadening has negligible effect on the frequency dependence of $\Sigma''$ (at sufficiently low energies). However, as we show in detail below,
 the renormalization of the energy splitting leads to a renormalization of the bare TLS-splitting distribution, decreasing its exponent from $\beta$ to $\beta - (1+\beta)\alpha$.
Thus, increasing the interaction strength transfers the spectral weight of the TLSs towards lower energies (at the limit $\alpha \to 1$, the spectral weight is pushed to zero energy, signaling a BKT transition of the TLSs to a localized phase). 
Consequently, the naive perturbative argument will hold for the renormalized splitting distribution, resulting in a tunable exponent as a function of $\alpha$.

\subsection{Summary of the single-bath spin-boson model}
\label{SBreview}
The single-bath spin-boson model (1bSB), or the Caldeira-Leggett model \cite{caldeira_influence_1981,weiss_quantum_2012}, is given by the Hamiltonian (note that the commonly used convention in the SB literature swaps $\sigma_x\leftrightarrow\sigma_z$ relative to our convention)
\begin{eqnarray} 
    H_{\text{1bSB}} &=& -h\sigma_z + g\sigma_x\phi + H_\phi. \
    \label{1bSB Hamiltonian}
\end{eqnarray}

The bosonic field $\phi=\sum_i (a_i+a_i^\dagger)$ is to be interpreted in terms of a bath of oscillators whose spectral function, dictated by the Hamiltonian $H_
\phi$, is assumed to be of power law form below some high energy cutoff $\omega_c$:
$\Pi(\omega\ll \omega_c)\propto |\omega|^s$. The cases where $s<1,s=1,s>1$ are respectively called the subohmic, ohmic, and superohmic baths. Throughout our work we will exclusively be interested in the ohmic case $\Pi(\omega) = \frac{\pi}{2}\alpha\omega$, using this as the definition of the dimensionless coupling constant $\alpha$. For extensive reviews, see, e.g., Refs.~\cite{leggett_dynamics_1987,weiss_quantum_2012}.

Historically, this model was proposed as a toy model for the study of quantum dissipation and decoherence \cite{caldeira_quantum_1983}.
For an ohmic bath, it was found that
the spin gradually loses its coherence as $\alpha$ is increased and becomes overdamped (in terms of the one-point function $\langle \sigma_x(t) \rangle$)  beyond $\alpha=1/2$ \cite{leggett_dynamics_1987,costi_thermodynamics_1999}. At $\alpha=1$, the spin undergoes a Berezinskii-Kosterlitz-Thouless (BKT) phase transition after which it becomes localized in one of the two $\hat{x}$ states \cite{leggett_dynamics_1987,weiss_quantum_2012}.

\begin{figure}[t]
\centering
\includegraphics[width=1\columnwidth]{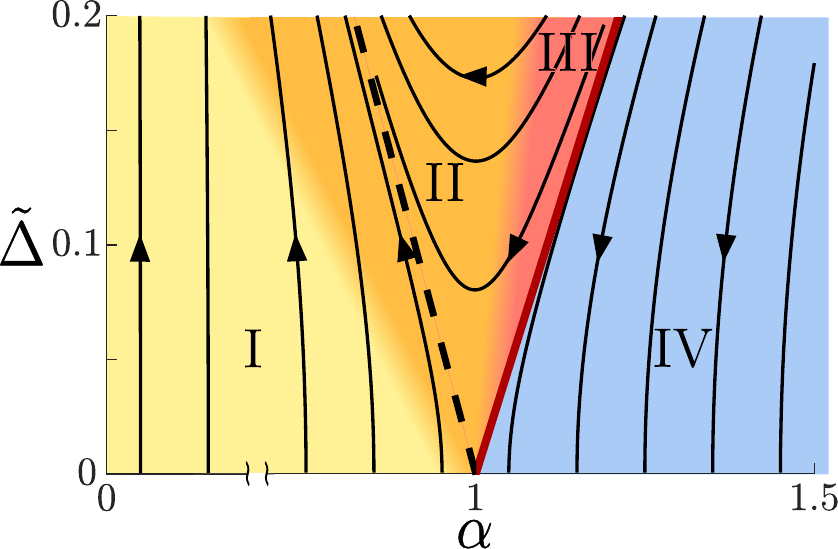}
\caption{\textbf{Schematic flow diagram of the 1bSB model.} We present the RG flow of Eqs.~\eqref{RGflow1} and \eqref{RGflow2}. The solid red line is the BKT speratrix between the dynamic phase ($\rm{I - III}$) and the localized phase ($\rm{IV}$), and the dashed black line represents the set of points which map into the isotropic Kondo model. For our purposes, we separate the dynamic phase into subregions according to the functional dependence of the renormalized scale $ h_R$ on the bare $\tilde{h}$. In region $\rm{I}$, the flow is nearly vertical (the beta function of $\alpha$ is small), and the renormalization of $ h_R$ is power law like. In region $\rm{II}$, the flow behaves similarly to that of the isotropic Kondo model, and so the renormalized scale  is exponential, with a weakly varying prefactor in the exponent. Finally, in region $\rm{III}$ the flow slows down significantly due to the vicinity to the BKT transition, and the renormalized scale is exponential and depends on the distance from the transition, vanishing exactly on the critical line. 
}
\label{fig:BKT flow}
\end{figure}

For our purposes, the most important corollary is that in the delocalized regime ($\alpha<1$), the TLS-splitting, $h$, is renormalized due to the high-frequency modes of the bath which must adjust to different positions whenever the $h\sigma_z$ term attemps to flip the TLS between the two $\hat{x}$ states (similarly to the Frank-Condon effect of electron-phonon coupling). This renormalization process, along with the BKT transition, are governed by the 
beta functions of $\alpha$ and the rescaled splitting $\tilde{h}\equiv h/\omega_c$, which, to  order $
\tilde{h}^2$, are given by
\begin{eqnarray}
    \frac{d\alpha}{d\ell} &=& -\alpha\tilde{h}^2,\label{RGflow1}\\
    \frac{d\tilde{h}}{d\ell} &=& (1-\alpha) \tilde{h},
    \label{RGflow2}
\end{eqnarray}
where $e^\ell$ is the renormalization group rescaling factor. The flow dictated by Eqs.~\eqref{RGflow1} and \eqref{RGflow2} on the $\alpha-\tilde{h}$ plane is shown in Fig.~\ref{fig:BKT flow}: 
For $\alpha$ near $1$ there exists a constant of flow, $x\equiv(1-\alpha)^2-\tilde{h}^2$, such that the BKT seperatrix corresponds to the rightmost of the two $x=0$ lines, with the localized (strong coupling) phase to the right of it.
The effective energy scales, 
 i.e. the renormalized splittings, $ h_R$, in the different regions of the phase diagram are given by 
\begin{equation}  
 h_R = c_\alpha \omega_c \times 
\left\{ \begin{array}{ccc}    \left(\frac{h}{\omega_c}\right)^{\frac{1}{1-\alpha}} & \rm{I} & {h}/{\omega_c}\ll (1-\alpha) \\
    e^{-b\frac{\omega_c}{h}} & \rm{II} & \begin{array}{c}
         {h}/{\omega_c}\approx (1-\alpha) \, {\rm{ or }}\,  \\
          {h}/{\omega_c}\gg|1-\alpha|
    \end{array}   \\
    e^{-\frac{\pi}{\sqrt{|x|}}} & \rm{III} & h/\omega_c \to \alpha-1 \\
    0 & \rm{IV} & \rm{else} 
\end{array}
\right.    
\label{Delta_r_renorm}
\end{equation}
where $c_
\alpha$ is a numerical prefactor which cannot be determined solely from the RG equations (but can be extracted using exact techniques such as bosonization or Bethe \textit{Ansatz} \cite{hur_entanglement_2008,camacho_exact_2019,filyov_method_1980,ponomarenko_resonant_1993}), and $b=b(\alpha,x)$ is a slowly varying function whose value is of order $1$ away from the BKT transition (both are given explicitly in App.~\ref{Appendix: rg 1bsb}).
The localization of the TLSs in region $\rm{IV}$ is due to the fact that the effective tunneling $ h_R$ between the the two $\hat{x}$ states
 flows to zero in that region.

 Using this information on the effective low-energy theory, we now turn to the study of correlation functions. As we apply the RG process, the operator $\sigma_x$ remains supported at low energies, such that correlation functions of $\sigma_x$ depend solely on parameters of the low-energy theory: $\alpha$ and $ h_R$, and not explicitly on $h$ or $\omega_c$. That is, at zero temperature, the $x$-susceptibility can be expressed in terms of a one-parameter scaling form \cite{costi_equilibrium_1996},
\begin{eqnarray}   
\chi_{x}''\left(\omega\right)&\equiv&  \frac{1}{\omega}f_\alpha\left(\frac{\omega}{ h_R}\right),
\label{eq: scaling form omega/Delta_r}
\end{eqnarray}
with $ h_R$ given by \eqref{Delta_r_renorm}, and $f_\alpha(x)$ being an $\alpha$-dependent scaling function. This relation between the frequency and the $ h_R$ dependence of the susceptibility will be crucial for obtaining exact results in the rest of this section.
Unlike $\sigma_x$, the correlation functions of the operators $\sigma_z,\sigma_y$ contain substantial spectral weight at high energies (of order $E_F$), and hence cannot be reduced to a single-parameter scaling form as in Eq.~\eqref{eq: scaling form omega/Delta_r}, and will depend on $\omega_c$ explicitly\footnote{The difference between $\sigma_x$ and $\sigma_y,\sigma_z$ is due to the fact that after integrating out the fast modes of the bath, the operators $\sigma_y,\sigma_z$ in the emergent low energy theory are highly dressed polaronic versions of the bare ones which move the high-energy bath modes along with the TLS. However, since the bath couples to the TLS in the $x-$basis, integrating it out does not dress $\sigma_x$, and thus the ``high energy" and ``low energy" $\sigma_x$ operators coincide. See \cite{guinea_dynamics_1985} for more details.}. This fact will be important when discussing the subleading contributions to the electronic self-energy in Sec.~\ref{multibath-model section}.

Another approach to the solution of the 1bSB model is via a mapping to the anisotropic Kondo model (AKM) or the resonant level model (RLM) \cite{anderson_exact_1970,guinea_bosonization_1985,leggett_dynamics_1987,sassetti_correlation_1990}. Remarkably, at $\alpha=1/2$, known as the ``Toulouse Point," the model maps to the non-interacting point of the RLM, and is thus exactly solvable. At this point the effective energy scale is $ h_R =\pi h^2/4\omega_c$, and the  temperature dependent susceptibility is given by
\begin{eqnarray}  \chi_x''\left(\omega,T\right)=\frac{4}{\pi^2}\frac{ h_R}{\omega^{2}+4 h_R^{2}}\left[\text{Im}\Phi\left(\omega,T\right)-\frac{2 h_R}{\omega}\text{Re}\Phi\left(\omega,T\right)\right], \nonumber \\
\label{TPchix}
\end{eqnarray}
where 
\begin{eqnarray} \Phi\left(\omega,T\right)=\psi\left(\frac{1}{2}+\frac{h_{R}}{2\pi T}\right)-\psi\left(\frac{1}{2}+\frac{h_{R}}{2\pi T}-i\frac{\omega}{2\pi T}\right)
\end{eqnarray} 
($\psi(z)$ being the Digamma function).
At zero temperature the scaling form takes the exact form 
\begin{equation}
       f_{1/2}\left(x\right)=\frac{4}{\pi^2}\frac{1}{x^{2}+4}\left[x\tan^{-1}\left(x\right)+\ln\left(1+x^{2}\right)\right].
\end{equation}

Having summarized the essential properties of the SB model, we shall now turn to evaluate the electronic properties of the $x-$model using the fact that each TLS is equivalent to a spin with a randomly distributed splitting $h$. Note also that the scaling behavior at $T=0$ can be extended to sufficiently small $T$ (to be defined below), as can be verified explicitly in the Toulouse point and at weak coupling, which enables us to estimate the finite-$T$ properties of the model in the following.

\subsection{Averaged TLS-susceptibility and electronic self energy}
\label{1bSB averaging section}

Once we have determined the TLS susceptibility $\chi_i\left(\omega\right)$, we can determine the fermionic self energy from Eq.\eqref{Dyson}. 
We compute the imaginary part of the retarded self energy  using 
\begin{eqnarray}
\Sigma''\left(\omega\right) & = & \frac{M}{N}\sum_{a=x,y,z}g_{a}^{2}\int_{-\infty}^{\infty}\frac{d\nu}{2\pi}\chi''_{a}\left(\nu\right)G''_{R}\left(\omega+\nu\right)\nonumber \\
 & \times & \left(\coth\left(\frac{\nu}{2T}\right)-\tanh\left(\frac{\nu+\omega}{2T}\right)\right).
\end{eqnarray}
If we use Eq.~\eqref{GF_av} for the fermion propagator we obtain that at $T=0$ (here $g_y=g_z=0$ and $g_x=g$)
\begin{eqnarray}
\Sigma''\left(\omega\right)=-\frac{M}{N}\rho_{F}g^{2}\int_{0}^{\omega}d\nu\left(\frac{1}{M}\sum_{i=1}^{M}
\chi_{x,i}''\left(\nu\right)\right).
\label{selfenergyfor averaging}
\end{eqnarray}
Since each of the $M$ TLSs contributing to this sum has a randomly distributed splitting $h_i$, the self energy of each electron will be a randomly distributed variable. However, since $M\gg1$ the central limit theorem guarantees that this random variable will be normally distributed, with width of order $M^{-1/2}$. We can thus replace this random variable by its expectation value (we will revisit this assumption, and consider when it breaks down for large but finite $M$, at the end of this section):
\begin{eqnarray}
    \frac{1}{M}\sum_{i=1}^{M}
\chi_{x,i}''\left(\omega\right) \to \overline{\chi_x''}(\omega) = \int \mathcal{P}_\beta(h) \chi_x''(\omega,h) dh.
\end{eqnarray}
Our task thus reduces to the calculation of the TLS susceptibility averaged over the distribution of splittings $\mathcal{P}_\beta(h)$. We note however that the distribution which is of interest to us is not that of the bare splittings, but of the renormalized splittings $ h_R$. We can instead define the renormalized distribution
\begin{equation}
    \mathcal{P}_r( h_R) = \left(\frac{d h_R}{dh}\right)^{-1}\mathcal{P}_\beta(h),
\end{equation}
which is nonzero up to the renormalized bandwidth $h_{c,R} \equiv  h_R(h_c)$.
Combined with Eq.~\eqref{eq: scaling form omega/Delta_r}, we may write the averaged susceptibility as 
\begin{equation}
    \overline{\chi_x''}(\omega) =   {\rm sgn}(\omega)\int_{|\omega|/h_{c,R}}^\infty \mathcal{P}_r\left(\frac{|\omega|}{x}\right) \frac{f_\alpha\left(x\right)}{x^2}dx.
    \label{average susceptibility recipe}
\end{equation}
We will now evaluate  this integral for the different functional forms of $ h_R(h)$ corresponding to the regions \rm{I-IV} in Fig.~\ref{fig:BKT flow} at zero temperature. 

\subsubsection{$\alpha<1-h_c/E_F$}
\label{alphalesonesubsc}

In region \rm{I} of Fig.~\ref{fig:BKT flow}, where the flow of $\alpha$ is weak, $ h_R$ is a power of $h$. Thus the effect of the renormalization would be to alter the exponent in the distribution:
\begin{equation}
    \mathcal{P}_r( h_R) =  \frac{\gamma h_R^{\gamma-1}}{h_{c,R}^\gamma}, \quad 
    \gamma \equiv (1+\beta)(1-\alpha).
    \label{Pr a<1}
\end{equation}
For energy scales far below the renormalized cutoff $\omega \ll h_{c,R}$, the result would not be sensitive to the exact form of the scaling function $f_\alpha$, and we retrieve results similar to those presented in the perturbative argument in Sec.~\ref{simple view section}, albeit with a modified exponent:
\begin{eqnarray}
    \overline{\chi_x''}(\omega) = {\rm sgn}(\omega) A_\alpha\gamma\frac{|\omega|^{\gamma-1}}{h_{c,R}^\gamma}, 
    \label{average susc gamma}
\end{eqnarray}
with $A_\alpha\equiv\int_{\omega/h_{c,R}}^\infty \frac{f_\alpha(x)}{ x^{\gamma+1}}dx$. 
Since the scaling function  $f_\alpha(x\to 0)\propto x^2$ 
(corresponding to the universal $1/t^2$ decay of the real-time correlation function at late times \cite{guinea_dynamics_1985,leggett_dynamics_1987,weiss_quantum_2012}
), we may continue the lower limit of the integral to $0$ (provided $\gamma<2$), such that $A_{\alpha} = \int_{0}^\infty \frac{f_\alpha(x)}{ x^{\gamma+1}}dx + \mathcal{O}\left(\left(\frac{|\omega|}{h_{c,R}}\right)^\gamma\right)$ \footnote{This result follows from dimensional analysis: If the upper integration limit can be continued to $\infty$, the only place where an energy scale appears is in the normalization of the distribution, $1/h_{c,R}^\gamma$, which sets the frequency dependence since the susceptibility is dimensionless.}.

Remarkably, observe that at low energies, the leading frequency dependence of the response of the whole collection of TLSs 
(which is the effective degree of freedom coupled to the electrons) 
is independent of broadening effects of the individual TLSs. Rather, it is governed solely by the renormalized distribution $\mathcal{P}_r$, while the functional form of the susceptibility of each individual TLS $f_\alpha(x)$ 
is absorbed into the
prefactor $A_\alpha$\footnote{Note that the low-frequency dependence of $\chi_x''$ serves as a ``cutoff'' for $\overline{\chi_x''}$: The exponent in the averaged susceptibility does not exceed that of the individual susceptibilities, namely, $\overline{\chi_x''}\propto |\omega|$ for all $\gamma>2$. This is equivalent to the cases where the integral in $A_\alpha$ diverges at the lower limit, resulting in a modification of the frequency dependence.}.  Hence, for $\omega\ll h_{c,R}$, we find that
\begin{eqnarray}    \Sigma''\left(\omega\right)&=&-A_\alpha \frac{M}{N}\rho_{F}g^{2}\left|\frac{\omega}{h_{c,R}}\right|^\gamma \\ &=&-\lambda A_\alpha h_{c,R}\left|\frac{\omega}{h_{c,R}}\right|^\gamma.
\label{alpha<1 self energy}
\end{eqnarray} 
We see that the self-energy depends on the parameters $\alpha$ and $\beta$ only via $\gamma$. In particular, for any initial $\beta\geq0$, the self energy realizes a MFL form
upon tuning the coupling to $\alpha=\frac{\beta}{1+\beta}$, and realizes any NFL exponent $\gamma<1$ by increasing $\alpha$ towards $1$. Note, however, that as we increase the coupling $\alpha$, the effective TLS bandwidth $h_{c,R}$ decreases such that smaller NFL exponents are restricted to narrower low-energy intervals; see Fig.~\ref{fig:main figure}.

The temperature dependence of $ \Sigma''$ at low $T$ and zero frequency follows from similar considerations. The contribution of an individual TLS to the self energy can be written as a scaling function:
\begin{eqnarray}
    \Sigma_1(\omega=0,T,h)=-\rho_F g^2 \frac{M}{N} f_\Sigma(T/h_R).
\end{eqnarray}
We can thus perform the averaging over $h_R$ at this stage, and analogously to Eq.~\eqref{average susceptibility recipe} we find that 
\begin{equation}
    \Sigma''\left(\omega=0,T\right)=A'_\alpha \frac{M}{N}\rho_Fg^{2}\left(\frac{T}{h_{c,R}}\right)^\gamma,
\end{equation}
with $A'_{\alpha} = \gamma \int_{0}^\infty \frac{f_\Sigma(x)}{ x^{1+\gamma}}dx$. Since $f_\Sigma(x\ll1)\propto x^2$ and $f_\Sigma(x\to \infty) \to 1$, this is well defined for $0<\gamma<2$. This suggests an $\omega/T$ scaling  of the form $\Sigma''\propto \max{\left(|\omega|,T\right)}^\gamma$.

\subsubsection{$\alpha>1+h_c/E_F$}
\label{alphagrtonesubsc}

For $\alpha>1+h_c/E_F$ all the TLSs have undergone the BKT transition and are in the localized phase, where the dominant TLS contribution at $T=0$ is an elastic scattering term. However, residual quantum fluctuations of the TLSs at finite frequencies provide a weak inelastic scattering mechanism which becomes the leading contribution to the temperature dependence of the dc resistivity.
The finite frequency behavior follows from scaling considerations (App.~\ref{Appendix 3-2a}, \cite{guinea_dynamics_1985}) and is given by 
\begin{eqnarray}
\label{alpha>1 self energy}
    \overline{\chi''_x} (\omega)&=& (1-\eta_\alpha)\delta(\omega) + 2\eta_\alpha (\alpha-1) \frac{E_F^{2-2\alpha} }{|\omega|^{3-2\alpha}} \\
    \eta_\alpha &=& \frac{2\alpha(1+\beta) }{(\alpha-1)(3+\beta)}\left(\frac{h_c}{E_F}\right)^2\ll 1
\end{eqnarray}
with $\eta_\alpha \propto (h_c/E_F)^2\ll 1$. We thus find in this regime that the leading inelastic contribution to the self energy is of the form 
\begin{equation}
    \Sigma''(\omega) = -\rho_F g^2 \frac{M}{N}\left(1-\eta_\alpha+ \eta_\alpha \left(\frac{|\omega|}{E_F}\right)^{2\alpha-2}\right).
\end{equation}
The low energy excitations of the system are thus those of a NFL for $1<\alpha<3/2$, a MFL at $\alpha=3/2$ and FL for $\alpha>3/2$. However, note that unlike in the regime $\alpha<1$, here the elastic contribution is much larger than the inelastic.
Note that while this behavior persists up to a large energy scale (a fraction of $E_F$), it is expected to be the dominant contribution to the self energy only below an energy scale of order $1/\alpha^{4-2\alpha}\left(h_c/E_F\right)^{\frac{\alpha-1}{2-\alpha}}$, defined as the scale at which conventional FL-like corrections to resistivity become comparable (i.e. assuming $\rho$ contains an additional $T^2/E_F$ contribution).


\subsubsection{$\alpha \approx 1$}
\label{alphaequalonesubsec}
In the ``critical" region $|1-\alpha|<h_c/E_F$ the above descriptions are not valid, since the flow of $h$ slows down and becomes comparable to the flow of $\alpha$. This slowdown leads to a logarithmic behavior of the self energy, which interpolates between the regimes described by Eq.~\eqref{alpha<1 self energy} and Eq.~\eqref{alpha>1 self energy}.
As an example of the behavior in region $\rm{II}$ of Fig.~\ref{fig:BKT flow}, we consider specifically the case $\alpha=1$. 
The renormalized scale is given by $ h_R = c_1 \omega_c \exp\left(-\frac{\pi\omega_c}{2h}\right)$, and as a result the renormalized distribution becomes logarithmic:
\begin{equation}
    \mathcal{P}_r( h_R) = \frac{(1+\beta)\log^{1+\beta}\left(\frac{\omega_c}{h_{c,R}}\right)}{ h_R\log^{2+\beta}\left(\frac{\omega_c}{ h_R}\right)},
    \label{a=1 dist}
\end{equation}
where we ignore subleading corrections (in $\log(\omega_c/h_{c,R})$), related to the prefactor $c_1$; see App.~\ref{Appendix: alpha=1}. Inserting Eq.~\eqref{a=1 dist} into Eq.~\eqref{average susceptibility recipe}, we obtain that
\begin{equation}
    \overline{\chi_{x}''}(\omega) =\frac{1}{\omega}(1+\beta)\log^{1+\beta}\left(\frac{\omega_c}{h_{c,R}}\right)\int_{\frac{|\omega|}{h_{c,R}}}^\infty \frac{f_1(x)}{x\log^{2+\beta}\left(\frac{\omega_c}{|\omega|}x\right)}dx.
    \label{Self energy alpha=1}
\end{equation}
Since the function $f_1(x)$ must decay faster than $1/x$ for $x\gg1$ (due to the sum rule Eq.~\eqref{sum rule 1}), and also since  $|\omega|/\omega_c\ll|\omega|/h_{c,R}$, we may ignore the $x$ inside the log in the denominator and then, as in the previous case, continue the lower limit of integration to $0$. The resulting self energy is given by
\begin{equation} \Sigma''\left(\omega\right)=-\frac{M}{N}\frac{\rho_Fg^{2}}{2}\left(\frac{\log\left(\frac{\omega_c}{h_{c,R}}\right)}{\log\left(\frac{\omega_c}{|\omega|}\right)}\right)^{1+\beta}.
\label{Sigma alpha=1}
\end{equation} Note that when $\alpha$ is not exactly $1$, the only difference would be that the factor of $\pi/2$ in the exponent of $ h_R$ will vary slightly. Repeating the above calculations, this will only alter the value of $h_{c,R}$, but not the functional form of the self energy.

For $1<\alpha<1+h_c/E_F$, we must distinguish the TLSs into those which are dynamical ($h/E_F>\alpha-1$), and those which are localized/frozen ($h/E_F\leq\alpha-1$). The contribution of the dynamical ones will be similar to that of the $\alpha=1$ case (shown explicitly in App.~\ref{Appendix: BKT seperatrix}, the exact result is somewhat more involved, but maintains the logarithmic form of \eqref{Sigma alpha=1}), while the frozen ones will contribute an elastic scattering term to leading order (i.e. a delta function peak around $\omega=0$), plus higher order Fermi-liquid terms which we ignore. Defining $m_\alpha \equiv \frac{\alpha-1}{h_c/\omega_c}$ as the fraction of frozen TLSs, the self energy will include both a constant elastic contribution along with the NFL contribution attained earlier. For simplicity, setting $\beta=1$ we find that the self energy will be:

\begin{equation} \Sigma''\left(\omega\right)=-\frac{M}{N}\rho_Fg^{2}\left(m_\alpha + \left(1-m_\alpha\right) B_\alpha\left(\frac{\log\left(\frac{\omega_c}{h_{c,R}}\right)}{\log\left(\frac{\omega_c}{|\omega|}\right)}\right)^{2}\right).
\end{equation}
As $\alpha$ approaches $1+h_c/\omega_c$ from below, both the relative weight, $1-m_\alpha$, of the inelastic contribution, as well as the energy scale, $h_{c,R}$, vanishes.


\section{The $xyz-$model}
\label{multibath-model section}
We consider a generalized variant of the model, where we allow electron-TLS coupling in arbitrary directions, i.e., $g_a^2 > 0$ for $a=x,y,z$ (keeping the field $\vec{h}$ parallel to the $z$ direction), which we dub `$xyz-$model'. Remarkably, we will show that throughout much of the parameter space (of $g_x-g_y-g_z$), the behavior is qualitatively similar to that of the $x-$model, namely, increasing the couplings will generically drive the model towards a BKT transition, leading to a tunable exponent in the electronic self-energy which depends on the distance from the transition.

To proceed we recall that the electronic self energy in the multi-channel model is given by
\begin{equation}
    \Sigma''(\omega) = - \frac{M}{N} \sum_{a=x,y,z} \rho_F g_a^2\int_0^\omega \overline{\chi''_a}(\nu) \frac{d\nu}{2\pi}.
\end{equation}
As before, the TLSs are decoupled such that the dynamics of each TLS are determined by solving an independent SB model coupled to three ohmic baths.
The corresponding multibath SB model (mbSB) for a single TLS is
\begin{equation}  
H_{\text{mbSB}} = -h\sigma_z +  \sum_{a=x,y,z} g_a\sigma^a_x\phi_a + \sum_{a=x,y,z}H_{\phi_a},
\label{mbSB model}
\end{equation}
where the bosonic field of \eqref{1bSB Hamiltonian} is generalized to three fields, $\phi_x,\phi_y,\phi_z$, corresponding to three independent baths which couple to the three spin directions. In our case, all three baths are ohmic and have the same cutoff (because $\omega_c = E_F$), such that the interaction strengths are measured via the three dimensionless couplings, $\alpha_{a}\equiv \rho_F^2 g_a^2/\pi^2$, $a=x,y,z$.   
Let us point out that the model \eqref{mbSB model} has two high-symmetry points: the $U(1)$ symmetric point, corresponding to $\alpha_x=\alpha_y$; and the $SU(2)$ symmetric point corresponding to $h=0,\alpha_x=\alpha_y=\alpha_z$ (for relevant works see, e.g., Refs.~\cite{zarand_quantum_2002,castro_neto_quantum_2003,novais_frustration_2005}). These points will not be of particular interest to us, as they are both unstable fixed points (see below) and require fine tuning.

Similarly to the 1bSB model, the low-energy properties of the model can be obtained from an analysis of the RG flow. Here, the RG equations can be derived perturbatively in two out the of three couplings; while one of the couplings, $\alpha_a$, is allowed to be arbitrarily large, the RG equations are valid to linear order in $\alpha_{b\ne a}$ \cite{zarand_quantum_2002,castro_neto_quantum_2003,novais_frustration_2005}. Loosely speaking, the RG analysis is valid near the axes in the $(\alpha_x,\alpha_y,\alpha_z)$-coordinate system.
The beta functions for the mbSB model are given by
\begin{eqnarray}  \frac{d\tilde{h}}{d\ell} &=& (1-\alpha_x-\alpha_y)\tilde{h}, \\
    \frac{d\alpha_a}{d\ell} &=& -\left(2\sum_{b\neq a}\alpha_b + \left(1-\delta_{az}\right)\tilde{h}^2\right)\alpha_a.
    \label{multibath RG}
\end{eqnarray}
In order to simplify the analysis, we first discuss the cases where only two of the baths are active (by setting $\alpha_z=0$ or $\alpha_y=0$), treating one bath as dominant and the other as a perturbation. We then generalize the discussion to the case where all three baths are active, where, as we will show, the physical pictures can be essentially reduced to the simplified cases (of the two active baths). Further, as the model does not contain any other stable fixed points, we expect that the RG analysis will qualitatively capture the physics for all couplings.

It is useful to think about the RG of the multi-bath cases as a two-step process: the first step describes the `fast' flow of the couplings, where the dominant bath assumes a weakly renormalized coupling while the other irrelevant baths flow to weak coupling; in the second step, the baths renormalize the TLS-fields according to the renormalized couplings. We now proceed to analyze the different cases, where the usefulness 2-step perspective becomes apparent.



\subsection{The $xy-$model}
\begin{figure}[t]
\centering
\includegraphics[width=0.45\textwidth]{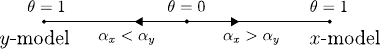} 
\caption{ Initial flow of the couplings in the $xy$-model. $\theta=\left(\tfrac{\alpha_x - \alpha_y}{\alpha_x + \alpha_y}\right)^2$  is the bath-anisotropy parameter. 
}
\label{fig:theta}
\end{figure}

We start by setting $\alpha_z=0$, i.e., the case where there are two active baths acting in the direction perpendicular to the field. The beta functions are given by
\begin{eqnarray} \frac{d\tilde{h}}{d\ell} &=& (1-\alpha_x-\alpha_y)\tilde{h}, \\ 
    \frac{d\alpha_x}{d\ell}&=&-2\alpha_y\alpha_x-\tilde{h}^2\alpha_x, \\
    \frac{d\alpha_y}{d\ell}&=&-2\alpha_x\alpha_y-\tilde{h}^2\alpha_y.\label{RG2bSB} 
\end{eqnarray}
It is insightful to define the bath anisotropy parameter, 
\begin{eqnarray}
    \theta\equiv \left(\frac{\alpha_x-\alpha_y}{\alpha_x+\alpha_y}\right)^2,
\end{eqnarray}
whose beta function is 
\begin{equation}
    \frac{d\theta}{d\ell} = (\alpha_x+\alpha_y) \theta (1-\theta).
\end{equation} 
The anisotropy is thus relevant whenever the couplings are not finely tuned to the $U(1)$ symmetric point $\theta=0$, and flows towards the maximally anistropic case $\theta=1$, i.e. where the larger of the two couplings dominates and the other one becomes irrelevant, as depicted in Fig.~\ref{fig:theta}. As we will now see, since the subdominant bath is irrelevant it can be integrated out easily, leading to a low energy description similar to that of the $x-$model with renormalized coupling.

Consider the case $\alpha_x\gg\alpha_y$. We focus on the regime $1-\alpha_x\gg \tilde{h}^2$ (i.e., far enough from the BKT transition), where simple analytical estimations can be made as the effect of $\tilde{h}$ on the flow of the couplings is negligible. Indeed, the equations can be solved by utilizing the fact that $\delta\alpha=\alpha_x-\alpha_y$ is an approximate constant along the flow. The resulting low-energy theory is described by the renormalized splitting, $ h_R$, and couplings, $\alpha_{x,R}=\delta\alpha\gg\alpha_{y,R}= ( h_R/\omega_c)^{\delta\alpha}\alpha_y$. The BKT transition is determined by the renormalized value of the dominant coupling, $\alpha_{x,R}$, such that the system becomes localized when $\alpha_{x,R}>1+\mathcal{O}(h/\omega_c)$, and below this value the effective energy scale assumes the familiar form $ h_R = \omega_c \left(h/\omega_c\right)^{1/(1-\alpha_{x,R})}$ (note that the exponent depends on $\alpha_{x,R}$ and not on the bare value). More details on the RG flow are shown in App.~\ref{Appendix: rg 2bSB}.

To proceed, we note that the low energy theory we have arrived at is nearly identical to that in the 1bSB, the one difference being a remaining weak coupling to the $y$ bath ($\alpha_{y,R}\ll1$) which we may now treat perturbatively. The operator $\sigma_x$ is only weakly dressed (it is renormalized only in the short first section of the flow when $\alpha_y$ is of order $1$), and thus we can once again conclude that its correlation functions will assume a one parameter scaling form, as in Eq.~\eqref{eq: scaling form omega/Delta_r}:
\begin{equation}
\chi_x''(\omega) = \frac{1}{\omega}f_{\alpha_{x,R}}\left(\frac{\omega}{ h_R}\right)+\delta\chi''_{x}(\omega),
\label{xy - chix}
\end{equation}
where $f_{\alpha_{x,R}}$ is a scaling function and the perturbative correction due to the coupling to the losing bath is of the form $ \delta \chi_x''(\omega) = 
\alpha_{y,R} \frac{1}{\omega}\tilde{f}_{\alpha_{x,R}}\left(\frac{\omega}{ h_R}\right)$, with a different scaling function $\tilde{f}_{\alpha_{x,R}}$. When averaging over the second term., the strong renormalization of the losing bath, $\alpha_{y,R}= ( h_R/\omega_c)^{\delta\alpha}\alpha_y$, effectively enhances the exponent in the renormalized distribution which results in a subleading frequency dependence of the averaged $\delta \chi_x''$; see App.~\ref{Appendix: subleading}. 

In contrast, the $y-$susceptibility does not assume a one parameter scaling form. Rather, it is suppressed by additional factors of $ h_R/\omega_c$, such that in the IR limit all spectral weight is shifted to frequencies of order $\omega_c$; see App.~\ref{Appendix: subleading}.

The electronic self-energy can be evaluated following the analysis of Sec.~\ref{1bSB averaging section}. 
Remarkably, the leading term in \eqref{xy - chix} assumes the same form as in the $x-$model:
\begin{equation}
     \Sigma''(\omega) =  -A_{\alpha_{x,R}}\rho_F g_x^2 \left|\frac{\omega}{h_{c,R}}\right|^\gamma,
     \label{selfenergy xy model}
\end{equation}
where $\gamma=(1+\beta)(1-\alpha_{x,R})$, and $A_{\delta \alpha_{x,R}}=\int_0^\infty \frac{f_{\alpha_{x,R}}(x)}{x^{1+\gamma}}dx$. The subleading correction due to $\delta \chi_x''$ is of the form $\delta\Sigma''(\omega) = -B \alpha_y \frac{\gamma}{\gamma+\delta\alpha} \rho_F g_x^2 \left|\frac{\omega}{h_{c,R}}\right|^\gamma   \left|\frac{\omega}{\omega_c}\right|^{\delta\alpha} $ with $B=\int_0^\infty \frac{\tilde{f}_{\delta\alpha}(x)}{x^{1+\gamma+\delta\alpha}}dx$. An additional subleading contribution to $\Sigma''$ is related to the coupling to the $y-$susceptibility, which we denote by $ \delta\Sigma''_{y}$. While an explicit evaluation of $ \delta\Sigma''_{y}$ is more challenging, the fact that $\delta\Sigma''_{y}$ is also subleading follows from the additional ``non-universal" factors of $h_R/\omega_c$ which it contains, similarly to the case of $\delta \chi_x''$.

As mentioned before, this RG-based analysis is perturbative in the strength of the weaker coupling, $\alpha_y$. However, at strong coupling (when $\alpha_x \gtrsim \alpha_y$) the value of $\alpha_{x,R}$ is no longer equal to $\delta\alpha$, and the BKT line $\alpha_{x,R}=1$ changes accordingly. The problem has been solved numerically for varying coupling strengths by \cite{kohler_nonequilibrium_2013}, who have found that at strong coupling the BKT line approaches the line $\alpha_x=\alpha_y$ asymptotically. This  behavior is depicted schematically in Fig.~\ref{fig:main figure multibath}.

Note that for $U(1)$-symmetric points,  $\alpha_x=\alpha_y=\alpha$, the coupling to the two baths is frustrated and the system flows to weak coupling~\cite{castro_neto_quantum_2003,novais_frustration_2005,belyansky_frustration-induced_2021}. This case will be discussed in an upcoming work~\cite{SCshort}. 


\subsection{The $xz-$model}
Consider now the case where $\alpha_x,\alpha_z>0$ and $\alpha_y=0$, dubbed $xz-$model. The major difference in this case compared to the $xy-$model is the fact that the $z$-bath is aligned with the ``field'' $h_z$, making the two baths inequivalent. The flow equations in this case are:
\begin{eqnarray} \frac{d\tilde{h}}{d\ell} &=& (1-\alpha_x)\tilde{h}, \\ 
    \frac{d\alpha_x}{d\ell}&=&-2\alpha_z\alpha_x-\tilde{h}^2\alpha_x, \\
    \frac{d\alpha_z}{d\ell}&=&-2\alpha_x\alpha_z.
    \label{RG2bSBz} 
\end{eqnarray}
As before, we neglect the effect of $\tilde{h}$ on the initial flow of the couplings, assuming that it is sufficiently small.

Let us start with the case $\alpha_x\gg \alpha_z$. Then,  
as the $x$-bath dominates, the flow is essentially identical to the $xy-$model with $z$ replacing $y$. 
The main difference 
is that $\sigma_z$, albeit being strongly dressed in the low energy theory, has a non-zero equilibrium expectation value, i.e., $\langle \sigma_z\rangle \ne 0 $. The $z-$susceptibility therefore contains a term proportional to $\left<\sigma_z\right>^2 \delta(\omega)$. Fortunately, $\left<\sigma_z\right>$ may be evaluated using the sum rule Eq.~\eqref{sum rule 2}, yielding
\begin{equation}
\left<\sigma_z\right>_\infty =  \left(\frac{ h_R}{\omega_c}\right)^{\delta\alpha} \left(a+b\frac{1-\left(\frac{ h_R}{\omega_c}\right)^{1-2\delta\alpha}}{1-2\delta\alpha}\right),
\end{equation}
with $a$ and $b$ being numerical constants which depend on $\alpha_x,\alpha_z$. Interestingly, this static piece contributes an elastic scattering term to the electronic self energy:
\begin{equation}
     \Sigma''_{el}(\omega) \propto -\rho_F g_z^2 \begin{cases}
 \left(\frac{h_{c,R}}{\omega_c}\right)^{2\delta\alpha} & \delta\alpha<1/2,\\
\frac{h_{c,R}}{\omega_c} \log^2\left(\frac{\omega_c}{h_{c,R}}\right) & \delta\alpha=1/2,\\
 \left(\frac{h_{c,R}}{\omega_c}\right)^{2-2\delta\alpha} & 1/2<\delta\alpha<1,
\end{cases} 
\label{elastic z bath term xz model}
\end{equation}
where $\mathcal{O}(1)$ numerical coefficients have been suppressed for clarity. The leading ``inelastic'' part of $\Sigma''$ due to the $x-$bath is identical to Eq.~\eqref{selfenergy xy model}, with appropriate renormalized value $\alpha_{x,R}$. It is worth recalling that the interaction-induced elastic term of Eq.~\eqref{elastic z bath term xz model} adds to the elastic scattering term due to onsite potential in the general model. We comment on this matter further in the discussion on the dc resistivity in Sec.~\ref{transport section}. 

We move on to the second scenario, where $\alpha_z\gg\alpha_x$. In this case, the $z-$bath dominates such that the coupling flows to a finite value, $\alpha_{z,R}=\delta\alpha=\alpha_z-\alpha_x$. Unlike the previous case, $h_z$ is only marginally renormalized, such that the low-energy parameters are given by 
\begin{eqnarray}
     h_R &=& \frac{\delta\alpha}{\alpha_z}h_z,\\
      \alpha_{x,R} &=& \alpha_x \left(\frac{ h_R}{\omega_c}\right)^{\delta\alpha}.
\end{eqnarray}
Note that when $\delta\alpha\to0$, $ h_R$ assumes the form shown in the $xy-$isotropic case, although the system is not $U(1)-$symmetric due to $h_z$.
For $\alpha_x=0$ the TLS is static, and the scattering is solely elastic. The effect of $\alpha_x$ is an addition of a weak inelastic term to $\chi_z''$, as well as a coupling of the electrons to $\chi''_x$. This will result in an inelastic contribution in the self energy such that $\Sigma'' = \Sigma''_{el} + \Sigma''_{inel}$ with
\begin{eqnarray}
\Sigma''_{el} &\approx&  -\frac{\rho_F g_z^2}{2\pi} \\
\Sigma''_{inel}(\omega)&\propto& -\frac{|\omega|^{\gamma+2\delta\alpha}}{h_{c,R}^\gamma\omega_c^{2\delta\alpha}}
\end{eqnarray}
where $\gamma=(1+\beta)$. While this is the leading inelastic contribution, it does not give rise to any MFL/NFL behavior for the considered splitting distributions with $\beta\geq 0$. For more details, see App.~\ref{Appendix: subleading}.

\subsection{The $xyz-$model}

Understanding the physical picture in the more general case where the electrons are coupled to all opeartors of the TLSs (i.e. $\alpha_a>0$ for $a=x,y,z$ and $h_x \equiv 0$), dubbed $xyz-$model, rests upon the fact that the anisotropy remains relevant such that one bath dominates over the others at low energies. The qualitative behavior is thus reduced to one of the two previous cases. In the case where the coupling to the $x-$ or $y-$ baths is the largest, the behavior is qualitatively similar to the $x-$model, albeit with a renormalized coupling, $\alpha_R(\vec{\alpha})$ (of the dominating bath), which depends on the initial values of the couplings. Specifically, the leading inelastic contribution to the self-energy satisfies
\begin{eqnarray}
    \Sigma''(\omega) - \Sigma''(0) \sim -|\omega|^{\gamma},\quad\gamma = (1+\beta_{z})(1-\alpha_R),
\end{eqnarray}
as we have demonstrated above. In addition, above a critical value, $\alpha_R(\vec{\alpha}) \geq 1$, the TLSs undergo a BKT transition to the localized phase, where most of the scattering is elastic. If, on the other hand, the $z-$bath dominates, the TLSs act essentially as static impurities, with an additional weak, FL-like inelastic contribution to the self energy (as in the $xz-$model with dominant $z-$bath).

\section{``Biased" model}

\label{biased section}

All of our above analyses relied on the assumption that the field, $\vec{h}$, is parallel to the $\hat{z}$ direction. Note that this cannot always be made the case by rotating a generic field $\vec{h}=(h_x,0,h_z)$ to point in this direction, since this would induce correlations between the the couplings $g_x,g_z$, and in turn will lead to a mbSB model with correlated baths. We thus keep the couplings uncorrelated and treat the case were $h_x,h_z\neq0$.

\subsection{Biased $x-$model}
\label{1b biased}

We start by considering the $x-$model with finite parallel fields $h_x>0$. We thus allow $\vec{h}$ to be randomly distributed in the $x-z$ plane with a joint distribution
$\mathcal{P}_{\beta_x,\beta_z}(h_x,h_z)\propto h_x^{\beta_x} h_z^{\beta_z}$ for $h_x,h_z<h_c$. This variant maps into the ``biased'' 1bSB, where a field parallel to the bath coupling, $h_x\sigma_x$, is added to the Hamiltonian \cite{leggett_dynamics_1987,weiss_quantum_2012}. The main difference in this case is that $h_x$ is unaffected by the bath (it commutes with the interaction), while $h_z$ is renormalized as before. In addition, the presence of a non-zero $h_x$ implies that $\langle \sigma_x(t\to \infty)\rangle \ne 0$, which leads to an ``elastic'' delta function term in $\chi_x''(\omega)$, namely, the $x-$susceptibility can be written as $\chi_x'' \equiv  \chi_{el}'' + \chi_{inel}''$, with

\begin{eqnarray}
    \chi_{x,el}''(\omega) &=& \left( \frac{2}{\pi} \tan^{-1}\left(\frac{h_x}{ h_R}\right) \right)^2 \delta(\omega), \\ \chi_{x,inel}''(\omega) &=&\frac{1}{\omega}f\left(\frac{\omega}{ h_R},\frac{\omega}{h_x}\right).
\end{eqnarray}
To leading order in $h_{c,R}/h_c$, we find that the low-energy self-energy is given by $\Sigma'' \equiv \Sigma''_{el}+\Sigma''_{inel}$, with 
\begin{eqnarray} 
\Sigma''_{el}\left(\omega\right)&=& -\frac{M}{N}\rho_F g^2, \\
\Sigma''_{inel}\left(\omega\right)&\propto& -\frac{M}{N}\rho_Fg^{2}\left|\frac{\omega}{h_c} \right|^{1+\beta_x}\left|\frac{\omega}{h_{c,R}}\right|^{(1+\beta_z)(1-\alpha)}. 
\label{biased self energy x model}
\end{eqnarray}
Note that since $h_c/ h_{c,R} \propto (\omega_c/h_c)^{\alpha/(1-\alpha)}\gg 1$, the elastic term contributes most of the spectral weight to $\chi_x$, and correspondingly, $\Sigma''_{el}\gg \Sigma''_{inel}$ at low energies. 
We provide an explicit calculation using the 2-parameter scaling-form along with exact evaluation at the TP in App.~\ref{biased case}. Notice that for the reasonable case of $\beta_x\geq0$ this results in a FL for $\alpha<1$ and approaches a MFL behaviour near $\alpha=1$.
\subsection{Biased $xyz-$model}

We now treat the most general variant of our model, where the couplings and fields are all allowed to point in generic directions. The behavior of the biased $xyz-$model can be similarly understood from the RG analysis,  where Eqs.~\eqref{multibath RG} are modified as \cite{khveshchenko_quantum_2004}
\begin{eqnarray}  \frac{d\tilde{h}_z}{d\ell} &=& (1-\alpha_x-\alpha_y)\tilde{h}_z,\\
\frac{d\tilde{h}_x}{d\ell} &=& (1-\alpha_z-\alpha_y)\tilde{h}_x,\\
    \frac{d\alpha_a}{d\ell} &=& -\left(\sum_{b\neq a}\alpha_b + \left(1-\delta_{az}\right)\tilde{h}_z^2+  \left(1-\delta_{ax}\right)\tilde{h}_x^2 \right)\alpha_a,\nonumber
    \\
    \label{biased xyz RG}
\end{eqnarray}
where $\tilde{h}_x \equiv h_x/\omega_c$. Importantly, there are no cross terms between the fields $h_x,h_z$ such that the qualitative behavior can be understood in terms of the approximately independent flow of the individual fields. Furthermore, since a bias along $\hat{y}$ is forbidden in order to respect time-reversal symmetry, the physical picture in the biased $xyz-$model is determined by whether or not the $y-$bath dominates. In the case where the $y-$bath dominates, both the `field', $h_z$, and the `bias', $h_x$, will be renormalized according to Eq.~\eqref{Delta_r_renorm} with $\alpha \to \alpha_{y,R}$. Consequently, the leading inelastic contribution to the self-energy will take the form
\begin{eqnarray}
    \Sigma''(\omega) &-& \Sigma''(0) \sim -|\omega|^{\gamma}\quad,\\ \gamma &=& (2+\beta_x + \beta_z)(1-\alpha_{y,R}),
\end{eqnarray}
along with elastic scattering terms as in Eq.~\eqref{elastic z bath term xz model}. In contrast, if the $x-$ or $z-$baths dominate, the behavior will be analogous the biased $x-$model with the appropriate renormalized couplings of the dominant bath; see Eq.~\eqref{biased self energy x model}. 


\section{Transport}
\label{transport section}
Let us begin by considering the electronic contribution to the electrical conductivity, and later incorporate the effect of the TLSs on the optical conductivity. Using the Kubo formula, the real part of the conductivity (associated with the electrons) is given by 
\begin{eqnarray}
    \sigma_{\rm el}\left(\Omega \right) = \frac{{\rm Im}\Pi^R_{J_x}\left(\Omega\right)}{\Omega},
\end{eqnarray}
where $\Pi^R_{J_x}$ is the retarded current correlator (along the $x$ direction). The current operator is given by $\boldsymbol{J}=\sum_{a}\int_{\boldsymbol{k}}\boldsymbol{v}_{\boldsymbol{k}}c^{\dagger}_{a\boldsymbol{k}}c_{a\boldsymbol{k}}$ and $\boldsymbol{v}_{\boldsymbol{k}} = \nabla_{\boldsymbol{k}}\varepsilon_{\boldsymbol{k}}$. The evaluation of $\Pi_{J_x}^R$ is greatly simplified since all vertex corrections vanish due to the spatial randomness of the couplings to the local TLSs  (similarly to Refs.~\cite{chowdhury_translationally_2018,patel_magnetotransport_2018}). The electronic optical conductivity is thus given by 
\begin{equation}
    \sigma_{\rm el}(\Omega) = \frac{1}{\Omega}\int_{\omega}\int_{\boldsymbol{k}}v_{\boldsymbol{k}}^2 \mathcal{A}_{\boldsymbol{k}}\left(\omega\right)\mathcal{A}_{\boldsymbol{k}}\left(\Omega+\omega \right)\left[f\left(\omega\right) - f\left(\Omega + \omega\right)\right].
\end{equation}
Here $\mathcal{A}_{\boldsymbol{k}}\left( \omega\right)\equiv - \frac{1}{\pi}{\rm Im}G^R_{\boldsymbol{k}}(\omega ) $ is the electronic spectral function and $f(\omega)$ denotes the Fermi distribution function. 

In the dc limit, the conductivity is given by
\begin{eqnarray}
    \sigma_{\rm el}(\Omega \to 0) = \frac{v_F^2\rho_F}{16 T}\int \frac{d\omega}{2\pi}\frac{1}{\left| \Sigma''(\omega)\right|}{\rm sech}^2\left(\frac{\omega}{2T}\right).
 \end{eqnarray}
Hence, the $T$-scaling of the dc resistivity follows the single-particle lifetime. It is instructive to first consider $\rho(T)$ in the $x-$model. For $\alpha < 1$, the low-$T$ behavior, $T\ll h_{c,R}$, is of the form 
\begin{eqnarray}
    \rho(T) = \rho_0 + AT^\gamma ,
\end{eqnarray}
where here we have restored the on-site disorder by setting $V^2>0$, corresponding to the residual resistivity term, $\rho_0$, and $\gamma = (1+\beta)(1-\alpha)$. Similarly, for $\alpha = 1$, we have that $\rho(T) - \rho_0 \propto 1/\left|\log\left(T\right)\right|^{1 + \beta}$; and for $\alpha> 1+h_c/\omega_c$, the TLSs are frozen at $T=0$, and thus contribute to the residual resistivity, with FL-like finite-$T$ corrections ($\rho(T) - \rho_0 \propto T^2$). In the intermediate regime, $1<\alpha<1+h_c/\omega_c$, the resistivity interpolates smoothly between these two behaviors. 
In the more general $xyz-$model, the resistivity follows the behavior of the $x-$model whenever the transverse couplings $\alpha_x$ or $\alpha_y$ dominate (as discussed extensively in the Sec.~\ref{multibath-model section}). Whereas, if the parallel coupling $\alpha_z$ dominates, the scattering is mainly elastic with weak FL-like temperature scaling. Similarly, the biased $xyz-$model follows analogous behavior to that of the $x-$model provided that the $y-$bath dominates, and to the biased $x-$model if the $x-$ or $z-$bath dominates.

\label{opticalconductivity}

We proceed to consider the optical conductivity. In addition to the contribution due to the itinerant electrons, we also assume that each TLS carries a randomly distributed dipole moment (recall the TLS are phenomenologically related to charged collective degrees of freedom) which depends on the state of the TLS:
\begin{equation}
    H_{\rm dipole} = \sum_{r,l} \vec{E}_r \cdot \left( \vec{d}^z_{r,l}\sigma^z_{r,l} + \vec{d}^x_{r,l}\sigma^x_{r,l} \right).
\end{equation}
Here $\vec{E}_r$ is the local electric field and $\vec{d}^{x,z}_{r,l}$ denote uncorrelated Gaussian random dipole moments of the TLS flavors, with variances $d_{x,z}^2$. In total, the (longitudinal) optical conductivity takes the two-component form 
\begin{eqnarray}
\sigma(\Omega) = \sigma_{\rm el}(\Omega) + \sigma_{\rm TLS}(\Omega). 
\end{eqnarray}
The electronic contribution is standard and follows straightforwardly from the form of the self energy. In particular, at low energies, where $-\Sigma''(\omega)=\frac{\Gamma}{2} +  c|\omega|^\gamma$ (i.e. we restore the elastic scattering term that does not affect any of the previous results), if the scattering is mainly inelastic ($\Gamma \ll c|\Omega|^\gamma$):
\begin{eqnarray}
    \sigma_{\rm el}(\Omega)\sim \begin{cases}
\frac{1}{\Omega^\gamma} & \gamma < 1, \\
\frac{1}{\Omega \log^2(1/\Omega)}  & \gamma = 1,\\
\frac{1}{\Omega^{2-\gamma}} & \gamma > 1.
\end{cases}
\end{eqnarray}
while if the scattering is mainly elastic ($\Gamma \gg c|\Omega|^\gamma$):
\begin{eqnarray}
    \sigma_{\rm el}(\Omega)\sim \frac{1}{\Gamma} - \frac{2^{\gamma+1}c}{(\gamma+1)\Gamma^2}|\Omega|^\gamma.
\end{eqnarray}
At higher energies, $\Omega \gtrsim h_{c,R}$, the TLS contribution to $\Sigma''$ saturates to a constant such that $\sigma_{\rm el}(\Omega \gtrsim h_{c,R}) \sim 1/\Omega$. 

The TLS contribution is given by 
\begin{eqnarray}
\sigma_{\rm TLS}\left(\Omega\right) & = \Omega \left( d_x^2 \overline{\chi''_x}(\Omega)+d_z^2 \overline{\chi''_z}(\Omega)\right).
\label{eq:dipole_opticalc}
\end{eqnarray}
Interestingly, the TLS contribution follows the frequency dependence of the inelastic part of $\Sigma''$ (provided that the $y-$bath is not dominant). In particular, if the dipole moments are not negligibly small, $\sigma_{\rm TLS}$ might constitute the leading frequency dependence, leading to a positive slope and non-monotonic behavior of the optical conductivity. Defining the energy scale $\Omega_*=\sqrt{\rho_F}v_F/d_a$, we find that if the scattering is dominantly  elastic and $\Gamma\gg\Omega_*$ then there will be an increasing optical conductivity around zero frequency. If inelastic scattering dominates, $\Gamma\ll\Omega_*$, the optical conductivity will always be decreasing around zero frequency, but will begin increasing for frequencies of order $\Omega_{\rm mIR} \sim  Z\Omega_*$ if the system is a FL (with $Z$ the quasiparticle weight), or $\Omega_{\rm mIR} \sim (\Omega_*/c)^{1/\gamma}$ if the system is a NFL (i.e. if $\gamma\leq 1$), leading to a so-called mid-IR peak around energies of order $h_{c,R}$ (assuming that $\Omega_{\rm mIR}<h_{c,R}$) \cite{bashan2023tunable}.

The  assumption that led to Eq.~\eqref{eq:dipole_opticalc} was that there are sufficiently many TLSs that carry a dipole moment and can therefore be optically excited. At the same time one expects that there are TLSs that locally come with a quadrupole moment. For example, they could locally distort a state of four-fold rotation symmetry to a lower symmetry. In this case one can excite the TLS via inelastic light scattering and the Raman response function\cite{Devereaux2007,Karahasanovic2015} will measure directly the TLS susceptibilities
\begin{equation} R_{\alpha,\beta}\left(\Omega\right)  =  \left(q^x_{\alpha,\beta}\right)^2 \overline{\chi''_x}(\Omega)+\left(q^x_{\alpha,\beta}\right)^2 \overline{\chi''_z}(\Omega).
\end{equation}
Here $q^\kappa_{\alpha,\beta}$ is the quadrupole moment due to the $\kappa$-component of the TLS pseuodspin. The individual tensor elements can be detected by an appropriate combination of the polarization of the incoming and scattered light.
Hence, the presence of TLS can, at least partially account for the broad Raman continuum that has been observed in many correlated electron materials \cite{Devereaux2007}.


It is intriguing to examine the MFL/NFL transport properties of our model through the viewpoint of Planckian dissipation and the putative bound on transport times \cite{hartnoll_planckian_2021}. Since there is no unique definition for the transport time, we consider two different approaches. Following Ref.~\cite{hartnoll_planckian_2021}, we can associate the transport time to the single-particle lifetime as the two are proportional in our model. In that case, the inverse transport time is Planckian in the sense that $1/\tau_{\rm tr} \sim T$ with an $\mathcal{O}(1)$ coefficient for the NFL phase while at the MFL point the coefficient is $\mathcal{O}(1/\ln(1/T))$. In particular, our model trivially satisfies a `Planckian bound' due to the Kramers-Kroning relations between the real and imaginary parts of $\Sigma$\footnote{Note that for MFLs or NFLs, the Kramers-Kroning relations relate the low energy regimes of the real and imaginary parts of the self energy, such that one is completely determined by the other (independent of the UV cutoff). This is in contrast to FLs where, since $\Sigma''(\omega)\sim  |\omega|^{\gamma}+\mathcal{O}(|\omega|^{\gamma})$ (with $\gamma>1$), the low and high energy sectors both contribute to the leading linear in $\omega$ behavior of ${\rm Re}\Sigma_{\rm ret}(\omega)$ (making Kramers Kroning relations not particularly useful if only low-energy information is accessible).}.
Alternatively the inverse transport time can be defined in terms of the energy scale for which the dc and ac conductivities become comparable \cite{chowdhury_translationally_2018}: $\sigma\left(\tau_{\rm tr}^{-1}(T),T=0 \right) \sim \sigma\left(\Omega=0,T \right)$. This procedure agrees with the single-particle lifetime result for NFLs while for the MFL case the transport time contains an additional log correction: $\tau_{\rm tr}^{-1} \sim T/\log^2(T)$.

Lastly, relying on the analysis of the weakly disordered MFL (or NFL) model in Ref.~\cite{tulipman_criterion_2022}, we note that the Wiedemann-Franz law is obeyed as $T\to 0$, regardless of the existence of well-defined Landau quasiparticles (in the absence of vertex correction, as we have here, the analysis is essentially identical).  

\section{Thermodynamics}
\label{thermodynamics section}
In this section, we study thermodynamic properties of the model. We mainly consider the $x-$model and further discuss the expected behavior in the $xyz-$model. It is worth noting that a direct evaluation of the free energy from the saddle point of the large-$N$ effective action is challenging due to its non-Gaussian nature. Instead, we obtain the specific heat from the internal energy, and corroborate our results with an alternative derivation of the specific heat from the entropy, where in particular we confirm the absence of $T=0$ residual entropy in our model.

Consider the internal energy density
\begin{eqnarray}
     U \equiv \frac{1}{N\mathcal{V
}} \langle H_{\rm el} + H_{\rm TLS} + H_{\rm int} \rangle,
\end{eqnarray}
where $H_{\rm el}$ and $H_{\rm TLS}$ correspond to the first two terms in \eqref{Hamiltonian}, respectively, and $\mathcal{V
}$ is the volume of the system. Let us henceforth suppress the factor $1/(N\mathcal{V})$ and assume $r=N/M=1$ for simplicity. By employing the equation of motion for the retarded and advanced electronic Green's functions (see App.~\ref{specific heat from Uint}), we may write
\begin{eqnarray}
    \langle H_{\rm el} + H_{\rm int} \rangle = \int_{\boldsymbol{k}}\int_{\omega}\omega n_F\left(\omega\right)\mathcal{A}_{\vec{k}} \left(\omega \right) \equiv U_{\rm el,0},
    \label{U_el0}
\end{eqnarray}
where $n_F(\omega)$ is the Fermi function. Note that due to the locality of the self energy, $U_{\rm el,0}$ corresponds to the internal energy of non-interacting electrons. To see this, we use the fact that $\int_{\boldsymbol{k}}\mathcal{A}\left(\omega,\boldsymbol{k}\right)=\rho_F$, hence $U_{\rm el,0} = \rho_F \int_{\omega} \omega n_F(\omega)$. The specific heat related to $U_{\rm el,0}$ is given by 
\begin{eqnarray}
c_{\rm el,0}&=&\rho_F\int_{\omega}\omega\frac{\partial n_F\left(\omega\right)}{\partial T}=\frac{\pi^{2}}{3}\rho_FT,
\end{eqnarray}
i.e., the specific heat of non-interacting electrons. Interestingly, all interaction effects are encoded in the renormalized TLS part of the internal energy, $U_{\rm TLS}=\langle H_{\rm TLS} \rangle$, which we will now evaluate. Using the sum rule Eq.~\eqref{sum rule 2}, we may express the TLS specific heat as
\begin{eqnarray}
  c_{\rm TLS}  = \frac{1}{2}\int_{\omega}\omega\frac{\partial \overline{ \chi''_x}(\omega,T)}{\partial T}.
  \label{cTLS}
\end{eqnarray}
As $h_{c,R}$ is the only energy scale, for $\omega,T\ll h_{c,R}$ one can write the TLS susceptibility as a 2-component scaling form, i.e. $\overline{\chi''}(\omega,T) = \tfrac{1}{\omega}F\left(\tfrac{\omega}{h_{c,R}},\tfrac{\omega}{T}\right)$. For concreteness, we assume the following scaling form:
\begin{eqnarray}
    \overline{\chi''_x}(\omega,T) = \overline{\chi''_x}(\omega,T=0) \times \left( \frac{|\omega|}{\sqrt{\omega^2+(aT)^2}}\right)^\varphi 
    \label{scaling chi}
\end{eqnarray}
with $a\sim\mathcal{O}(1)$ some numerical coefficient and scaling exponent $\varphi>0$,\footnote{Exact results for $\alpha\ll1$ and $\alpha=1/2$ indicate that $\varphi =1$. } which both affect the result only by a numerical prefactor and not the $T$ dependence.

\begin{figure}[t]
\centering

\includegraphics[width=1\columnwidth]{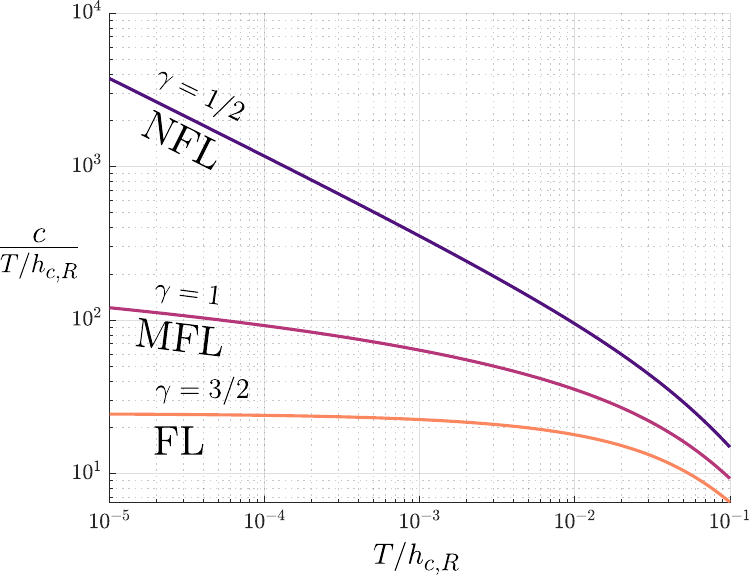}
\caption{Specific heat, as extracted from the internal energy at the Toulouse point, for values of $\gamma$ for FL, MFL and NFL behavior (corresponding to $\beta=2,1,0$). As expected from the above scaling arguments \eqref{Entropy Scaling}, at low temperatures the ratio $c/T$ approaches a constant for the FL, has logarithmic divergence for the MFL, and polynomial divergence with exponent $1-\gamma$ for the NFL.}
\label{fig:Cv}
\end{figure}

Using Eq.~\eqref{cTLS}, we obtain that 
\begin{eqnarray}
    c_{\rm TLS} \sim  \begin{cases}
\frac{T}{h_{c,R}} & \gamma>1,\\
\frac{T}{h_{c,R}}\log\frac{h_{c,R}}{T} & \gamma=1,\\
\left(\frac{T}{h_{c,R}}\right)^{\gamma} & \gamma<1.
\end{cases}
\label{c_TLS results}
\end{eqnarray}
In addition, Eq.~\eqref{cTLS} can be evaluated numerically at the Toulouse point where the exact temperature dependence of $\chi''(\omega,T)$ is known analytically.
The results, confirming the $T$-dependence found in Eq.~\eqref{c_TLS results}, are shown in Fig.~\ref{fig:Cv} for $h$ distributions corresponding at the Toulouse point to a FL, MFL and NFL.

We corroborate the above discussion with an alternative derivation of the specific heat from the entropy. To do so, we consider the addition of a single TLS per site to the theory with $M=0$. In the language of the SB model, the excess entropy added to the system, defined by $\delta S \equiv S(M=1) - S(M=0)$, is known as the `impurity contribution' \cite{costi_thermodynamics_1999}. Importantly, $\delta S$ is determined by the spectral function of the bath, the renormalized splitting and the temperature. Hence, since the TLSs are decoupled for any $M$ provided that $N\gg1$ (as $\overline{g_{ijl} g_{ijl'}}=0$ for $l'\ne l$), and the particle-hole bath is ohmic, we may write the entropy of $M$ TLSs by adding the impurity contributions of the individual TLSs.

Considering the $x-$model, the impurity contribution of a single TLS for $\alpha<1$ is given by \cite{guinea_dynamics_1985,gorlich_specific_1988,costi_scaling_1998,costi_thermodynamics_1999}
\begin{equation}
    \delta S\left(x\right) = \begin{cases}
\frac{\alpha\pi}{3}x + \mathcal{O}\left(x^3\right) & x\ll 1,\\
\log 2 & x\gg 1,\\
\end{cases}
\end{equation}
where $x=T/ h_R$. The entropy of the full system (i.e. in the large-$M,N$ limit) can therefore be written as
\begin{eqnarray}
    S(T) = S_0(T) + \Delta S(T),
    \label{entropy impurity contribution}
\end{eqnarray}
where $S_0 = S(M=0)$ denotes the contribution of the non-interacting electrons and
\begin{eqnarray}
    \Delta S(T) =  M\mathcal{V}\int_0^{h_{c,R}} \mathcal{P}_r( h_R) \delta S\left(\frac{T}{ h_R}\right) d h_R.
    \label{S scaling form}
\end{eqnarray}
To evaluate $\Delta S$, we divide the integral over $ h_R$ to $ h_R<T$ and $ h_R>T$, denoted by $S_<$ and $S_>$, respectively, and, substituting \eqref{Pr a<1} in \eqref{S scaling form}, we obtain
\begin{eqnarray}
    S_<(T) &\approx& \frac{\pi\alpha\gamma}{3h_{c,R}^\gamma}\frac{T}{h_{c,R}}\frac{h_{c,R}^{\gamma-1}-T^{\gamma-1}}{\gamma-1}, \\  S_>(T) &\approx& \frac{\pi\alpha}{3}a_{\gamma} \left(\frac{T}{h_{c,R}}\right)^\gamma,
    \label{Entropy Scaling}
\end{eqnarray}
where $a_{\gamma} \equiv  \int_0^1 \delta S\left(\frac{1}{x}\right) x^{\gamma-1}dx\approx 1/\gamma$. Hence, for $T\ll h_{c,R}$,
\begin{eqnarray}
    \Delta S(T) \sim \begin{cases}
\frac{T}{h_{c,R}} & \gamma>1,\\
\frac{T}{h_{c,R}}\log\frac{h_{c,R}}{T} & \gamma=1,\\
\left(\frac{T}{h_{c,R}}\right)^{\gamma} & \gamma<1.
\end{cases}
    \label{Total entropy low T Scaling}
\end{eqnarray}
Since $S_0\sim T$, the total entropy, $S$, obeys the same scaling as $\Delta S$, which also holds for the specific heat, in agreement with Eq.~\eqref{c_TLS results}. 

It is also worth noting that there is no residual extensive entropy at $T=0$, in contrast to theories of MFLs constructed from variants of the Sachdev-Ye-Kitaev model \cite{chowdhury_sachdev-ye-kitaev_2022}. 

Physically, the $T-$scaling of the impurity contribution in Eq.~\eqref{entropy impurity contribution} stems from scattering of the low-energy modes of the bath by the TLS. Hence in the cases where $\alpha_x$ or $\alpha_y$ dominate, the impurity contribution is expected to follow Eq.~\eqref{entropy impurity contribution} since the low energy theory is identical to that of the $x-$model (up to weak perturbations). It therefore follows that $S(T)$ satisfies Eq.~\eqref{Total entropy low T Scaling}. Moreover, when $\alpha_z$ dominates, the system realizes a FL (with additional static impurities) and weak renormalization of the splittings, such that $S(T)\propto T$.

\section{Superconductivity}
\label{superconductivity section}

In order to study the superconducting instability, 
we introduce spinful electrons to the Hamiltonian Eq.~(\ref{Hamiltonian},\ref{H int}), namely, we let $c^\dagger_{i,\vec{r}}\to c^\dagger_{i,s,\vec{r}}$ with $s=\{\uparrow,\downarrow\}$. Note that since the TLSs have no spin structure (assuming that the underlying glass is non-magnetic), the couplings $\vec{g}$ do not depend on spin index $s$, and the interaction term is diagonal in spin space. The mapping of the spinful variant of the Hamiltonian onto a spin-boson problem is similar to the spinless case, but one must also consider the anomalous bilocal field, analogously to Eqs.~(\ref{Gfermion},\ref{Dtls}), defined as
\begin{equation}
    \label{Ffermion}
F_{\vec{r},\vec{r'}}(\tau,\tau')=\frac{1}{N}\sum_i c_{i,\downarrow,\vec{r}}(\tau) c_{i,\uparrow,\vec{r'}}(\tau'),
\end{equation}
and enforced via the anomalous self-energy $\Phi_{\vec{r},\vec{r'}}(\tau,\tau')$ \cite{esterlis_cooper_2019,HAUCK2020168120}; see App.~\ref{appendix superconductivity}.

We study the critical temperature $T_{c}$ as a function of the couplings $\vec{\alpha}$. Approaching the SC state from the normal state, where $\Phi=0$, we obtain $T_{c}$ as the solution for the linearized Eliashberg equation (in imaginary frequency) for the local anomalous self-energy $\Phi(i\omega)$:
\begin{eqnarray}
    \label{Phi eliashberg}
    \Phi(i\omega) = T\sum_{\omega'} \frac{D_\Phi (i\omega-i\omega')}{\left|\omega' +i\Sigma(i\omega')\right|} \Phi(i\omega'),
\end{eqnarray} where the bosonic propagator is
\begin{equation}
    \label{D Phi}
    D_\Phi(i\omega)  =  \sum_{a=x,y,z} \lambda_{a} h_{c,R} \chi_a(i\omega) \times (-1)^{\delta_{a,y}}.
\end{equation}
 Notice that while the interactions via $g_x$ and $g_z$ mediate pairing, $g_y$ is pair breaking, as it couples to a current in fermion-flavor space (i.e., to an anti-symmetric fermionic bilinear).
By rewriting Eq.~\eqref{Phi eliashberg} as $\Phi(i\omega) = \sum_{\omega'} K(i\omega,i\omega') \Phi(i\omega')$, we see that $T_{c}$ is determined as the minimal temperature for which the largest eigenvalue of the kernel $K$ is equal to $1$. 

In the following we first obtain $T_{c}$ in the spinful variant of the $x$-model as a representative example and later comment on the behavior of $T_{c}$ in other variants. Specifically, we estimate $T_{c}$ analytically by focusing on the ``weak'' ($\lambda\equiv\lambda_x\ll1$) and ``strong'' ($\lambda\gg1$) coupling regimes, where $T_{c}$ is much smaller or larger than the characteristic energy scale $h_{c,R}$, respectively. Note that ``weak'' and ``strong'' coupling regimes do not necessarily correspond to small or large values of $\alpha$. For simplicity we further assume that $h_c/E_F\ll |1-\alpha|$ (to avoid the complications resulting from the slowdown of the RG flow of $h$ around $\alpha=1$) and mainly focus on the parametric dependence of $T_c$, ignoring various $\mathcal{O}(1)$ coefficients.

\subsubsection{Weak coupling $(\lambda \ll 1)$}
In the weak coupling regime, we rely on the $\alpha<1$ and $T=0$ form of the susceptibility given in Eq.~\eqref{average susc gamma}, and assume an $\omega/T$ scaling with exponent $\varphi>0$ (similar to Eq.~\eqref{scaling chi})
\begin{equation}
    \overline{\chi''_x}(\omega,T) = {\rm sgn}(\omega) \gamma A_\alpha \frac{|\omega|^{\gamma-1}}{h_{c,R}^\gamma } \times \min\left(1,\frac{|\omega|}{T} \right)^\varphi.
\end{equation}
We perform analytical continuation to imaginary frequencies (see App.~\ref{appendix superconductivity}) and obtain 
\begin{eqnarray}
    D_\Phi(i\omega_n) &\propto&  \frac{\gamma\lambda}{\gamma-1}\left(1-\left|\frac{\omega_n}{h_{c,R}}\right|^{\gamma-1}\right).
\end{eqnarray}
 In the FL phase $\gamma>1$, the leading piece of $D_{\Phi}$ at low $T$ is constant, resulting in the conventional BCS form of $T_{c}$. However, in the MFL point or NFL phase, where $\gamma\leq 1$, $D_{\Phi}(i\omega)$ diverges logarithmically or with exponent $\gamma-1$, respectively. Consequently, $T_{c}$ crosses over from a BCS form to an algebraic, quantum critical form. Explicitly, by solving the Eliashberg equation in this regime, we obtain
\begin{eqnarray} T_{c}\propto h_{c,R}\begin{cases}
\exp\left(-\frac{\gamma-1}{\gamma \lambda}\right) & ,\ensuremath{\gamma>1+\mathcal{O}(\sqrt{\lambda)}}\\
\exp\left(-\frac{1}{\sqrt{\lambda}}\right)& ,\ensuremath{\gamma=1}\\
\left(\frac{\gamma \lambda}{1-\gamma}\right)^{\frac{1}{1-\gamma}} & ,\gamma<1-\mathcal{O}(\sqrt{\lambda)})
\end{cases}
\end{eqnarray}
The results for $\gamma\leq1$ are similar to those found by \cite{DTSC1998,ChubukovGamma1,ChubukovGamma2} for other cases of quantum critical pairing. 
For consistency, we must require that $T_c\ll h_{c,R}$, which translates into $\lambda \ll 1$. Due to the vanishing of $h_{c,R}$ as $\alpha\to1$, the problem will eventually cross over to the strong coupling regime, where $T_c\gg h_{c,R}$, beyond some intermediate value $\alpha<1$.

\subsubsection{Strong coupling $(\lambda \gg 1)$}
We now consider the transition to superconductivity at temperatures $T\gg h_{c,R}$. In this regime, we obtain the finite-$T$ TLS-susceptibility via a combination of scaling arguments with known results. For details, see App.~\ref{appendix superconductivity}. We find that 
\begin{eqnarray}
\overline{\chi''_x}\left(\omega,T\right)\propto\frac{h_{c}^{2}}{E_{F}^{2\alpha}}\frac{\left(\max\left(|\omega|,a_{\alpha}T\right)\right)^{2-2\alpha}}{\omega}
\end{eqnarray}
where $a_\alpha = \mathcal{O}(1)$. Similarly to the weak coupling limit, we obtain the parametric form of $T_{c}$ by performing the analytical continuation and solving the linearized Eliashberg equation; see App.~\ref{appendix superconductivity}. Remarkably, in an analogous fashion to the ``weak coupling'' regime, we find that $T_{c}$ exhibits a series of crossovers from a quantum critical, to a `marginal BCS' \cite{DTSC1998}, to a conventional BCS form as the coupling is \textit{increased} (rather than decreased, as in the ``weak coupling'' case). Explicitly, we have that 
\begin{eqnarray}
    T_{c}\propto E_{F}\begin{cases}
\left(\frac{\alpha^2}{3-2\alpha}\epsilon\right)^{\frac{1}{3-2\alpha}} & ,\alpha<3/2-\mathcal{O}(\sqrt{\epsilon})\\
\exp\left(-\frac{1}{\alpha\sqrt{\epsilon}}\right) & ,\alpha=3/2\\
\exp\left(-\frac{2\alpha-3}{\alpha^{2}\epsilon}\right) & ,\alpha>3/2+\mathcal{O}(\sqrt{\epsilon})
\end{cases}
\end{eqnarray}
where we defined the small parameter $\epsilon\equiv\frac{M}{N}\left(\frac{h_{c}}{E_{F}}\right)^{2}\ll1$. Interestingly, $T_{c}$ decreases up to $\alpha = 3$, where it has a local minimum. For larger values of $\alpha$, it increases and approaches the limiting form $T_{c}\propto E_F \exp(-1/\alpha\epsilon)$. We expect our results to hold as long as $\alpha \ll E_F/h_c$ (such that $\alpha\epsilon\ll1$). For consistency of the strong coupling analysis we must require that $T_{c} \gg h_{c,R}$. While this is trivially fulfilled when $\alpha>1$, for $\alpha<1$ this results in the requirement $\lambda\gg 1$, which is, as expected, complimentary to the weak coupling condition. 

Intuitively, the reduction of $T_{c}$ for larger values of the coupling corresponds to the fact that at finite $T$, while the TLSs are nearly frozen (i.e. resemble classical impurities), they preserve their quantum mechanical nature and can thus mediate pairing. The accessible low-energy spectral weight for pairing diminishes with the coupling strength and therefore suppresses superconductivity (this trend is reversed beyond $\alpha=3$, where the increase in the coupling strength is more significant than the shift of the remaining spectral weight to high frequencies). 

\subsubsection{Superconductivity in other model variants}

Following from the discussion of the $x$-model, we comment on the expected behavior of $T_{c}$ in generic variants of the model. Let us first consider cases with $g_y=0$ (i.e., without pair-breaking interactions). In the ``weak coupling'' regime, in the sense defined above, $T_{c}$ is determined by the behavior of the dominant bath, such that it qualitatively follows that of the $x$-model if $\alpha_x$ is dominant, or otherwise assumes a conventional BCS form (see Fig.~\ref{fig:main figure multibath}). In the ``strong coupling'' limit, however, the behavior of $T_{c}$ is non-universal, namely, it is determined by the susceptibility of the least irrelevant operator. Mapping out the quantitative form of $T_{c}(\vec{\alpha})$ necessitate the exact renormalized exponents of the TLS-susceptibilities and is beyond the scope of our work (given the leading exponents, the analysis is identical to that of the $x$-model). However, recalling that in both cases where $\alpha_x$ or $\alpha_z$ dominates, the strong coupling behavior approaches a BCS form, we expect that at intermediate couplings, $T_{c}$ will smoothly interpolate from a quantum critical to a BCS form as $\alpha_z$ is increased for fixed $\alpha_x$. Lastly, introducing pair-breaking interactions, i.e. a non-zero $\alpha_y$, suppresses $T_{c}$ \cite{HAUCK2020168120}.

\section{$1/N$ Corrections}

In this section, we discuss two perturbative corrections that arise at leading order in $1/N$: the validity of the self-averaging assumption and the effect of electron-mediated TLS-TLS interactions, i.e. RKKY-like interactions. 


\label{finite N section}

\subsection{Validity of self averaging}
\label{1/M variance}

An important assumption of our above analysis lies in the self averaging of the model, which allows us to replace the sum over many TLS-susceptibilities by its mean value [with respect to $\mathcal{P}_r(h)$], due to the fact that its variance is suppressed by a factor of $1/M$. While this assumption is clearly valid in the limit $M\to \infty$, for any finite (but still large) $M$ the standard deviation may dominate over the mean at sufficiently low energies due to its different frequency dependence. Indeed, consider the variance of the average TLS auto-correlation function in imaginary time:
\begin{eqnarray}
    \text{Var}\left(\frac{1}{M}\sum_{i=1}^M \left<\sigma_x^i(\tau)\sigma_x^i(0)\right>\right) = \frac{1}{M}\int \mathcal{P}_r( h_R) S(\tau)^2 d h_R. \nonumber 
    \\
\end{eqnarray}
For simplicity let us focus on the regime of interest $\alpha<1$ in the $x-$model. By dimensional considerations, at long times the dimensionless integral must be proportional to $(h_{c,R} \tau)^{-\gamma}$ (assuming that there is no obstruction to taking the upper integration limit to $\infty$, which is the case for $\gamma<4$). As a result, by taking the square root and transforming to the frequency domain, we obtain the root-mean-square the TLS-susceptibility:
\begin{equation} \sqrt{\overline{\left(\chi_x''\right)^2}}(\omega) \sim \frac{1}{\sqrt{M}} \frac{|\omega|^{\gamma/2-1}}{h_{c,R}^{\gamma/2}}.
\label{rmschix}
\end{equation}
Comparing Eq.~\eqref{rmschix} to the mean in Eq.~\eqref{average susc gamma}, we conclude that statistical fluctuations can be  neglected above a parametrically small energy scale:
$\omega \sim M^{-1/\gamma}§{h_{c,R}}$.
For energies below this scale, the self-averaging assumption is no longer valid and a more systematic treatment of the $1/M$ (and $1/N$) fluctuations is needed to determine the behavior of the model. 

\subsection{RKKY interactions}

Another effect arising when $N$ is taken to be large but finite, is the emergence of RKKY-like interactions between the different TLSs, mediated by the itinerant electrons. We analyze this perturbative effect in the spirit of Ref.~\cite{VladAbsence}. We shall consider the $x$-model for simplicity, the generalization to other variants is straightforward. 

Including the RKKY-like term,
\begin{eqnarray}
    &&H_{\rm RKKY} = \sum_{jk}\frac{g_{ijk,\vec{r}}g_{i'j'k',\vec{r}'}}{N^2 g^2} \nonumber \\
     &&\times\Pi_{jk}(\vec{r} - \vec{r}',\tau-\tau')\sigma^x_{i,\vec{r}}(\tau)\sigma^x_{i',\vec{r}'}(\tau'),
\end{eqnarray}
each TLS will now feel the effect of a sub-ohmic bath arising from the RKKY coupling to other TLSs, in addition to the ohmic particle-hole bath. Following the analysis of Ref.~\cite{VladAbsence}, this contribution to the bath will be proportional to $\overline{\chi_x(i\omega)}$, and thus the full bath will be of the form
\begin{eqnarray}
    \Pi(i\omega) = \alpha|\omega|+ \frac{\lambda^2}{N} h_{c,R}^{2-\gamma} |\omega|^{\gamma-1} 
\end{eqnarray}
with sub-ohmic exponent $2-\gamma$.

In the limit of large yet finite $M,N$, the sub-ohmic contribution to the bath may be neglected above the small energy scale $\omega\sim \left(\tfrac{\lambda^2}{\alpha N}\right)^{1/(2-\gamma)}h_{c,R}$. However, even for very large $N$ this energy scale will eventually approach $h_{c,R}$ near $\alpha\to1$ since $\lambda$ diverges as $h_{c,R}\to0$.
Below this scale, the self-consistent approximation of a TLS-induced sub-ohmic bath acting on itself breaks down, and a more systematic analysis is needed to determine the behavior at very low energies. 
The low-energy behavior in similar cases \cite{Sengupta1995} suggests that this state remains non-trivial in the sense that $\gamma$ is expected to remain less than 2. 

Lastly, note that the subohmic nature of the TLS-induced bath considered above is a result of perturbing around the $N,M\to \infty$ saddle point. In a more realistic finite-but-large-$M$ setting, we expect the subohmic behaviour to crossover to ohmic below a small energy scale, corresponding to the lowest renormalized splitting of the nearby TLSs. In this case, a qualitative change in the behaviour of the TLSs is less obvious, and the system might remain stable to the weak RKKY-like interactions even at low energies.


\section{Discussion and Outlook}
\label{discussion section}

In this work, we have studied a class of large $N$ models of itinerant electrons interacting with local two-level systems via spatially random couplings. These models, inspired by the possibility of metallic glassiness in strongly correlated materials, exhibit a remarkably rich phenomenology at low energies. Most strikingly our theory hosts a robust extended NFL phase in a considerable part of parameter space. At the crossover from FL to NFL our theory realizes a MFL 
that shows strange metallic behavior with $T$-linear resistivity and $T\log(1/T)$ specific heat. Note that the MFL/NFL behavior does not necessitate the existence of a quantum critical point.
Physically, the departure from FL behavior is rooted in the fact that the characteristic energy of each TLS is algebraically suppressed by the interaction, thus providing significant spectral weight of low-energy excitations which constitute an efficient scattering mechanism for the electronic degrees of freedom. These abundant low-energy excitations further manifest in a rich phenomenology of the critical transition temperature to the superconducting ground state of the system. 

The physical picture of the simplest variant of our theory (the $x$-model), studied in Ref.~\cite{bashan2023tunable}, qualitatively persists upon relaxing several simplifying assumptions, such as allowing for interactions with different operators of the TLSs and introducing arbitrary TLS-fields. 
Aiming at more realistic models, we further considered the effects of relaxing additional simplifications, such as $1/N$ corrections, spatial correlations in $\vec{g}_{ijl,\vec{r}}$ and the self averaging assumption. While these 
{tend to suppress} the NFL behaviour found in this work below some energy scale suppressed by powers of $N$, there are physical reasons to think that this scale remains small in a realistic setting. Specifically, recalling that TLSs in physical systems are extended objects, the interaction would retain a high degree of connectivity (i.e. each TLS would interact with many electrons and vice versa), which in turn could preserve the self averaging property, and frustrate effects of RKKY-like interactions.

It is interesting to ask what is the relation between the interaction strengths ($\alpha_{x,y,z}$) to actual physical knobs in realistic systems. This is a complicated question as the microscopic origin of such TLSs is not well understood.
However, there have been many studies attempting to provide a microscopic theoretical framework for understanding these objects~\cite{Lubchenko2001,Burnett2014, Khomenko2020, Khomenko2020,Khomenko2021,Ji2022, Mocanu2023}. It is possible that as the system approaches a 
glassy charge or spin ordering transition, 
the shape, size and other properties of these TLSs change, affecting the magnitude of their coupling to electrons, or the relative sizes of the couplings to the $x,y,z$ operators. Thus, tuning a physical knob of the system could be parameterized as a nontrivial path in the space of couplings, leading to a nontrivial variation of the exponent in the electronic self energy.

To this end, another issue concerns the density of states of TLSs, which is parametrically larger than that of the electrons (i.e. $h_{c,R}^{-1} \gg \rho_F$). A direct consequence is the seemingly enhanced coupling $\lambda \equiv \frac{M}{N}\frac{\rho_F}{h_{c,R}}\alpha$ that appears in the electronic self energy. It appears, however, natural to expect that $\alpha \sim \lambda$ at least up to some intermediate coupling strength. This is the case if $M/N \sim h^{-1}_{c,R}/\rho_F \ll 1$, i.e., if the TLSs are sparse compared to the electrons. Physically, this seems plausible based on the mesoscopic considerations mentioned above.

Non-Fermi liquid behavior is ultimately tied to an anomalous spectrum
of gapless excitations. Such a spectrum is usually believed to emerge
from  collective modes with soft long wavelength fluctuations.
As we showed in this paper, it can also be the result of quantum fluctuations
of modes that are localized in a region of size $l$, where each mode
has an excitation gap $E_{{\rm min}}\sim h_{R}$ but is governed by
a singular distribution function ${\cal P}\left(h_{R}\right)\propto h_{R}^{\gamma-1}$
with $\gamma>0$. Even if the correlation function for a given $h_{R}$
decays rapidly
in time, $\chi_{h_{R}}\left(\tau\right)\sim\exp\left(-h_{R}\tau\right)$, the average $\chi_{{\rm av}}\left(\tau\right)=\int dh_{R}{\cal P}\left(h_{R}\right)\chi_{h_{R}}\left(\tau\right)$
then decays like a powerlaw $\propto\tau^{-\gamma}$ and the system becomes critical. 
For the static
susceptibility, $\chi_{{\rm av}}\left(T\right)=\int^{1/T}d\tau\chi_{{\rm av}}\left(\tau\right)$, 
it follows that $\chi_{{\rm av}}\left(T\right)\propto T^{\gamma-1}$ and $C\propto T^{\gamma}$
for the heat capacity. Non-Fermi liquid behavior occurs for $\gamma<1$.

Such a singular distribution function was also obtained from quantum
Griffiths behavior~\cite{Thill1995}. Let us therefore compare and contrast our results
with the ones that follow from quantum Griffiths physics, where rare,
large droplets of size $l$ occur with probability $p_{l}\propto e^{-cl^{d}}$
and possess an exponentially small gap $h_{l}\propto e^{-bl^{d}}$~\cite{Thill1995}.
This yields a power-law for 
\begin{equation}
{\cal P}\left(h_{R}\right)=\int dl^{d}p_{l}\delta\left(h_{R}-h_{l}\right)\propto h_{R}^{\gamma-1},    
\end{equation}
with non-universal exponent $\gamma=b/c$. Exponentially small gaps
occur for the random transverse field Ising model~\cite{Thill1995}. However, as soon
as one includes the coupling to conduction electrons, large droplets
will freeze by the Caldeira-Leggett mechanism, and one rather finds
super-paramagnetic behavior of classical droplets~\cite{Millis2001,Millis2002}. On the other hand,
for systems with a continuous order parameter symmetry power law quantum
Griffiths behavior becomes possible even in the presence of particle-hole
excitations~\cite{Vojta2005}. This behavior was also seen in recent numerical simulations~\cite{patel2023localization}.
In contrast to this quantum Griffiths behavior, in our approach we consider the coupling of TLSs of characteristic size
 of several lattice spacings to conduction electrons. While isolated TLSs
are governed by ${\cal P}\left(h\right)\propto h^{\beta}$
that is, on its
own, not sufficiently singular ($\beta > 0$), strong local
quantum fluctuations due to the coupling to conduction-electrons renormalize
the excitation gap $h\rightarrow h_{R}\sim h^{1/(1-\alpha)}$, which
reduces the exponent $\beta+1\rightarrow \beta_{R}+1 =  \left(\beta+1)(1-\alpha\right)$.

While our theory does not aim to realistically describe any specific material, the existence of a tunable non-Fermi liquid phase in a controlled microscopic theory could shed new light on some aspects of strange metallicity. It provides a novel viewpoint on the widely observed extended strange metal regime \cite{cooper_anomalous_2009,hussey_generic_2013,hussey_dichotomy_2011,hussey_tale_2018,lozano_strange-metal_2023} that does not rely on a putative quantum critical point. Further, while conventional wisdom typically interprets the resistivity in terms of a $T-$linear and a $T^2$ components, i.e. $\rho - \rho_0 = AT + BT^2$ \cite{park_isotropic_2008,cooper_anomalous_2009,doiron-leyraud_correlation_2009,hussey_dichotomy_2011,jin_link_2011,hussey_tale_2018,legros_universal_2019,berben_compartmentalizing_2022}, our theory offers an alternative interpretation\footnote{Following the procedure in Ref.~\cite{hussey_dichotomy_2011}, the derivative $d\rho /d T$ can distinguish the two behaviors provided that the $T\to 0$ resistivity is measured with sufficient accuracy.} where the exponent is a continuous parameter, $\rho - \rho_0 = CT^{\gamma}$. Interestingly, this interpretation (also known as power-law liquid) has been shown to be consistent with experimental data of strange metals \cite{dagan_evidence_2004,reber_power_2015,zhao_quantum-critical_2019,smit_momentum-dependent_2021,harada_revised_2022}.

Several natural questions remain open. Aiming to better understand more realistic scenarios, a systematic study of our model for finite-$N$ is called for, either by analytical or numerical methods. In addition, the behavior deep inside the superconducting state might exhibit interesting new physics, as the electrons constituting the Ohmic bath in the normal state are becoming gapped, which has non-trivial effects on the TLSs and vice versa. More broadly, one may consider various other physical systems containing a coexistence of electrons and two-level systems, where the framework developed in this work can be applied.



\textit{Acknowledgements.---}
We thank G. Grissonnanche, S. A. Kivelson, C. Murthy, A. Pandey, B. Ramshaw, and B. Spivak for numerous discussions and for a collaboration on prior unpublished work. 
We are grateful to Natan Andrei, Girsh Blumberg, Andrey Chubukov, Rafael Fernandes, Tobias Holder, Yuval Oreg, and Alexander Shnirman for helpful discussions. J.S. was supported by the German Research Foundation (DFG) through CRC TRR 288 ``ElastoQMat,'' project B01 and a Weston Visiting Professorship at the Weizmann Institute of Science. E.B. was supported by the European Research Council (ERC) under grant HQMAT (Grant Agreement No. 817799) and by the Israel-US Binational Science Foundation (BSF). 
Some of this work was performed at KITP, supported in part by the National Science Foundation under PHY-1748958.

\bibliography{TLSs.bib}

\appendix
\onecolumngrid

\section{Diagrammatic approach to mapping}
\label{diagrammatic mapping}
\begin{figure}[H]
\centering
\includegraphics[width=0.9\textwidth]{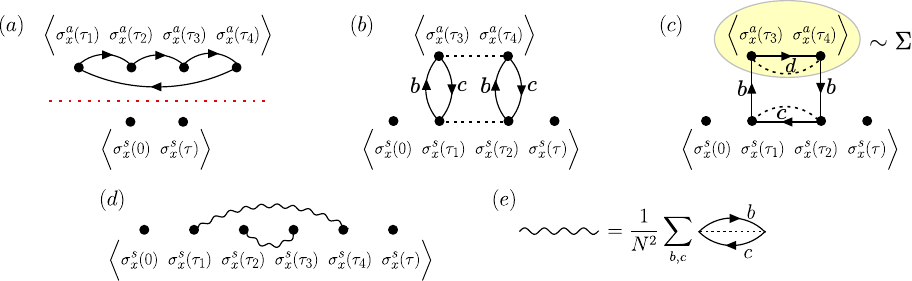} 
\caption{Examples of diagrams considered in the mapping. a) If all insertions are of other TLSs, the bubble disconnects from the ``external vertices" and cancels with the vacuum diagrams.  b) Other insertions of different TLSs become subleading in $N$. c) The contribution of insertions of a different TLS $a\neq s$ can be absorbed into the full electron green function. d) Example of a contributing diagram, with the wavy lines representing particle hole pairs (as defined in e)). Here electrons are denoted by solid line, contractions over realizations of $g_{abc}$ by dashed lines and TLS operators, who do not admit a direct diagrammatic expansion, by full circles.
}
\label{fig:casesformapping}
\end{figure}

We now present an alternative approach for the mapping to the spin-boson model, where we demonstrate, by a perturbative expansion of the interaction, that the electrons constitute an Ohmic bath to the TLSs. Importantly, due to the spatial randomness of the couplings, the bath is that of non-interacting particle-hole pairs, i.e., it is independent of the electronic self energy. Consider the correlation function for a single TLS of flavor $s$:
$S^x_{s}\left(\tau\right)\equiv\left\langle T_{\tau}\left\{\sigma_s^{x}\left(\tau\right)\sigma_s^{x}\left(0\right)\right\} \right\rangle$. The expansion in interaction vertices reads (we suppress the spatial index $\boldsymbol{r}$ since all operators act on the same site): 
\begin{eqnarray}
S_{s}^{x}\left(\tau\right)=\left\langle T_{\tau}\left\{ \sigma_{s}^{x}\left(\tau\right)\sigma_{s}^{x}\left(0\right)\sum_{n=0}^{\infty}\frac{\left(-1\right)^{n}}{n!}\left[\prod_{i=1...n}\left(\int d\tau_{i}\sum_{abc}\frac{g_{abc}}{N}\sigma_{a}^{x}\left(\tau_{i}\right)c_{b}^{\dagger}\left(\tau_{i}\right)c_{c}\left(\tau_{i}\right)\right)\right]\right\} \right\rangle ,
\end{eqnarray}
which decouples into a sum of terms of the form 
\begin{eqnarray}
I_n &=& \int_{\tau_{1},\tau_{2}...\tau_{n}}\sum_{a_{1}b_{1}c_{1}a_{2}b_{2}c_{2}...}\left(\frac{g_{a_{1}b_{1}c_{1}}}{N}\frac{g_{a_{2}b_{2}c_{2}}}{N}...\right)\left\langle T_{\tau}\left\{ \sigma_{s}^{x}\left(\tau\right)\sigma_{s}^{x}\left(0\right)\left(\sigma_{a_{1}}^{x}\left(\tau_{i}\right)...\sigma_{a_{n}}^{x}\left(\tau_{n}\right)\right)\right\} \right\rangle \nonumber\\
&\times&\left\langle T_{\tau}\left\{ c_{b_{1}}^{\dagger}\left(\tau_{i_{1}}\right)c_{c_{1}}\left(\tau_{i_{1}}\right)c_{b_{2}}^{\dagger}\left(\tau_{i_{2}}\right)c_{c_{2}}\left(\tau_{i_{2}}\right)...\right\} \right\rangle 
\end{eqnarray}
By integrating over the realizations of $g_{abc}$, we note that  (i) terms where all interaction TLS indices $a_{1},...,a_{n}\ne s$ are ``disconnected'' and
cancel with the vacuum diagrams; (ii) if only some (but not all) $a_{i}=s$, the contribution is either
subleading in $1/N$ or corresponds to a self-energy insertion for
the electrons (see Fig.~\ref{fig:casesformapping}). Thus if we treat the electrons self consistently as being fully dressed, the only relevant insertions of the interaction are those in which $a_i=s$.

Hence $I_n$ can be written as (all TLS indices $=s$)
\begin{eqnarray}\nonumber
I_n = \int_{\tau_{1},\tau_{2}...}\sum_{b_{1}c_{1}b_{2}c_{2}...}\left(\frac{g_{sb_{1}c_{1}}}{N}\frac{g_{sb_{2}c_{2}}}{N}...\right)\left\langle T_{\tau}\left\{ \sigma_{s}^{x}\left(\tau\right)\sigma_{s}^{x}\left(0\right)\left(\sigma_{s}^{x}\left(\tau_{i}\right)...\sigma_{s}^{x}\left(\tau_{n}\right)\right)\right\} \right\rangle \\
\times\left\langle T_{\tau}\left\{ c_{b_{1}}^{\dagger}\left(\tau_{i_{1}}\right)c_{c_{1}}\left(\tau_{i_{1}}\right)c_{b_{2}}^{\dagger}\left(\tau_{i_{2}}\right)c_{c_{2}}\left(\tau_{i_{2}}\right)...\right\} \right\rangle 
\end{eqnarray}
Considering the electronic part (including the couplings $g_{sb_ic_i}$), we see that the leading order contribution in $1/N$ corresponds to terms with
$b_{1}\ne c_{1}\ne b_{2}...$ (all indices are distinct), namely,
\begin{eqnarray} 
C_n &\equiv&\sum_{b_{1}c_{1}b_{2}c_{2}...}\frac{1}{N^{n}}\left(g_{sb_{1}c_{1}}^{2}g_{sb_{2}c_{2}}^{2}...\right)\left\langle T_{\tau}\left\{ c_{b_{1}}^{\dagger}\left(\tau_{i_{1}}\right)c_{c_{1}}\left(\tau_{i_{1}}\right)c_{c_{1}}^{\dagger}\left(\tau_{i_{2}}\right)c_{b_{1}}\left(\tau_{i_{2}}\right)\right\} \right\rangle \\
 & \times&\left\langle T_{\tau}\left\{ c_{b_{2}}^{\dagger}\left(\tau_{i_{1}}\right)c_{c_{2}}\left(\tau_{i_{3}}\right)c_{c_{2}}^{\dagger}\left(\tau_{i_{4}}\right)c_{b_{2}}\left(\tau_{i_{4}}\right)\right\} \right\rangle ...+{\rm permutations} + \mathcal{O}\left( \frac{1}{N} \right).
\end{eqnarray}
The particle-hole pairs obey the bosonic Wick's theorem since the couplings $g_{abc}$ obey it. Therefore, 
\begin{eqnarray}
C_{n}=\tilde{J}\left(\tau_{i_{1}}-\tau_{i_{2}}\right)\tilde{J}\left(\tau_{i_{3}}-\tau_{i_{4}}\right)...
+{\rm permutations}
\end{eqnarray}
where we have denoted 
\begin{eqnarray}
    \tilde{J}\left(\tau_{i_{1}}-\tau_{i_{2}}\right)\equiv g^{2}G\left(\tau_{i_{1}}-\tau_{i_{2}}\right)G\left(\tau_{i_{2}}-\tau_{i_{1}}\right)
\end{eqnarray}
As mentioned earlier, the fact that the bath is Ohmic follows from spatial randomness of the couplings, which translates to the TLS being coupled to the local particle-hole correlators, with $G\left(\tau\right)=\int_{\boldsymbol{k}}G\left(\tau,\boldsymbol{k}\right)$. Upon analytical continuation, the spectral function of the bath is given by
\begin{eqnarray}
\tilde{J}\left(\omega\right)=\frac{g^{2}\rho_F^{2}}{2\pi}\omega\equiv\eta\omega,
\end{eqnarray}
such that the dimensionless coupling strength (using the conventions of Ref.~\cite{weiss_quantum_2012}) is given by 
\begin{eqnarray}
\alpha\equiv\frac{2}{\pi}\eta=\frac{g^{2}\rho_F^{2}}{\pi^{2}}.
\end{eqnarray}
Generalizing the derivation to the $xyz-$model is straightforward.

\section{Details of RG flow of 1bSB}
\label{Appendix: rg 1bsb}
We present here 
the calculation of the renormalized scale $ h_R$ in the different regimes $\rm{I-III}$ in the 1bSB. As a reminder, the flow equations are, to order $\tilde{h}^2$:
\begin{eqnarray}
    \frac{d\alpha}{d\ell} &=& -\tilde{h}^2\alpha \\
    \frac{d\tilde{h}}{d\ell} &=& \left(1-\alpha\right)\tilde{h} 
\end{eqnarray}
We allow the couplings to flow until there is only one energy scale in the problem, i.e. until the cutoff $\Lambda = \omega_c*\exp(-\ell)$ and TLS energy $ h_R = \tilde{h}\Lambda'$ become equal, which is given by $\tilde{h}(\ell^*)=1$.

We start with the regime where, $\left(1-\alpha\right)\gg\tilde{h}^2$, so that the flow of $\alpha$ is much slower than the flow of $\tilde{h}$. We can thus solve the flow of $\tilde{h}$ while treating $\alpha$ as a constant, and the weak change in $\alpha$ at the end of the flow (when $\tilde{h}$ approaches 1) will only change the result by a multiplicative factor, which is absorbed into the definition of the prefactor $c_{\alpha}$ in Eq.~\eqref{deltacalpha}. Therefore, allowing $\tilde{h}$ to flow until it reaches the value $1$ we find that
\begin{equation}
    (1-\alpha)\ell^* = \log{\frac{\omega_c}{h}}  
    \label{constant alpha flow}
\end{equation}
Inserting Eq.~\eqref{constant alpha flow} into the definition of $ h_R$, we find that 
\begin{equation}
     h_R = c_\alpha \omega_c \left(\frac{h}{\omega_c}\right)^\frac{1}{1-\alpha}
    \label{deltacalpha}
\end{equation}
As mentioned in the main text, this prefactor $c_\alpha$ cannot be determined merely from the RG flow. However, it can be extracted using exact techniques such as bosonization or Bethe ansatz \cite{hur_entanglement_2008,camacho_exact_2019,filyov_method_1980,ponomarenko_resonant_1993}, and is given by 
\begin{equation}
    c_\alpha =\left(\Gamma(1-\alpha) \exp\left(\alpha\log{\alpha} + (1-\alpha)\log(1-\alpha)\right) \right)^{\frac{1}{1-\alpha}}
\end{equation}
which satisfies the two known limits $c_0=1,c_{1/2}=\pi/4$.
We now study the regime near the BKT transition, $\alpha\approx1$. In this regime, we define $J=1-\alpha$ such that $|J|\ll1$. The RG equations then approximately become:
\begin{eqnarray}
    \frac{dJ}{d\ell} &=& \tilde{h}^2\\
    \frac{d\tilde{h}}{d\ell} &=& \tilde{h}J 
\end{eqnarray}
Note that the combination $x_0\equiv \tilde{h}^2-J^2$ obeys:
\begin{equation}
    \frac{1}{2}\frac{dx_0}{d\ell}=\tilde{h}\frac{d\tilde{h}}{d\ell}-J\frac{dJ}{d\ell}  =0
\end{equation}
so $x_0$ is constant along flow lines. Using this relation the equations can thus be solved easily
\begin{eqnarray}
\frac{d\tilde{h}}{d\ell} & =&\tilde{h}\sqrt{\tilde{h}^{2}-x_{0}}\\
\ell^* & =&\int_{\tilde{h}_{0}}^{1}\frac{d\tilde{h}}{\tilde{h}\sqrt{\tilde{h}^{2}-x_{0}}}.
\end{eqnarray}
where $\tilde{h}_0 = h/\omega_c$. We now separate to the cases where $x_0>0$ and $x_0<0$. If $x_0<0$, we have that
\begin{eqnarray}
\ell^*&=&\frac{\text{asinh}\left(\frac{\sqrt{-x_{0}}}{\tilde{h}_{0}}\right)-\text{asinh}\left(\sqrt{-x_{0}}\right)}{\sqrt{-x_{0}}} \\
 h_R &=& \omega_{c}\left(\frac{\frac{\sqrt{-x_{0}}}{\tilde{h}_{0}}+\sqrt{1-\frac{x_{0}}{\tilde{h}_{0}^{2}}}}{\sqrt{-x_{0}}+\sqrt{1-x_{0}}}\right)^{-\frac{1}{\sqrt{-x_{0}}}}
\end{eqnarray}
If $\tilde{h}_{0}^2\ll J$ this expression simplifies to the power law given earlier. On the other hand, for $x_0\to0$ this expression becomes the familiar Kondo scale of the isotropic Kondo model $ h_R \propto \omega_c \exp(-{1}/{\tilde{h}_{0}})$.
For $x_0>0$, we obtain that
\begin{equation}
\label{l star BKT}
\ell^*=\frac{\text{atan}\left(\sqrt{\frac{1-x_{0}}{x_{0}}}\right)- {\rm{sgn}}(J)\text{atan}\left(\sqrt{\frac{\tilde{h}_{0}^{2}-x_{0}}{x_{0}}}\right)}{\sqrt{x_{0}}}.
\end{equation}
Taking $x_0\to 0$ also gives the isotropic Kondo result $\ell = 1/\tilde{h}_{0}$. Setting $J=0$ we find that $\ell= \pi/2\tilde{h}_{0}$. Therefore, in this regime we can approximately think of the renormalized scale as taking the form $ h_R \propto \omega_c \exp(- b(J,\tilde{h}_{0})/\tilde{h}_{0})$ with $b$ being a slowly varying function of order $1$.
However, when the flow approaches the BKT line $J=-\tilde{h}$ this approximation does not hold, and instead the renormalized scale is set by the distance from the transition:
\begin{equation}
\label{Delta r BKT}
     h_R \propto \omega_c \exp\left(-\frac{\pi}{\sqrt{x_{0}}}\right).
\end{equation}

\section{Sum rules for the 1bSB}
This is based on a short analysis first derived in \cite{guinea_dynamics_1985}.
We define the correlation function
\begin{equation}
    \left<\sigma_x(t) \sigma_x(0)\right> = \int_\omega e^{i\omega t }A_x(\omega)
\end{equation}
This is related to the dynamical susceptibility by the fluctuation dissipation theorem \cite{auerbach_interacting_1998}:
\begin{equation}
    A_x(\omega)=\left(1+\coth\left(\frac{\beta\omega}{2}\right)\right)\chi''_x(\omega)
\end{equation}
Additionally, the following equation of motion derives from the Hamiltonian Eq.~\eqref{1bSB Hamiltonian}:
\begin{eqnarray}
    i\frac{d\sigma_x}{dt}=-2h\sigma_y
\end{eqnarray}
Thus, by Fourier transforming $\left<\sigma_x(t) \sigma_x(0)\right>,\left<\frac{d\sigma_x}{dt} (t) \sigma_x(0)\right>$ and $\left<\frac{d\sigma_x }{dt}(t)\frac{d\sigma_x}{dt}(0)\right>$, setting $t=0$ and using the antisymmetry of $\chi''_x(\omega)$ we obtain the three sum rules:
\begin{eqnarray}
    1&=&\int_\omega \chi''_x(\omega)  \coth\left(\frac{\beta\omega}{2}\right) \label{sum rule 1}\\ 
    2h\left<\sigma_z\right> &=&\int_\omega \omega\chi''_x(\omega)  \label{sum rule 2} \\ 
    4h^2&=& \int_\omega \omega^2 \chi''_x(\omega) \coth\left(\frac{\beta\omega}{2}\right)  \label{sum rule 3}
\end{eqnarray}

\section{Explicit calculation of $\overline{\chi''_x}$ for the $x-$model}
The average (imaginary part of the) susceptibility is given by
\begin{eqnarray}
    \overline{\chi''_x} &=& \int_0^{h_c} \chi''_x(\omega,h) \mathcal{P}_\beta(h)dh  \\
    &=& \int_0^{h_{c,R}} \frac{1}{\omega} f_\alpha \left(\frac{\omega}{ h_R}\right) \mathcal{P}_r( h_R) d h_R \\
    &=& {\rm{sgn}}(\omega) \int_{|\omega|/h_{c,R}}^\infty \frac{f_\alpha \left(x\right)}{x^2} \mathcal{P}_r(|\omega|/x) dx 
\end{eqnarray}
The result of this integral thus depends on the renormalized distribution $\mathcal{P}_r$.
\subsection{$\alpha<1$}
Starting with $ h_R=c_\alpha \omega_c (h/\omega_c)^{1/(1-\alpha)}\Rightarrow h = ( h_R/c_\alpha)^{1-\alpha}\omega_c^{-\alpha}$:
\begin{eqnarray}
    \mathcal{P}( h_R) &=& \mathcal{P}(h) \left(\frac{dh}{d h_R}\right)\\
    &=& \mathcal{N}  h_R^{\beta(1-\alpha)-\alpha}
\end{eqnarray}
Since the distribution is cut off at $h_{c,R}= h_R(h_c)$, the normalization constant must be $\mathcal{N}=\gamma/h_{c,R}^\gamma$, with $\gamma=\beta(1-\alpha)-\alpha+1=(1+\beta)(1-\alpha)$. Inserting this into the averaged susceptibility gives:
\begin{eqnarray}
    \overline{\chi''_x} &=& {\rm{sgn}}(\omega) \int_{|\omega|/h_{c,R}}^\infty \frac{f_\alpha \left(x\right)}{x^2} \frac{\gamma|\omega|^{\gamma-1}}{h_{c,R}^\gamma x^{\gamma-1}} dx \\
    & =& \frac{1}{\omega} \left|\frac{\omega}{h_{c,R}}\right|^\gamma \times \gamma \int_{|\omega|/h_{c,R}}^\infty \frac{f_\alpha \left(x\right)}{x^{\gamma-1}} dx
\end{eqnarray}
Using that fact that at long times $\chi(t) \propto 1/t^2$, we see that $\chi''_x(\omega/ h_R)\propto \omega \Rightarrow f_\alpha(x\ll1)\propto x^2$. Near the lower integration limit the integrand is $\propto 1/x^{\gamma-3}$. If $\gamma<2$ then the integral converges when taking $\omega/h_{c,R}\to0$, and can thus be considered as a constant. (If $\gamma>2$ then the integral diverges and the resulting frequency dependence is $\overline{\chi''_x}\propto \omega$. This is because the averaged susceptibility cannot decay faster than the susceptibility of the TLSs with highest $h$.)
\subsection{$\alpha\approx1$}
\label{Appendix: alpha=1}
In this case we use $ h_R=c_\alpha\omega_c\exp(-b\omega_c/h)
\Rightarrow h=\frac{b\omega_{c}}{2\log\left(\frac{c_\alpha \omega_{c}}{ h_R}\right)}$, which gives the renormalized distribution:
\begin{eqnarray}
\label{P Delta_r alpha=1}
    \mathcal{P}( h_R)
    &=& \frac{\mathcal{N}}{ h_R\log^{2+\beta}\left(\frac{c_\alpha\omega_c}{ h_R}\right)}
\end{eqnarray}
and the normalization can be found to be $\mathcal{N}=(1+\beta)\log^{1+\beta}(\omega_c/h_{c,R})$. We neglect for simplicity the factor of $c_\alpha \sim \mathcal{O}(1)$ inside the logarithm. The averaged susceptibility is then given by
\begin{eqnarray}
    \overline{\chi''_x} &=& {\rm{sgn}}(\omega) (1+\beta)\log^{1+\beta}(\omega_c/h_{c,R}) \int_{|\omega|/h_{c,R}}^\infty \frac{f_\alpha \left(x\right)}{x \log^{2+\beta}\left(x\omega_c/|\omega|\right)} dx 
\end{eqnarray}
In order to simplify the integral, we rely on the fact that $f_\alpha(x\gg1)\propto 1/x^{4-2\alpha}$ \cite{guinea_dynamics_1985} and $f(x\ll1)\propto x^2$, such that most of the weight of the integral is around $x\sim\mathcal{O}(1)$, for which $|\log(x)|\ll \log(\omega_c/|\omega|)$. Thus we may neglect the $x$ dependence inside the log, giving the form of the susceptibility presented in the main text (using the sum rule Eq.~\eqref{sum rule 1} for $\int_0^\infty f(x)/x dx=1/2$).

\subsection{Through the BKT transition ($1<\alpha<1+h_c/E_F$)}

\label{Appendix: BKT seperatrix}
\begin{figure}[H]
\centering
\includegraphics[width=1\textwidth]{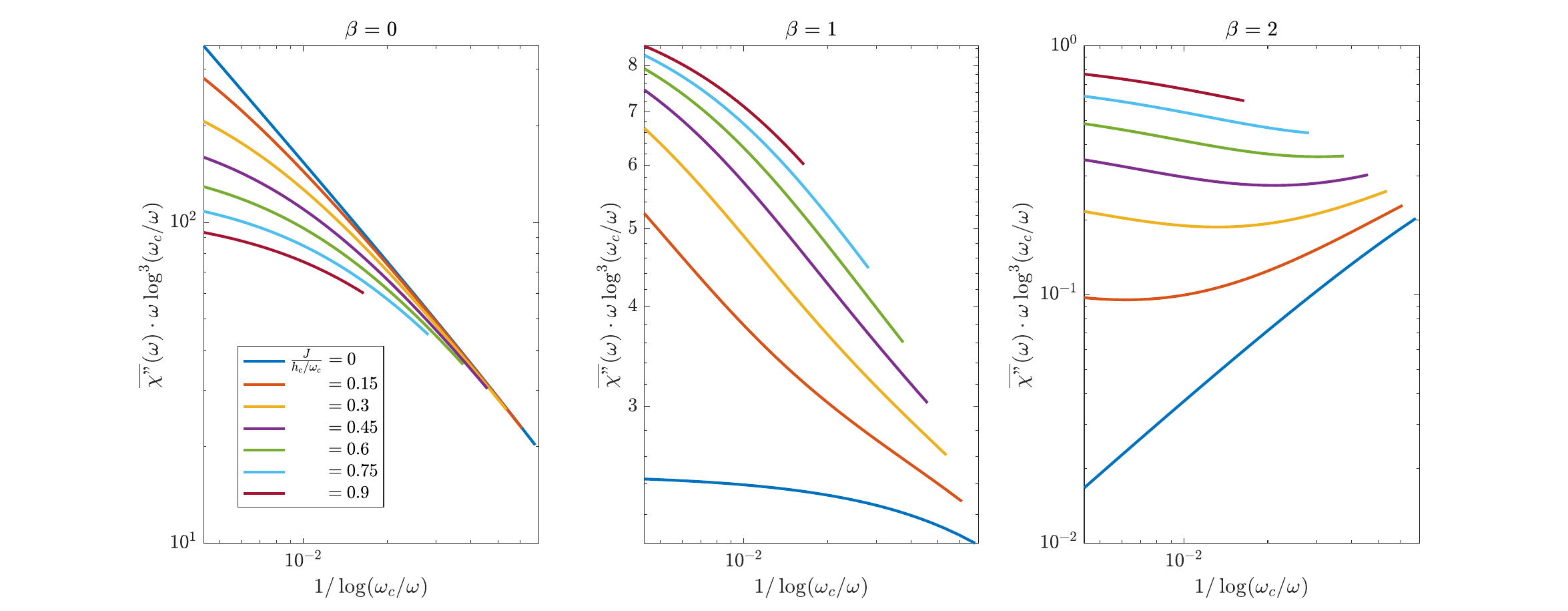} 
\caption{Averaged susceptibility of TLSs around the BKT transition, for $\beta=0,1,2$ and varying values of $1<\alpha<1+h_c/\omega_c$. As expected, for $\alpha=1$ ($J=0$) the susceptibility is $\propto 1/\omega \log^{\beta+2}(\omega_c/\omega)$, while as $\alpha\to1+h_c/\omega_c$ ($J\to h_c/\omega_c)$ this changes smoothly into $\propto 1/\omega \log^{3}(\omega_c/\omega)$. Note that the change in the cutoff of the values in the $x$ axis with increasing $J$ is due to the lowering of $h_{c,R}$.}
\label{fig:bkt appendix}
\end{figure}

The  behavior around the BKT transition is slightly more convoluted, since when $x_0\to 0^-$ the dependence of $ h_R$ on $\tilde{h}$ is slightly different. However, if we work close enough to the transition, we can just change variables to $y_0(\tilde{h})=\sqrt{-x_0}$, and use the form \eqref{Delta r BKT} which explicitly depends only on $y_0$. Changing variables we thus find that:
\begin{equation}
    \mathcal{P}(y_0) = \frac{1+\beta}{h_c^{1+\beta}} y_0 \left(y_0^2+J^2\right)^{\frac{-1+\beta}{2}}
\end{equation}
with the cutoff $y_c = \sqrt{(h_c/w_c)^2-J^2}$, and note that the range $y\in(0,y_c)$ covers only the range $h\in(|J|,h_c)$, since the TLSs with $h<|J|$ are in the localized phase.
Thus, if $y_c\ll J$ we can approximate $\mathcal{P}(y_0)\propto y_0$ for any $\beta$, and we will therefore find that the distribution of $\mathcal{P}( h_R)$ will be identical to \eqref{P Delta_r alpha=1} with $\beta=1$. Thus, while for $\alpha=1$ the exponent of the log will be $1+\beta$, it will change smoothly to $2$ near the end of the transition. 
We evaluate this numerically for any value of $1<\alpha<1+h_c/\omega_c$ using the form given in \eqref{l star BKT}, and for the sake of the computation using the simplification $\chi''(\omega)=\delta(|\omega|-2 h_R)$ (since the results should not depend on the actual function $f(x)$ but rather on the form of the distribution $\mathcal{P}_r$). The results, confirming the analysis presented in this subsection and the previous one, are shown in Fig.~\ref{fig:bkt appendix}.

\subsection{Localized phase ($\alpha>1+h_c/E_F$)}
\label{Appendix 3-2a}
In the localized phase, where $h_R=0$, most of the weight of $\chi''_x(\omega)$ lies in a delta function at zero frequency. However, there are still weak residual quantum fluctuations. The form of these fluctuations can be found using a simple scaling analysis: we write the susceptibility at finite frequency as some function $\chi''_x(\omega)=\frac{1}{\omega}F(h/E_F,|\omega|/E_F)$. Reducing the cutoff to $E_F\to E_F/b$, the field rescales to $h/E_F\to h/E_F/b^(1-a)$. Since the result must be independent of $b$, we can set $b=|\omega|/E_F$ and find that 
\begin{equation}
    \chi''_x=\frac{1}{\omega}F\left(\frac{h}{E_F}\left(\frac{E_F}{|\omega|}\right)^{1-\alpha},1\right) = \frac{1}{\omega}F\left(\frac{h(\omega)}{|\omega|},1\right)
\end{equation}
with $h(\omega)=h(|\omega|/E_F)^\alpha$ the frequency dependant energy scale. Since $h(\omega)\ll\omega$, we may expand to second order using Fermi's golden rule, and find 
\begin{equation}
    \chi''_x(\omega) \propto \frac{1}{\omega} \left(\frac{h(\omega)}{\omega}\right)^2 \propto  {\rm sign}(\omega)  \frac{h^2}{E_F^{2\alpha}|\omega|^{3-2\alpha}}
\end{equation}
The constant of proportionality may be set using the sum rule Eq.~\eqref{sum rule 3}, and then averaging over $h$ we obtain Eq.~\eqref{alpha>1 self energy} of the main text.
\subsection{Biased case}
\label{biased case}

For the biased case, the $\left\langle \sigma_{x}\right\rangle $
has an equilibrium value, so that $\left\langle \sigma_{x}\left(t\to\infty\right)\sigma_{x}\left(0\right)\right\rangle \to\left\langle \sigma_{x}\right\rangle ^{2}$.
We therefore decompose 
\begin{eqnarray}
\chi''_{x}\left(\omega\right)=\left\langle \sigma_{x}\right\rangle ^{2}\delta\left(\omega\right)+\chi''_{inel}\left(\omega\right)
\end{eqnarray}

The equilibrium value, which contributes to the elastic scattering rate,
is given by \cite{sassetti_correlation_1990}
\begin{eqnarray}
\left\langle \sigma_{x}\right\rangle =\frac{2}{\pi}\text{atan}\left(\frac{h_{x}}{ h_R}\right)
\end{eqnarray}

Since $h_{c}/h_{c,R}\propto\left(\omega_{c}/h_{c}\right)^{\alpha/\left(1-\alpha\right)}\gg1$,
when averaging over $h_{x},  h_R$ will not have much effect, and
thus 
\begin{eqnarray}
\overline{\left\langle \sigma_{x}\right\rangle ^{2}}=1-\mathcal{O}\left(\frac{h_{c,R}}{h_{c}}\right)
\end{eqnarray}
The inelastic contribution $\chi_{inel}''$ will now have a two-parameter
scaling form:
\begin{eqnarray}
\chi''_{inel}\left(\omega\right)=\frac{1}{\omega}f_{\alpha}\left(\frac{\omega}{ h_R},\frac{h_{x}}{ h_R}\right)
\end{eqnarray}

For $\alpha<1$, the distributions are $\mathcal{P}\left( h_R\right)=\frac{\gamma}{h_{c,R}^{\gamma}} h_R^{\gamma-1},\mathcal{P}\left(\epsilon\right)=\frac{1+\beta_{x}}{h_{c}^{1+\beta_{x}}}h_{x}^{\beta_{x}}.$
Thus:
\begin{align*}
\overline{\chi''}_{inel}\left(\omega\right) & =\int\mathcal{P}\left( h_R,h_{x}\right)\frac{1}{\omega}f_{\alpha}\left(\frac{\omega}{ h_R},\frac{h_{x}}{ h_R}\right)d h_Rdh_{x}\\
 & =\left(1+\beta_{x}\right)\gamma\frac{1}{h_{c,R}^{\gamma}h_{c}^{\beta_{x}+1}}\int h_R^{\gamma+\beta_{x}}y^{\beta_{x}}f_{\alpha}\left(\frac{\omega}{ h_R},y\right)\frac{d h_R}{\omega}dy\\
 & =\left(1+\beta_{x}\right)\gamma\frac{\omega^{\gamma+\beta_{x}}}{h_{c,R}^{\gamma}h_{c}^{\beta_{x}+1}}\int_{\omega/h_{c,R}}^{\infty}\frac{dx}{x^{\gamma+\beta_{x}+2}}\int_{0}^{h_{c}x/\omega}dyf\left(x,y\right)
\end{align*}

Since $h_{c}/h_{c,R}\gg1$, the upper limit of the $y$ integral
is large for any value of $x>\omega/h_{c,R}$. Therefore defining
$\tilde{f}_{\alpha}\left(x\right)=\int_{0}^{\infty}f_{\alpha}\left(x,y\right)dy$,
we can rewrite the susceptibility in a form similar to earlier:
\begin{align*}
\overline{\chi''_{inel}}\left(\omega\right) & \approx\left(1+\beta_{x}\right)\gamma\frac{\omega^{\gamma+\beta_{x}}}{h_{c,R}^{\gamma}h_{c}^{1+\beta_{x}}}A_{\alpha}\\
A_{\alpha} & =\int_{0}^{\infty}\frac{\tilde{f}_{\alpha}\left(x\right)}{x^{\gamma+\beta_{x}+2}}dx
\end{align*}

Here we have assumed that the upper limit of the $y$ integral and
the lower limit of the $x$ integral can be continued to $\infty$
safely. In this case, the validity of this assumption is not as clear
as it was in the unbiased case. We verify this by an explicit calculation
at the TP. There, the scaling function is given exactly by {[}Ulrich
Maura{]}:

\begin{eqnarray}
f\left(x,y\right)=\frac{4}{\pi}\frac{1}{x^{2}+4}\left(x\text{atan}\left(x+y\right)+x\text{atan}\left(x-y\right)+\ln\left(\frac{\left(1+\left(x+y\right)^{2}\right)\left(1+\left(x-y\right)^{2}\right)}{\left(1+y^{2}\right)^{2}}\right)\right)
\end{eqnarray}

For large $y$, $f\left(x,y\right)\propto\frac{1}{y^{2}}$, so the
integral over $y$ indeed converges (this should generically be the
case since the single-TLS susceptibility is an analytic symmetric
function of $y$ which vanishes for $y\to\infty$). In this case the
integral can be evaluated exactly, and we find that:
\begin{eqnarray}
\tilde{f}\left(x\right)=\frac{4x^{2}}{x^{2}+4}
\end{eqnarray}

We confirm that $\tilde{f}\left(x\ll1\right)\propto x^{2}$, just
as in the 1bSB, and our approximation is justified as long as $\gamma+\beta_{x}\leq1$.

Near the critical point $\alpha\to1$, the splitting distribution
takes the form $\mathcal{P}\left( h_R\right)\propto\frac{1}{ h_R\left(\log\omega_{c}/ h_R\right)^{2+\beta_{z}}}$.
When integrating over $y$, the effective distribution will change
to $\mathcal{P}\left( h_R\right)\propto\frac{ h_R^{\beta_{x}}}{\left(\log\omega_{c}/ h_R\right)^{2+\beta_{z}}}$.
Therefore the self energy will be of the form:
\begin{eqnarray}
\Sigma''_{inel}\left(\omega\right)\propto\frac{\omega^{1+\beta_{x}}}{\left(\log\omega_{c}/ h_R\right)^{2+\beta_{z}}}
\end{eqnarray}

Note that for the physical case $\beta_{x}=\beta_{z}=0$ this will
result in MFL-like behavior around $\alpha\approx1$.

\section{RG flow of 2bSB}
\label{Appendix: rg 2bSB}
We now discuss the details of the RG flow of the 2-bath SB model. We will mainly consider the region of interest which is analogous to the $\alpha<1$ region in the 1bSB, where the effect of $h$ on the flow of the couplings is negligible and the renormalization of $ h_R$ is a power law.
Therefore we begin by examining the effect of the two couplings on each other. For example, for the $xy-$model the RG equations will be (as in Eq.~\eqref{RG2bSB})
\begin{equation}
    \frac{d\alpha_x}{d\ell}=\frac{d\alpha_y}{d\ell}=-2\alpha_x\alpha_y +\mathcal{O}(\tilde{h}^2).
\end{equation}
We can simplify these equations by using the constant of flow $\delta\alpha=\alpha_x-\alpha_y$, which is approximately conserved along flow. We assume $\delta\alpha>0$ without loss of generality. We can then simply integrate the equations:
\begin{equation}
    2\ell = \int^{\alpha_x^0}_{\alpha_x(\ell)} \frac{d\alpha_x}{\alpha_x(\alpha_x-\delta\alpha)} = \frac{\log\left(\frac{r}{1-\delta\alpha/\alpha_x}\right)}{\delta\alpha} 
\end{equation}
where $r=\alpha_y^0/\alpha_x^0$, and $\alpha_a^0$ are the bare couplings. We thus find that:
\begin{eqnarray}
    \alpha_x(\ell) &=& \frac{\delta\alpha}{1-r e^{-2\delta\alpha\ell}} \\
    \alpha_y(\ell) &=& \frac{r\delta\alpha}{e^{2\delta\alpha\ell}-r}
\end{eqnarray}
Assuming that the initial $h/\omega_c\ll1$ is small enough, the flow will reach $\delta\alpha\ell\ll1$, at which point the dominant coupling, which in this case is $\alpha_x$, saturates at the value $\alpha_{x,R}=\delta\alpha$, while the subleading coupling continues to decrease: $\alpha_{y,R} = r\delta\alpha (\Lambda'/\omega_c)^{\delta\alpha}$. Since the flow stops when $ h_R=\Lambda'$ then in the low energy theory $\alpha_{y,R} = r\delta\alpha( h_R/\omega_c)^{\delta\alpha}$. Once this point has been reached, we can examine the beta function of $\tilde{h}$:
\begin{equation}
    \frac{d\tilde{h}}{d\ell}= (1-\alpha_x-\alpha_y) \tilde{h}
    \label{beta fun h xy}
\end{equation}
 As mentioned above, after some ``time'' $\ell\sim\delta\alpha^{-1}$ (which importantly does not depend on the initial value of $h/\omega_c$), $\alpha_x$ will saturate, while $\alpha_y$ becomes negligible. Thus at this point the flow is identical to the flow of the 1bSB, with $\alpha = \delta\alpha$. We therefore find that if $\delta\alpha>1$ the tunneling flows to zero and the TLS becomes localized, while for $\delta\alpha<1$ the renormalized tunneling assumes the familiar form $ h_R \propto \omega_c (h/\omega_c)^{1/(1-\delta\alpha)}$. Note that in this case the proportionality constant will depend on the time it took $\alpha_x$ to saturate, which is a quantity which depends on $\alpha_x^0,\alpha_y^0$ and not on $h,\omega_c$. This can be found exactly by insetring $\alpha_{x,y}(\ell)$ into \eqref{beta fun h xy} and integrating:
\begin{eqnarray}
    \log{\frac{\omega_c}{h}} &=& \int_0^{\ell^*} \left(1+\delta\alpha\frac{1+re^{-2\delta\alpha\ell}}{1-re^{-2\delta\alpha\ell}} \right)d\ell = (1-\delta\alpha)\ell^* - \log\left(\frac{1-re^{-2\delta\alpha\ell^*}}{1-r}\right) \\ 
    \Rightarrow h_R &\propto&  (1-r)^{\frac{2}{1-\delta\alpha}}\omega_c\left(\frac{h}{\omega_c}\right)^{\frac{1}{1-\delta\alpha}}
\end{eqnarray}
    
The flow of the couplings in the $xz-$model is identical. However, if the dominant coupling is $\alpha_z$ then after $\ell \gtrsim \delta\alpha^{-1}$ the flow of $h$ will slow down, and thus $h$ will only by renormalized  by a multiplicative factor. We find in this case (analogous to only inserting $\alpha_y(\ell)$ into \eqref{beta fun h xy}):
\begin{eqnarray}
    \log{\frac{\omega_c}{h}} &=& \int_0^{\ell^*} \left(1+\delta\alpha\frac{re^{-2\delta\alpha\ell}}{1-re^{-2\delta\alpha\ell}} \right)d\ell = \ell^* - \log\left(\frac{1-re^{-2\delta\alpha\ell^*}}{1-r}\right) \\ 
    \Rightarrow h_R &=&  \frac{\delta\alpha}{\alpha_z} h
\end{eqnarray}
    
\section{Subleading corrections in $xyz-$model}
\label{Appendix: subleading}
Following the methods of Ref.~\cite{guinea_dynamics_1985}, we characterize the magnitude of the different subleading corrections in the multi-bath case. There are two types of subleading corrections: the susceptibilities of the subdominant baths which appear in the self energy, and perturbative corrections to the susceptiblity of the dominant bath due to the weak coupling to the losing baths.
We will study these in the $xy-$model and in the $xz-$model, and the generalization to the $xyz-$model is straightforward since the couplings to the subleading baths are perturbative in the low energy theory.

\subsection{$xy-$model}
As usual we will assume without loss of generality that $\alpha_x>\alpha_y$. As mentioned above, there are two types of corrections to the self energy.
We start with that due to perturbative corrections to $\chi_x$. As presented in \cite{guinea_dynamics_1985}, if only the $x$ bath was present after integrating out the high energy modes, we could expand the ground and excited states as (performing perturbation theory in the low-energy modes):
\begin{eqnarray}
    \left|g\right>_0 &=&  \left|\tilde{\downarrow }\right> + \frac{\phi_x}{ h_R} \left|\tilde{\uparrow} \right> +\frac{1}{2} \left(\frac{\phi_x}{ h_R}\right)^2\left|\tilde{\downarrow} \right> + \cdots \\
    \left| {\omega_i,x} \right>_0 &=&  b_{x,i}^\dagger \left| g \right> + \cdots
    \label{guinea perturbation theory}
\end{eqnarray}
where $\phi_\alpha = \sum_i \frac{\sqrt{\Pi^a(\omega_i)}}{ h_R} (b_{a,i}^\dagger+b_{a,i})$, $\Pi^a(\omega)\propto \alpha_a^r $ and $b_{a,i}^\dagger,b_{a,i}$ are respectively the bath operator, bath spectral function, and boson creation and annihilation operators of the $a$ bath. Importantly, the states $\left|\tilde{\uparrow},\tilde{\downarrow}\right>=1/\sqrt{2}\left(\left|\tilde{+}\right>\pm\left|\tilde{-}\right>\right)$ are superpositions of the \textit{high-frequency-moded dressed} $x$ states $\left|\tilde{\pm}\right>$. 
This gives the expected $\chi_x(\omega\ll h_R)\propto \alpha_{x,R} \omega/ h_R^2$ at low frequencies, but for general frequencies should be treated in a non-perturbative manner in $\alpha_{x,R}$. However, since $\alpha_{y,R}$ is small, we can add it perturbatively only to first order:
\begin{eqnarray}
    \left|g\right> &\approx& \left|g\right>_0 -i\frac{\phi_y}{ h_R}\left(1+\left(\frac{\phi_x}{ h_R}\right)^2+\cdots\right)\left|\tilde{\uparrow} \right> 
\end{eqnarray}
and the relevant excited states will involve insertions of one $y$ boson with multiple $x$ bosons. Using the spectral decomposition for $\chi_x$:
\begin{eqnarray}
    \chi_x(\omega) = \sum_n \left|\left<n\left|\sigma_x\right|g\right> \right|^2 \delta(E_n-\omega)
\end{eqnarray}
we find that the leading correction will come from  matrix elements of the form $\left<{\omega_1,\cdots\omega_{2k},x;\omega_j,y}\left|\sigma_x\right|g\right>$. While the summation over the many orders of $\phi_x$ is nontrivial, we know that it must produce a scaling function that only depends on $\omega/ h_R$, and we may thus write:
\begin{eqnarray}
    \chi_x''(\omega,\alpha_{y,R}) &=& \chi_x''(\omega,0) + \alpha_{y,R} \frac{1}{\omega} \tilde{f}_{\alpha_{x,R}}\left(\frac{\omega}{ h_R}\right) \\
    &=&  \frac{1}{\omega} f_{\alpha_{x,R}}\left(\frac{\omega}{ h_R}\right) + \alpha_y \left(\frac{ h_R}{\omega_c}\right)^{\alpha_{x,R}} \frac{1}{\omega} \tilde{f}_{\alpha_{x,R}}\left(\frac{\omega}{ h_R}\right)
\end{eqnarray}
where in the second line we inserted the expression for $\alpha_y$. While averaging over the first term will give the usual contribution, in the second term we can treat the distribution as effectively having an increased exponent $\tilde{\mathcal{P}}_r\sim h_R^{\gamma-1+\alpha_{x,R}}$, which will in turn produce a term with a subleading frequency dependence in the averaged susceptibility $\propto \omega^{\gamma-1+\alpha_{x,R}}$.

We now consider the susceptibility $\chi_y$, whose spectral decomposition is 

\begin{eqnarray}
    \chi_y(\omega) = \sum_n \left|\left<n\left|\sigma_y\right|g\right> \right|^2 \delta(E_n-\omega).
\end{eqnarray}
Using the fact that the bare $\sigma_y$ flips the TLS without properly adjusting the high-energy bosons, we have that
\begin{eqnarray}
\left<\tilde{\uparrow}\left|\sigma_y\right|\tilde{\downarrow}\right> \propto \frac{ h_R}{h} \propto \left(\frac{ h_R}{\omega_c}\right)^{\alpha_{x,R}}.
    \label{nonuniversal matrix element}
\end{eqnarray}
For small frequencies we may use the perturbative form $\eqref{guinea perturbation theory}$ and find that $\chi_y''\propto \left(\frac{ h_R}{\omega_c}\right)^{2\alpha_{x,R}}\chi_x''$, which will naively translate into a frequency dependence $\omega^{\gamma-1+2\alpha_{x,R}}$ in the averaged susceptibility. However, the averaged susceptibility depends on the full $\chi''_y$, and since for intermediate and high frequencies this perturbation theory is not applicable, we cannot fully determine the nonuniversal prefactor, and can only argue that $\overline{\chi_y''}\propto \omega^{\gamma-1+\epsilon}$ with $\epsilon>0$.

\subsection{$xz-$model}
We begin by studying the similar case where $\alpha_x>\alpha_z$. 
The susceptibility $\chi_x''$ will now acquire similar corrections due to $\alpha_z$. However, the matrix elements with single insertions of $\sigma_z\phi_z$ vanish, and we must instead go to second order in $\sigma_z\phi_z$. This means that the corresponding correction to the averaged susceptibility will be $\propto \omega^{\gamma-1+2\alpha_{x,R}}$.
The susceptibility $\chi_z''$ will be nonuniversal due to considerations identical to \eqref{nonuniversal matrix element}, and will thus be suppressed by a prefactor $\left(\frac{ h_R}{\omega_c}\right)^{2\alpha_{x,R}}$ at low frequencies, although we do not know the generalization of it to higher frequencies. However, in addition this susceptibility includes a static delta function peak due to the equilibrium value of $\left<\sigma_z\right>_\infty$. For small $\alpha_z\ll1$ this can be calculated using the sum rule  Eq.~\eqref{sum rule 2}:
\begin{equation}
\left<\sigma_z\right> = \frac{1}{h} \int_0^{\omega_c} f(\omega/ h_R)d\omega = \left(\frac{ h_R}{\omega_c} \right) ^{\alpha_x}  \times \left( \int_0^{\omega_c/ h_R} f(x) dx\right) 
\end{equation}
We must therefore find if this integral converges or diverges when the upper limit is taken to $\omega_c/ h_R\to\infty$. Ref.~\cite{guinea_dynamics_1985} shows that $f(x\gg1) \propto 1/x^{2-2\alpha}$, so that for $\alpha<1/2$ this integral converges to some constant $=A_\alpha$ while for $\alpha>1/2$ this integral diverges as a power law $= A_\alpha (\omega_c/ h_R)^{2\alpha-1}$ (for $\alpha=1/2$ it diverges logarithmicaly $\propto\log(\omega_c/ h_R)$). Thus we can write
\begin{equation}
    \chi_z''\approx  \delta(\omega) \times A_{\alpha_x} \left(\frac{h_{c,R}}{\omega_c}\right)^{\min(\alpha_x,1-\alpha_x)} +\cdots
\end{equation}
with the dots referring to the subleading frequency-dependent terms.

We now turn to the case where $\alpha_z>\alpha_x$. Here, with no $\alpha_x$ the ground state is exactly a coherent state of all the bosons centered around the location corresponding to $\left|\downarrow\right>$. Acting with the high-frequency mode dressed $\tilde{\sigma}_x$ will only agitate the low energy modes, and thus we can write $\left<\tilde{\uparrow}\left|\tilde{\sigma}_x\right|\tilde{\downarrow}\right> = s_x \sim \mathcal{O}(1)$. Therefore incorporating the effects of $\alpha_{x,R}$ perturbatively will modify the ground state as:
\begin{eqnarray}
    \left|g\right> =  \left(1+\frac{1}{2}s_x^2\left(\frac{\phi_x}{ h_R}\right)^2\right)\left|\tilde{\uparrow}\right> + s_x\frac{\phi_x}{ h_R}\left|\tilde{\downarrow}\right> + \cdots
\end{eqnarray}
In terms of the resulting modification to $\chi''_z$, we can easily see that the elastic peak will decrease by a small amount proportional to $\alpha_{x,R}$, and the inelastic part will be modified by a term proportional to $(\alpha_{x,R})^2 \omega/ h_R^2$ (assuming that $\omega/ h_R \ll 1/\alpha_{x,R}$), which in turn will give a correction to $\overline{\chi''_z}$ proportional to $\omega^{\gamma-1+2\alpha_{z,R}}$.
The susceptilibty $\chi_x''$ is simply a delta function time the factor corresponding to $\eqref{nonuniversal matrix element}$:
\begin{eqnarray}
    \chi_x''(\omega) &=& \left(\frac{ h_R}{\omega_c}\right)^{2\alpha_{z,R}} \delta(\omega\pm 2 h_R) \\
    \overline{\chi_x''}(\omega) &\propto& \omega^{\gamma-1+2\alpha_x^z}
\end{eqnarray}
And we can thus conclude that the frequency dependence of the self energy in this case will be 
\begin{eqnarray}
    \Sigma''(\omega)-\Sigma''(0) \propto \frac{|\omega|^{\gamma+2\alpha_{z,R}}}{h_{c,R}^\gamma \omega_c^{2\alpha_{z,R}}}
\end{eqnarray}
Note that when averaging over $ h_R$ we neglect the contribution from TLS whose splitting obeys $\omega/ h_R \gg (\omega_c/ h_R)^{\alpha_{z,R}} \rightarrow  h_R \ll \omega (\omega/\omega_c)^{\alpha_{z,R}/(1-\alpha_{z,R})}$, for which the perturbation theory breaks down.
\section{Derivation of specific heat from internal energy}
\label{specific heat from Uint}
Here we derive Eq.~\eqref{U_el0} that enables us to obtain the specific heat of the model following Ref.~\cite{gasser_greensche_2002}. Considering the Hamiltonian \eqref{Hamiltonian}:
\begin{eqnarray}
    H=\sum_{\alpha}\varepsilon_{\alpha}c_{\alpha}^{\dagger}c_{\alpha}+\sum_{\alpha\beta\gamma,l}g_{\alpha\beta\gamma}^{l}c_{\alpha}^{\dagger}c_{\beta}\sigma_{\gamma}^{l}+\sum_{\gamma}h_{\gamma}\sigma_{\gamma}^{z}.
\end{eqnarray}
The single-particle quantum numbers stand for combinations of momenta and flavor indices and we also supress factors of $M$ and $N$ for brevity. It is useful to introduce an arbitrary retarded fermionic Green's function
\begin{eqnarray}
    \left\langle \left\langle A,B\right\rangle \right\rangle _{t}^{r}\equiv-i\theta\left(t\right)\left\langle \left[A\left(t\right),B\right]_{+}\right\rangle 
\end{eqnarray}
and its Fourier transform $\left\langle \left\langle A,B\right\rangle \right\rangle _{\omega}^{r}$. Here $\left[A,B\right]_{+}$ is the anti-commutator. The advanced Green's function is given by $\left\langle \left\langle A,B\right\rangle \right\rangle _{\omega}^{a}=\left(\left\langle \left\langle A,B\right\rangle \right\rangle _{\omega}^{r}\right)^{*} $where $\left(\cdot\right)^{*}$ denotes complex conjugation. We use the fact that the retarded and advanced Green's functions both obey the equation of motion: 
\begin{eqnarray}
    \omega\left\langle \left\langle A,B\right\rangle \right\rangle _{\omega}=\left\langle \left[A,B\right]_{+}\right\rangle _{\omega}+\left\langle \left\langle \left[A,H\right]_{-},B\right\rangle \right\rangle _{\omega}.
    \label{eomforU1}
\end{eqnarray}
We use Eq.~\eqref{eomforU1} to obtain the equation for motion of the retarded/advanced fermionic Green's function:
\begin{eqnarray}
   \left(\omega-\varepsilon_{\alpha}\right)\left\langle \left\langle c_{\alpha},c_{\alpha}^{\dagger}\right\rangle \right\rangle _{\omega}=1+\sum_{\alpha\beta\gamma,l}g_{\alpha\beta\gamma}^{l}\left\langle \left\langle c_{\beta}\sigma_{\gamma}^{l},c_{\alpha}^{\dagger}\right\rangle \right\rangle _{\omega}. 
   \label{eomforU}
\end{eqnarray}
We proceed to consider the internal energy $U\equiv\left\langle H\right\rangle $: 
\begin{eqnarray}   U=\sum_{\alpha}\varepsilon_{\alpha}\left\langle c_{\alpha}^{\dagger}c_{\alpha}\right\rangle +\sum_{\alpha\beta\gamma,l}g_{\alpha\beta\gamma}^{l}\left\langle c_{\alpha}^{\dagger}c_{\beta}\sigma_{\gamma}^{l}\right\rangle +\sum_{\gamma}h_{\gamma}\left\langle \sigma_{\gamma}^{z}\right\rangle 
\end{eqnarray}
Using the identity (that follows from the spectral representation)
\begin{eqnarray}
    \left\langle AB\right\rangle =i\int_{-\infty}^{\infty}\frac{d\omega}{2\pi}\frac{\left\langle \left\langle A,B\right\rangle \right\rangle _{\omega}^{r}-\left\langle \left\langle A,B\right\rangle \right\rangle _{\omega}^{a}}{e^{\beta\omega}+1},
\end{eqnarray}
and Eq.~\eqref{eomforU} we may express the first two terms in $U$ as
\begin{eqnarray}
\sum_{\alpha}\varepsilon_{\alpha}\left\langle c_{\alpha}^{\dagger}c_{\alpha}\right\rangle +\sum_{\alpha\beta\gamma,l}g_{\alpha\beta\gamma}^{l}\left\langle c_{\alpha}^{\dagger}c_{\beta}\sigma_{\gamma}^{l}\right\rangle =-\int\frac{d\omega}{\pi}\sum_{\alpha}\text{Im}\left\langle \left\langle c_{\alpha},c_{\alpha}^{\dagger}\right\rangle \right\rangle _{\omega}^{r}.
\end{eqnarray}
The right-hand-side can be written in terms of the electronic spectral function: $\int_{\boldsymbol{k}}\mathcal{A}_{\boldsymbol{k}}\left(\omega\right)=-\frac{1}{\pi}\sum_{\alpha}\text{Im}\left\langle \left\langle c_{\alpha},c_{\alpha}^{\dagger}\right\rangle \right\rangle _{\omega}^{r}$. Inserting this form to the expression for the internal energy, we obtain the form given in Eq.~\eqref{U_el0}.

\section{Superconductivity}
\label{appendix superconductivity}

\subsection{Effective action}

We generalize our model to spin-$1/2$ fermions by altering the interaction term in Eq.~\eqref{Hamiltonian} to 
\begin{equation}
H_{{\rm int}}=\frac{1}{{N}}\sum_{\boldsymbol{r},s,ijl}\boldsymbol{g}_{ijl,\boldsymbol{r}}\cdot\boldsymbol{\sigma}_{l,\boldsymbol{r}}c_{i\boldsymbol{r}s}^{\dagger}c_{j\boldsymbol{r}s}.
\end{equation}
Here, $c_{i\boldsymbol{k}\alpha}^{\dagger}$ is the fermionic creation
operator for momentum $\boldsymbol{k},$ spin $s=\uparrow,\downarrow$, and flavor
index $i=1,\cdots,N/2$ (so that the total number of electron flavors remains $N$), and the rest of the definitions are identical to the case in the main text.

After averaging over the coupling constants via replica trick,
introducing the bilocal fields (including the new pairing field $F$)
\begin{eqnarray}
G_{\boldsymbol{r},\boldsymbol{r}'s}\left(\tau,\tau'\right) & = & \frac{1}{N}\sum_{i}\bar{c}_{i\boldsymbol{r}\alpha}\left(\tau\right)c_{i\boldsymbol{r}'s}\left(\tau'\right),\nonumber \\
F_{\boldsymbol{r},\boldsymbol{r}'}\left(\tau,\tau'\right) & = & \frac{1}{N}\sum_{i}c_{i\boldsymbol{r}\downarrow}\left(\tau\right)c_{i\boldsymbol{r}'\uparrow}\left(\tau'\right),\nonumber \\
\chi_{a,\boldsymbol{r}}\left(\tau,\tau'\right) & = & \frac{1}{M}\sum_{l}\sigma_{l,\boldsymbol{r}}^{a}\left(\tau\right)\sigma_{l,\boldsymbol{r}}^{a}\left(\tau'\right),
\end{eqnarray}
and integrating over the fermions, we obtain the effective action:
\begin{eqnarray}
S & = & -N{\rm tr}\log\left(\hat{G}_{0}^{-1}-\hat{\Sigma}\right)-N\int\sum_{\boldsymbol{r},s}G_{\boldsymbol{r},\boldsymbol{r'}s}\left(\tau,\tau'\right)\Sigma_{\boldsymbol{r'},\boldsymbol{r}s}\left(\tau',\tau\right)\nonumber \\
 & - & N\int\sum_{\boldsymbol{r},\sigma}\left(F_{\boldsymbol{r},\boldsymbol{r'}}^{\dagger}\left(\tau,\tau'\right)\Phi_{\boldsymbol{r'},\boldsymbol{r}}\left(\tau',\tau\right)+F_{\boldsymbol{r},\boldsymbol{r'}}\left(\tau,\tau'\right)\Phi_{\boldsymbol{r'},\boldsymbol{r}}^{\dagger}\left(\tau',\tau\right)\right)\nonumber \\
 & - & M\sum_{a}\sum_{\boldsymbol{r}}g_{a}^{2}\int\chi_{a,\boldsymbol{r}}\left(\tau,\tau'\right)\nonumber \\
 & \times & \left[\text{\text{\ensuremath{\sum_{s}}\ensuremath{G_{\boldsymbol{r},\boldsymbol{r}s}\left(\tau,\tau'\right)G_{\boldsymbol{r},\boldsymbol{r}s}\left(\tau',\tau\right)}}}-(-1)^{\delta_{a,y}}2F_{\boldsymbol{r},\boldsymbol{r}}^{\dagger}\left(\tau,\tau'\right)F_{\boldsymbol{r},\boldsymbol{r}}\left(\tau',\tau\right)\right]\nonumber \\
 & + & M\int\sum_{\boldsymbol{r}s}\chi_{a,\boldsymbol{r}}\left(\tau,\tau'\right)\Pi_{a,\boldsymbol{r}}\left(\tau',\tau\right)+\sum_{\boldsymbol{r}}\sum_{l=1}^{M}S_{{\rm TLS}}\left[\boldsymbol{\sigma}_{l,\boldsymbol{r}}\right],\label{eq:action_bilinear}
\end{eqnarray}
where 
\begin{eqnarray}
S_{{\rm TLS}}\left[\boldsymbol{\sigma}\right] & = & S_{{\rm Berry}}\left[\boldsymbol{\sigma}\right]-\int d\tau\boldsymbol{h}_{l,\boldsymbol{r}}\cdot\boldsymbol{\sigma}\left(\tau\right)\nonumber \\
 & - & \int d\tau d\tau'\sum_{a}\Pi_{a,\boldsymbol{r}}\left(\tau'-\tau\right)\sigma^{a}\left(\tau\right)\sigma^{a}\left(\tau'\right)
\end{eqnarray}
is the action of a spin-boson problem with multiple baths. 

In the first term we use a $2\times2$ Nambu-Gor'kov formulation,
sufficient for singlet pairing
\begin{equation}
\hat{\Sigma}_{\boldsymbol{r}\boldsymbol{r'}}\left(\tau,\tau'\right)=\left(\begin{array}{cc}
\Sigma_{\boldsymbol{r}\boldsymbol{r'}\uparrow}\left(\tau,\tau'\right) & \Phi_{\boldsymbol{r}\boldsymbol{r'}}\left(\tau,\tau'\right)\\
\Phi_{\boldsymbol{r}\boldsymbol{r'}}^{\dagger}\left(\tau,\tau'\right) & -\Sigma_{\boldsymbol{r}'\boldsymbol{r}\downarrow}\left(\tau',\tau\right)
\end{array}\right).
\end{equation}
Generalizations to triplet pairing are straightforward but can only
play a role for odd-frequency pairing. We use a similar expression
for the propagator
\begin{equation}
\hat{G}_{\boldsymbol{r}\boldsymbol{r'}}\left(\tau,\tau'\right)=\left(\begin{array}{cc}
G_{\boldsymbol{r}\boldsymbol{r'}\uparrow}\left(\tau,\tau'\right) & F_{\boldsymbol{r}\boldsymbol{r'}}\left(\tau,\tau'\right)\\
F_{\boldsymbol{r}\boldsymbol{r'}}^{\dagger}\left(\tau,\tau'\right) & -G_{\boldsymbol{r}'\boldsymbol{r}\downarrow}\left(\tau',\tau\right)
\end{array}\right).
\end{equation}
The bare propagator in frequency and momentum space is 
\begin{equation}
\hat{G}_{0\boldsymbol{k}}\left(i\omega\right)^{-1}=\left(\begin{array}{cc}
i\omega-\varepsilon_{\boldsymbol{k}} & 0\\
0 & i\omega+\varepsilon_{\boldsymbol{k}}
\end{array}\right).
\end{equation}
In the limit of large $M$ and $N$, with fixed ratio $M/N$, we can
analyze the saddle point limit. We consider a saddle point that does
not break time-reversal symmetry $G_{\boldsymbol{r}\boldsymbol{r'}\uparrow}\left(\tau,\tau'\right)=G_{\boldsymbol{r}\boldsymbol{r'}\downarrow}\left(\tau,\tau'\right)$
and drop the spin index. Performing the variation with respect to
$\hat{\Sigma}$ gives 
\begin{equation}
\hat{G}_{\boldsymbol{r},\boldsymbol{r}'}\left(i\omega\right)=\left.\left(\hat{G}_{0}^{-1}\left(i\omega\right)-\hat{\Sigma}\left(i\omega\right)\right)^{-1}\right|_{\boldsymbol{r},\boldsymbol{r}'}.
\end{equation}
The variation w.r.t. $G$ and $F$ yield
\begin{eqnarray}
\Sigma_{\boldsymbol{r},\boldsymbol{r}'}\left(\tau\right) & = & \delta_{\boldsymbol{r},\boldsymbol{r}'}\frac{M}{N}\sum_{a}g_{a}^{2}G_{\boldsymbol{r},\boldsymbol{r}}\left(\tau\right)\chi_{a,\boldsymbol{r}}\left(\tau\right)\\
\Phi_{\boldsymbol{r},\boldsymbol{r}'}\left(\tau\right) & = & -\delta_{\boldsymbol{r},\boldsymbol{r}'}\frac{M}{N}\sum_{a}(-1)^{\delta_{a,y}}g_{a}^{2}F_{\boldsymbol{r},\boldsymbol{r}}\left(\tau\right)\chi_{a,\boldsymbol{r}}\left(\tau\right)
\end{eqnarray}
These two equations resemble the ones that occur for electrons that
couple to bosonic modes with propagator $\chi_{a,\boldsymbol{r}}\left(\tau\right)$
via a Yukawa coupling. The stationary point that follows from the
variation with respect to $\chi$ is

\begin{eqnarray}
\Pi_{a,\boldsymbol{r}}\left(\tau\right) & = & -2g_{a}^{2}\left[G_{\boldsymbol{r},\boldsymbol{r}}\left(\tau\right)G_{\boldsymbol{r},\boldsymbol{r}}\left(-\tau\right)-(-1)^{\delta_{a,y}}F_{\boldsymbol{r},\boldsymbol{r}}^{\dagger}\left(\tau\right)F_{\boldsymbol{r},\boldsymbol{r}}\left(-\tau\right)\right],
\end{eqnarray}
an expression that is also analogous to the self energy of a bosonic
problem.

The TLS-correlation function $\left\langle \sigma_{l,\boldsymbol{r}}^{a}\left(\tau\right)\sigma_{l,\boldsymbol{r}}^{a}\left(\tau'\right)\right\rangle $
is determined from the solution of the spin boson problem.

\subsection{Linearized Eliashberg equations}

As long as we are only interested in the onset of pairing and the
superconducting phase transition is of second order we can focus on
the linearized gap equation. In this case we can neglect the feedback
of superconductivity on the ohmic bath. The solution of the
spin-boson problem then yields the local propagator $\chi_{a}\left(\omega\right)$.
The equation for the momentum-independent normal self energy is:
\begin{eqnarray}
\Sigma\left(i\omega\right) & = & \frac{M}{N}T\sum_{\omega',a}g_{a}^{2}G\left(i\omega'\right)\chi_a\left(i\omega-i\omega'\right)\nonumber \\
 & = & \frac{M}{N}T\sum_{\omega',a}g_{a}^{2}\rho_{F}\int d\varepsilon_{\boldsymbol{k}}\frac{1}{i\omega'-\varepsilon_{\boldsymbol{k}}-\Sigma\left(\omega'\right)}\chi_{a}\left(i\omega-i\omega'\right).
\end{eqnarray}
The linearized equation for the $s$-wave anomalous self energy is
(assuming particle-hole symmetry for simplicity, we use the fact that $i\Sigma\left(i\omega'\right)$ is real): 
\begin{eqnarray}
\Phi\left(i\omega\right) & = & -\frac{M}{N}T\sum_{\omega',a}g_{a}^{2}(-1)^{\delta_{a,y}}F\left(i\omega'\right)\chi_{a}\left(i\omega-i\omega'\right)\nonumber \\
 & = & \frac{M}{N}T\sum_{\omega',a}\int d\varepsilon_{\boldsymbol{k}}\frac{\rho_{F}g_{a}^{2}(-1)^{\delta_{a,y}}\Phi\left(i\omega'\right)\chi_{a}\left(i\omega-i\omega'\right)}{\left(i\omega'-\varepsilon_{\boldsymbol{k}}-\Sigma\left(i\omega'\right)\right)\left(-i\omega'-\varepsilon_{\boldsymbol{k}}+\Sigma\left(i\omega'\right)\right)}\nonumber \\
 & = & \frac{M}{N}T\sum_{\omega',a}\int d\varepsilon_{\boldsymbol{k}}\frac{\rho_{F}g_{a}^{2}(-1)^{\delta_{a,y}}\Phi\left(i\omega'\right)\chi_{a}\left(i\omega-i\omega'\right)}{\left(i\omega'+i\Sigma\left(i\omega'\right)\right)^{2}+\varepsilon_{\boldsymbol{k}}^{2}}.
\end{eqnarray}
If we perform the integration over $\varepsilon_{\boldsymbol{k}}$,
we get for both self energies the Eliashberg equations.
\begin{eqnarray}
\Sigma\left(i\omega\right) & = & -iT\sum_{\omega'}{\rm sign}\left(i\omega'\right)D_{\Sigma}\left(i\omega-i\omega'\right),\nonumber \\
\Phi\left(i\omega\right) & = & T\sum_{\omega'}\frac{\Phi\left(i\omega'\right)}{\left|\omega'+i\Sigma\left(i\omega'\right)\right|}D_{\Phi}\left(i\omega-i\omega'\right),
\end{eqnarray}
with 
\begin{eqnarray}
D_{\Sigma}\left(i\omega\right) & = & \frac{M}{N}\sum_{a}\rho_{F}g_{a}^{2}\chi_{a}\left(i\omega\right),\nonumber \\
D_{\Phi}\left(i\omega\right) & = & \frac{M}{N}\sum_{a}(-1)^{\delta_{a,y}}\rho_{F}g_{a}^{2}\chi_{a}\left(i\omega\right).
\end{eqnarray}
The contribution due to the coupling $g_{y}$ is pair breaking and
sufficiently large $g_{y}$ can partially or fully destroy superconductivity.

For the solution of the linearized gap equation we introduce 
\begin{equation}
\Delta\left(i\omega_{n}\right)=\frac{\omega_n\Phi\left(i\omega_{n}\right)}{\omega_n+i\Sigma(i\omega_n)},
\end{equation}
 which yields a closed equation 
\begin{equation}
\Delta\left(i\omega_{n}\right)=T\sum_{n'} {\rm sign}(i\omega_{n'})\left(\frac{\Delta\left(i\omega_{n'}\right)}{\omega_{n'}}D_\Sigma(i\omega_n-i\omega_{n'})-\frac{\Delta\left(i\omega_{n}\right)}{\omega_{n}}D_\Phi(i\omega_n-i\omega_{n'}))\right).
\end{equation}
One nicely finds that if  $D_{\Sigma}\left(i\omega\right)=D_{\Phi}\left(i\omega\right)\equiv D(i\omega)$ (
i.e. $g_{y}=0$) 
the zeroth bosonic Matsubara frequency does
not contribute to the solution of the coupled equation. Static fluctuations
are irrelevant for the pairing problem, in agreement with Anderson's
theorem. From now on we will assume that $g_y=0$.

\begin{figure}[t]
\centering
\includegraphics[width=0.9\columnwidth]{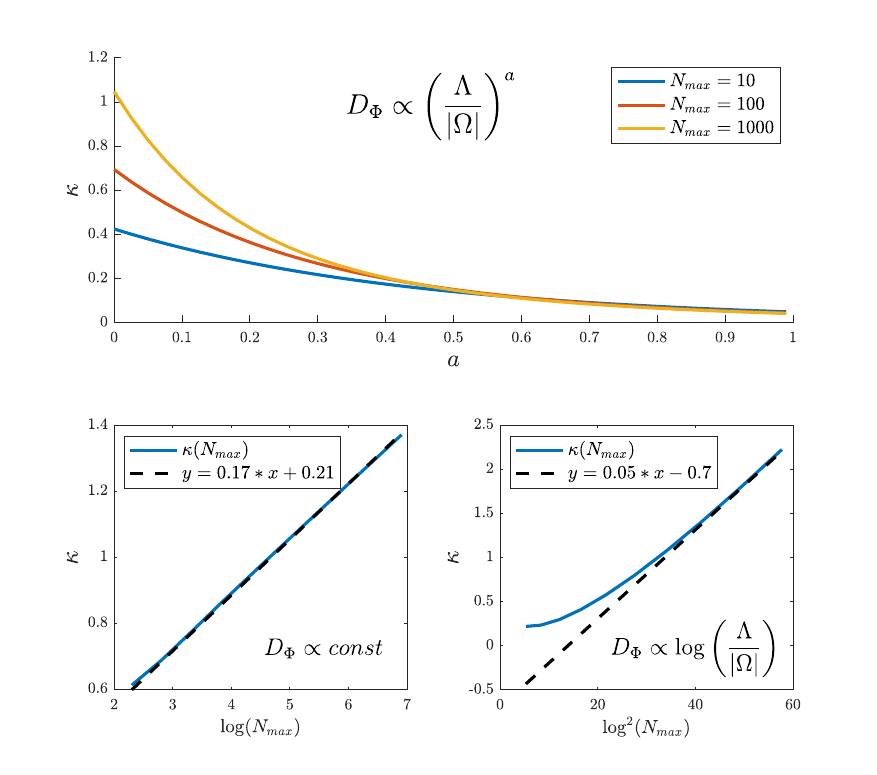}
\caption{Numerical calculation of the largest eigenvalue $\kappa(a,N_{max})$ as a function of $a$ (top) and for the special cases $a=0$ (bottom left) and $a=0'$ (i.e. logarithmic $D_\Phi$) (bottom right). For any finite value of $a$ this approaches a constant as $N_{\max}\to\infty$, while for $a=0,0'$ it scales as $\log(N_max),\log^2(N_max)$ respectively at large $N_{max}$. }
\label{fig:eigenvalue eliashber}
\end{figure}

Hence, in what follows, we can just skip $n'=n$ in the sum. Then
we do not have any problem with a potentially divergent $D\left(0\right)$
for $\gamma\leq1$. 
\begin{equation}
\Delta\left(i\omega_{n}\right)=T\sum_{n'\neq n}\left(\Delta\left(i\omega_{n'}\right)-\frac{\omega_{n'}}{\omega_{n}}\Delta\left(i\omega_{n}\right)\right)\frac{D\left(i\omega_{n}-i\omega_{n'}\right)}{\left|\omega_{n'}\right|}.
\end{equation}
For even frequency pairing we have $\Delta\left(i\omega_{n}\right)=\Delta\left(-i\omega_{n}\right)$.
Hence we can write (we only consider $\omega_{n}>0$)
\begin{eqnarray}
\Delta\left(i\omega_{n}\right) & = & T\sum_{n'\geq0}\Delta\left(i\omega_{n'}\right)\frac{\left(1-\delta_{n,n'}\right)D\left(i\omega_{n}-\omega_{n'}\right)+D\left(i\omega_{n}+i\omega_{n'}\right)}{\omega_{n'}}\nonumber \\
 & - & \Delta\left(i\omega_{n}\right)T\sum_{n'\geq0}\frac{\left(1-\delta_{n,n'}\right)D\left(i\omega_{n}-i\omega_{n'}\right)-D\left(i\omega_{n}+i\omega_{n'}\right)}{\omega_{n}}
\end{eqnarray}
Next we introduce 
\begin{equation}
\Psi_{n}=\frac{\Delta\left(i\omega_{n}\right)}{\left|\omega_{n}\right|^{1/2}}
\end{equation}
and obtain
\begin{eqnarray}
\Psi_{n} & = & T\sum_{n'\geq0}\Psi_{n'}\frac{\left(1-\delta_{n,n'}\right)D\left(i\omega_{n}-i\omega_{n'}\right)+D\left(i\omega_{n}+i\omega_{n'}\right)}{\sqrt{\omega_{n}\omega_{n'}}}\nonumber \\
 & - & \Psi_{n}T\sum_{n'\geq0}\frac{\left(1-\delta_{n,n'}\right)D\left(i\omega_{n}-i\omega_{n'}\right)-D\left(i\omega_{n}+i\omega_{n'}\right)}{\omega_{n}}
\end{eqnarray}
We can write this as a matrix equation, where $N_{\max}$ is the maximum
number of Matsubara frequencies included:
\begin{eqnarray}
    \Psi_n &=& \sum_{n'=0}^{\infty} K_{n,n'}\Psi_n'\\ 
    K_{n,n'} &=& \delta_{n,n'}K_n^{diag}+(1-\delta_{n,n'})\frac{D(i\omega_n-\omega_{n'})+D(i\omega_n+\omega_{n'})}{\pi\sqrt{(2n+1)(2n'+1)}}\\
    K_n^{diag} &=& \sum_{m=0}^{\infty} \frac{(1-\delta_{n,m})D(i\omega_n-\omega_{m})+(1+\delta_{n,m})D(i\omega_n+\omega_{m})}{\pi(2n+1)}
    \end{eqnarray}
Given this matrix equation, we first consider the behavior of $T_{c}$ for a power law form of the pairing propagator (with some high energy cutoff $\Lambda$, so that the Matsubara sum runs up to $N_{max}=\Lambda/T$)
\begin{eqnarray}
D(i\omega_n)=A \left|\frac{\Lambda}{\omega_n}\right|^{a}\equiv A\left(\frac{\Lambda}{T}\right)^{a} d_n 
\end{eqnarray} 
where we introduced the rescaled propagator $d_n= \frac{1}{(2\pi n)^a}$. Since $K_{n,n'}$ is linear in $D$, we may also define its rescaled version $K_{n,n'}= A\left(\frac{\Lambda}{T}\right)^{a} k_{n,n'}$ where $k_{n,n'}$ is defined by replacing $D(i\omega_n)$ with $d_n$ in the definition of $K_{n,n'}$. Defining $\kappa(a,N_{max})$ as the largest eigenvalue of $k_{n,n'}$, the Eliashberg equation simplifies to:
\begin{eqnarray}
\label{Tc from eigenvalue}
    1=A\left(\frac{\Lambda}{T_{c}}\right)^{a}\kappa(a,N_{max})
\end{eqnarray}
The qualitative behavior of the eigenvalue can be estimated by inspecting the diagonal elements of $k$, namely, a simple power counting indicates whether the sum converges or not, which determines the depedence on the cutoff $N_{max}$. In particular, for $a=0$, $\kappa \approx \kappa_0\log(N_{max})$ and we find the BCS solution $T_{c} \propto \Lambda \exp(-1/A\kappa_0)$. For $a>0$ the series converges for $n,n'\to\infty$ and $\kappa=\kappa_a$ depends only on the exponent $a$, giving the critical temperature $T_{c} = \Lambda (A\kappa_a)^{1/a}$. A numerical calculation of $\kappa_a$ is shown in Fig.~\ref{fig:eigenvalue eliashber}.
Finally, consider the case where $D(i\omega_n)=A\log(\Lambda/|\omega_n|)$ (corresponding to a MFL behavior of the electrons). Using $d_n=\log(N_{max}/|n|)$ one can show numerically that $\kappa\approx\kappa_0' \log^2(N_{max})$ such that $T_{c}\propto\Lambda\exp(-1/\sqrt{A\kappa_0})$. An analytic derivation of this result using the Eliashberg equation is given in \cite{Schmalian2005}.
\subsection{Analytic continuation of the bosonic propagator}
We start with the imaginary part of the bosonic propagator, assuming a power-law form:
\begin{equation}
    D''(\omega) = {\rm sign}(\omega)A\left|\frac{\Lambda}{\omega}\right|^a
\end{equation}
 Analytically continuing to Matsubara frequencies (for $a<2$), we obtain
\begin{eqnarray}
D(i\Omega) &=& -\frac{1}{\pi}\int_{-\infty}^{\infty}  \frac{d\omega}{i\Omega-\omega} D''(\omega)\\
    &=& \frac{2}{\pi}\int_{0}^{\infty} \frac{\omega d\omega}{\Omega^2+\omega^2} D''(\omega) \\
    &=& \frac{2A\Lambda^a}{\pi}\int_{0}^{\Lambda} \frac{\omega^{1-a}d\omega}{\Omega^2+\omega^2}  \\ 
    &=& \frac{2A}{(2-a)\pi} \frac{\Lambda^{2}}{|\Omega|^2} {}_2F_1\left(1,-\frac{a}{2},1-\frac{a}{2},-\frac{\Lambda^2}{\Omega^2}\right)
\end{eqnarray}
where ${}_2F_1$ is the Gaussian hypergeometric function. In order to obtain anayltical results, we approximate:
\begin{eqnarray}
    D(i\Omega) &\approx& \frac{2A\Lambda^a}{\pi}\left(\int_{0}^{\Omega}\frac{\omega^{1-a}}{\Omega^2}d\omega + \int_{\Omega}^{\Lambda}\omega^{-1-a}d\omega \right) \\ 
    &=& \frac{2A}{\pi}\left( \frac{1}{2-a} \left|\frac{\Lambda}{\Omega}\right|^a - \frac{1}{a}\left(1-\left|\frac{\Lambda}{\Omega}\right|^a\right) \right)
\end{eqnarray}
which coincides with the appropriate limits of the function ${}_2F_1$. Note that the leading behavior for $|\Omega|\ll \Lambda$ is of the form 
\begin{eqnarray}   D(i\Omega) \approx \frac{2A}{\pi} \times \begin{cases}
        \frac{1}{|a|} & a<0\\
        \log\left(\frac{\Lambda}{|\Omega|}\right) &  a = 0\\
        \frac{2}{a(2-a)} \left|\frac
        {\Lambda}{\Omega}\right|^{a} & 0<a<2
    \end{cases}
\end{eqnarray}
We now use the above to obtain $T_{c}$ in the $x$-model.
\subsection{Detailed analysis of the $x-$model}
We analyze the scaling of $T_{c}$ for two parameter regimes: ``weak coupling'', for which $T_{c}\ll h_{c,R}$,
and ``strong coupling'' for which $T_{c}\gg h_{c,R}$. 
We will find approximate solutions by analytically continuing the TLS susceptibility $\overline{\chi_x}$ and identifying the most singular contribution to $D_\Phi(i\Omega)$, from which we obtain $T_{c}$ using Eq.~\eqref{Tc from eigenvalue}.

\subsubsection{Weak coupling}

Here we study the susceptibility for $T,\omega\ll h_{c,R}$. From
dimensional considerations, we write the averaged susceptibility as
a scaling function:
\begin{eqnarray}
\chi_x''\left(\omega,T\right)=\frac{A_\alpha\gamma}{\omega}\left|\frac{\omega}{h_{c,R}}\right|^{\gamma}\min\left(1,\left(\frac{\omega}{bT}\right)^{\delta}\right)
\end{eqnarray}
with some $\delta>0$ and $b\sim\mathcal{O}(1)$. Analyzing the susceptibility at weak coupling
and at the Toulouse point suggests that $\delta=1$, although as we
will see the exact value of $\delta$ does not qualitatively change $T_{c}$. The corresponding
Matsubara frequency correlator is given by 
\begin{eqnarray}
\chi_x\left(i\Omega_{n}\right) & \approx \frac{2\gamma A_\alpha}{\pi h_{c,R}}\left(
\frac{1}{\gamma-1}-\frac{2}{\gamma^2-1}\left|\frac{\Omega}{h_{c,R}}\right|^{\gamma-1}-\frac{\delta}{(\gamma+1)(\gamma+1+\delta)}\frac{T^{\gamma}}{\Omega^2h_{c,R}^{\gamma-1}}\right)
\end{eqnarray}
The most singular contribution to $D_\Phi$ is therefore (the temperature dependent term is not significant and can be ignored, since $\Omega \gtrsim T$)
\begin{eqnarray}
D_\Phi(i\Omega) \approx 2\pi  A_\alpha \lambda \times \begin{cases}
        \frac{\gamma}{\gamma-1} & \gamma>1\\
        \log\left(\frac{h_{c,R}}{|\Omega|}\right) &  \gamma = 1\\
        \frac{2\gamma}{1-\gamma^2} \left|\frac
        {h_{c,R}}{\Omega}\right|^{1-\gamma} & \gamma<1
    \end{cases}
\end{eqnarray}
Using Eq.\eqref{Tc from eigenvalue} we find that
\begin{eqnarray}
    T_{c}/h_{c,R} \propto  \begin{cases}
        \exp\left(-\frac{\gamma-1}{2\pi A_\alpha \gamma \kappa_0 \lambda}\right) & \gamma>1+\mathcal{O}(\sqrt{\lambda})\\
        \exp\left(-\frac{1}{\sqrt{2\pi A_\alpha \kappa_0' \lambda}}\right) &  \gamma = 1\\
        \left(\frac{4\pi A_\alpha \gamma \kappa_{\gamma-1} }{1-\gamma^2} \lambda\right)^{\frac{1}{1-\gamma}} & \gamma<1-\mathcal{O}(\sqrt{\lambda})
    \end{cases}
\end{eqnarray}
where the requirement $|\gamma-1 |>\mathcal{O}(\sqrt{\lambda})$ in the BCS and quantum critical regimes is necessary for self consistency.
 Additionally, demanding that
$T_{c}\ll h_{c,R}$, which is assumed in taking the low-$T$ form of the TLS-susceptibility, requires $\lambda\ll1$, i.e.
\begin{eqnarray}
1\ll\alpha\frac{M}{N}\frac{E_{F}}{h_{c,R}}\propto\alpha\frac{M}{N}\left(\frac{E_{F}}{h_{c}}\right)^{\frac{1}{1-\alpha}}.
\end{eqnarray}
This condition will always break down at some $\alpha<1$. For $M/N\sim \mathcal{O}(1)$ this will happen at very small values of $\alpha$: $\alpha \propto h_c/E_F$, while in the limit where TLSs are extremely sparse: $\frac{M}{N}\ll\frac{h_{c}}{E_{F}}$,
this occurs at 
$\alpha\approx1-\frac{\log\left(\frac{E_{F}}{h_{c}}\right)}{\log\left(\frac{N}{M}\right)}$.

\subsubsection{Strong coupling}

We now turn to the regime $T\gg h_{c,R}$. Note that for $\alpha>1$ this is always the case since $h_{c,R}=0$.
For frequencies $\omega\ll T$ the TLS correlation function decays exponentially with rate \cite{BrayMoore82},\cite{leggett_dynamics_1987} (Eq 5.29), 
\begin{eqnarray}
\Gamma = c\frac{h^{2}}{T}\left(\frac{T}{E_{F}}\right)^{2\alpha}
\end{eqnarray}
with $c$ some $\alpha$ dependent prefactor. Note that $T\gg \Gamma$ for $T\gg h_R$. In this regime, the TLS-susceptibility can be approximated as
\begin{eqnarray}
    \chi''_x\left(\omega\right)=
    \frac{1}{2\pi}\frac{\Gamma}{T}\frac{\omega}{\Gamma^{2}+\omega^{2}} 
\end{eqnarray}
(The prefactor $\Gamma/T$ is due to the sum rule  Eq.~\eqref{sum rule 1}).
For $\omega\gg T,h_R$ the analysis of App.~\ref{Appendix 3-2a} can be extended for  $\alpha\neq 1$ (or more precisely $|\alpha-1|>h_c/E_F$). Overall for $\omega\gg h_{c,R}$ one finds that 
\begin{eqnarray}
\chi''_x\left(\omega,T\right)=4\alpha\frac{h^{2}}{E_{F}^{2\alpha}}\frac{\left(\max\left(\omega,bT\right)\right)^{2-2\alpha}}{\omega}
\end{eqnarray}
with $b\sim \mathcal{O}(1)$.
Analytically continuing and seperating the different frequency regimes, we define
\begin{eqnarray}
\chi_x\left(i\Omega\right)=\frac{2}{\pi}\left(\int_{0}^{\Gamma}+\int_{\Gamma}^{bT}+\int_{bT}^{\Omega}+\int_{\Omega}^{E_{F}}\right)d\omega\frac{\omega\chi''_x\left(\omega\right)}{\omega^{2}+\Omega^{2}}\equiv\chi_{1}+\chi_{2}+\chi_{3}+\chi_{4}.
\end{eqnarray}
Thus, for $\Omega>aT\gg\Gamma$
\begin{eqnarray}
\chi_{1} & \approx& \frac{1}{3\pi^2}\frac{\Gamma^2}{T\Omega^2} \\ 
\chi_{2} & \approx& \frac{1}{\pi^2} \frac{\Gamma}{\Omega^2}\left(1-\frac{\Gamma}{bT}\right)\\
\chi_{3} & \approx &\frac{2\alpha}{\pi}\frac{h^2}{E_F^{2\alpha}}\frac{|\Omega|^{2\alpha-3}}{2\alpha-1}\left(1-\left(\frac{bT}{|\Omega|}\right)^{2\alpha-1}\right)\\
\chi_{4} & \approx &\frac{2\alpha}{\pi}\frac{h^2}{E_F^3}\frac{1}{2\alpha-3}\left(1-\left(\frac{|\Omega|}{E_F}\right)^{2\alpha-3}\right)
\end{eqnarray}
The most singular contribution to $D_\Phi$ is given by
\begin{eqnarray}
\label{D_phi 3-2a}
    D_\Phi(i\Omega) = 2\pi \epsilon \alpha^2 b_\beta \times \begin{cases}
        \frac{2\alpha}{1-2\alpha}\frac{E_F^{3-2\alpha}}{T^{1-2\alpha}\Omega^2} & \alpha<1/2 \\
        \frac{E_F^2}{\Omega^2}\log\left(\frac{|\Omega|}{T}\right) & \alpha = 1/2 \\
        \frac{2}{(2\alpha-1)(3-2\alpha)} \left|\frac{E_F}{\Omega}\right|^{3-2\alpha} & 1/2<\alpha<3/2 
        \\
        \log \left(\frac{E_F}{|\Omega|}\right) & \alpha = 3/2 \\
        \frac{1}{2\alpha-3} & \alpha>3/2
    \end{cases}
\end{eqnarray}
where $b_\beta=\left(\frac{1+\beta}{3+\beta}\right)$ comes from averaging over $h^2$. Inserting these into Eq.~\eqref{Tc from eigenvalue}, we obtain
\begin{eqnarray}
    T_{c}/E_F \propto  \begin{cases}
    \left(\frac{4\pi\alpha^2\kappa_{3-2\alpha}b_\beta}{(2\alpha-1)(3-2\alpha)}\epsilon\right)^{\frac{1}{3-2\alpha}} & \alpha <3/2 - \mathcal{O}(\sqrt{\epsilon}) \\
    \exp\left(-\frac{1}{\alpha\sqrt{2\pi b_\beta \kappa'_0 \epsilon}}\right) & \alpha = 3/2 \\
    \exp\left(-\frac{2\alpha-3}{2\pi\alpha^2\beta\kappa_0\epsilon}\right) & \alpha > 3/2 + \mathcal{O}(\sqrt{\epsilon})
    \end{cases}
\end{eqnarray}
Note that for $\alpha\leq 1/2$ the prefactor in the parentheses changes, according to the corresponding expression in Eq.~\eqref{D_phi 3-2a}. However, the dependence on the small parameter $\epsilon$ remains $\epsilon^{1/(3-2\alpha)}$ for all $\alpha<3/2$.

Once again, for $\alpha<1$ consistency requires that $T_{c}\gg h_{c,R}$,
which translates into 
\begin{eqnarray}
1\ll\alpha\frac{M}{N}\left(\frac{E_{F}}{h_{c}}\right)^{\frac{1}{1-\alpha}}\iff\lambda\gg1
\end{eqnarray}
which is complementary to the requirement for weak coupling. 

\end{document}